%% file: main.tex
\documentclass[twocolumn,tighten]{aastex631}
\usepackage{CJKutf8}
\usepackage{amsmath}
\usepackage{enumitem}
\usepackage{booktabs,pbox}
\usepackage{bm}

\graphicspath{{./figures/}{../figures/}{./images/}{../images/}}

\newcommand{\sbullet}{\mathbin{\vcenter{\hbox{\scalebox{0.7}{$\bullet$}}}}}

\newcommand{\sbsc}[1]{_\mathrm{#1}}
\newcommand{\spsc}[1]{^\mathrm{#1}}
\newcommand{\brkt}[1]{\left<#1\right>}

\DeclareRobustCommand{\ion}[2]{%
\relax\ifmmode
\ifx\testbx\f@series
{\mathbf{#1\,\mathsc{#2}}}\else
{\mathrm{#1\,\mathsc{#2}}}\fi
\else\textup{#1\,{\mdseries\textsc{#2}}}%
\fi}

\newcommand{\lasso}{\textit{lasso}}

\newcommand{\HI}{\ion{H}{i}}
\newcommand{\HII}{\ion{H}{ii}}
\newcommand{\Htwo}{\mbox{$\mathrm{H}_2$}}
\newcommand{\CO}[2]{\mbox{$\mathrm{CO}\,(#1\text{--}#2)$}}

\newcommand{\Mstar}{M_\star}
\newcommand{\Rdisk}{r\sbsc{disk}}
\newcommand{\Reff}{r\sbsc{e}}
\newcommand{\Riso}{r\sbsc{25}}

\newcommand{\rgal}{r\sbsc{gal}}
\newcommand{\Vcirc}{V\sbsc{circ}}
\newcommand{\Omegacirc}{\Omega\sbsc{circ}}
\newcommand{\OortA}{A\sbsc{Oort}}

\newcommand{\Rsub}[1]{R\sbsc{#1}}

\newcommand{\LCOsub}[1]{L\sbsc{CO,\,#1}}

\newcommand{\Msub}[1]{M\sbsc{#1}}

\newcommand{\rhosub}[1]{\rho\sbsc{#1}}
\newcommand{\ICO}{I\sbsc{CO}}
\newcommand{\ICOsub}[1]{I\sbsc{CO,\,#1}}

\newcommand{\sigCOsub}[1]{\sigma\sbsc{CO,\,#1}}
\newcommand{\Sigmol}{\Sigma\sbsc{mol}}
\newcommand{\Sigsub}[1]{\Sigma\sbsc{#1}}
\newcommand{\sigmol}{\sigma\sbsc{mol}}
\newcommand{\sigsub}[1]{\sigma\sbsc{#1}}
\newcommand{\Pturb}{P\sbsc{turb}}
\newcommand{\Pturbsub}[1]{P\sbsc{turb,\,#1}}
\newcommand{\alphavir}{\alpha\sbsc{vir}}
\newcommand{\alphavirsub}[1]{\alpha\sbsc{vir,\,#1}}
\newcommand{\alphaCO}{\alpha\sbsc{CO}}
\newcommand{\alphaCOline}[2]{\alpha\sbsc{CO(#1\text{--}#2)}}

\newcommand{\Ihi}{I\sbsc{21cm}}
\newcommand{\Sigatom}{\Sigma\sbsc{atom}}

\newcommand{\Siggas}{\Sigma\sbsc{gas}}

\newcommand{\Iwiseone}{I\sbsc{3.4\,\mu m}}
\newcommand{\Iiracone}{I\sbsc{3.6\,\mu m}}
\newcommand{\MtoLwiseone}{\Upsilon_{3.4\,\mu m}}
\newcommand{\Sigstar}{\Sigma\sbsc{\star}}

\newcommand{\fgas}{f\sbsc{gas}}
\newcommand{\fmol}{f\sbsc{mol}}
\newcommand{\tdepsub}[1]{t\sbsc{dep,\,#1}}

\newcommand{\SigSFR}{\Sigma\sbsc{SFR}}

\newcommand{\uI}{\mbox{$\rm MJy~sr^{-1}$}}
\newcommand{\uIco}{\mbox{$\rm K~km~s^{-1}$}}
\newcommand{\uLco}{\mbox{$\rm K~km~s^{-1}~pc^2$}}
\newcommand{\uIhi}{\mbox{$\rm K~km~s^{-1}$}}
\newcommand{\uIha}{\mbox{$\rm erg~s^{-1}\ cm^{-2}\ arcsec^{-2}$}}

\newcommand{\ualphaCO}{\mbox{$\rm M_\odot~pc^{-2}\ (K~km~s^{-1})^{-1}$}}
\newcommand{\uV}{\mbox{$\rm km~s^{-1}$}}
\newcommand{\uM}{\mbox{$\rm M_\odot$}}
\newcommand{\uSig}{\mbox{$\rm M_\odot~pc^{-2}$}}

\newcommand{\uSigSFR}{\mbox{$\rm M_\odot~yr^{-1}~kpc^{-2}$}}
\newcommand{\uP}{\mbox{$\rm K~cm^{-3}$}}
\newcommand{\ut}{\mbox{$\rm Myr$}}
\newcommand{\uOmega}{\mbox{$\rm Gyr^{-1}$}}
\newcommand{\uOortA}{\mbox{$\rm km~s^{-1}~kpc^{-1}$}}

\newcommand{\Dbeam}{D\sbsc{beam}}
\newcommand{\Abeam}{A\sbsc{beam}}

\newcommand{\Farea}{f\sbsc{area}}
\newcommand{\Fflux}{f\sbsc{flux}}

\newcommand{\galnumtot}{80}
\newcommand{\galnumhi}{53}
\newcommand{\galnumirac}{61}
\newcommand{\galnumha}{60}
\newcommand{\galnumrc}{62}
\newcommand{\galnumfin}{42}
\newcommand{\apernumtot}{$46{,}628$}
\newcommand{\apernumwpix}{$3{,}383$}
\newcommand{\apernumwpixdisk}{$2{,}724$}
\newcommand{\apernumwobj}{$2{,}784$}
\newcommand{\apernumfin}{$871$}

\newcommand{\apersize}{1.5~kpc}
\newcommand{\aperarea}{1.95~kpc$^2$}
\newcommand{\annuluswidth}{500~pc}

\shorttitle{Molecular Clouds in Environmental Context}
\shortauthors{Sun et PHANGS Collaboration}

\begin{document}


\title{Molecular Cloud Populations in the Context of Their Host Galaxy Environments:\\ A Multiwavelength Perspective}



\newcommand{\McMaster}{\affiliation{Department of Physics and Astronomy, McMaster University, 1280 Main Street West, Hamilton, ON L8S 4M1, Canada}}

\newcommand{\CITA}{\affiliation{Canadian Institute for Theoretical Astrophysics (CITA), University of Toronto, 60 St George Street, Toronto, ON M5S 3H8, Canada}}

\newcommand{\OSU}{\affiliation{Department of Astronomy, The Ohio State University, 140 West 18th Avenue, Columbus, OH 43210, USA}}

\newcommand{\CCAPP}{\affiliation{Center for Cosmology and Astroparticle Physics (CCAPP), 191 West Woodruff Avenue, Columbus, OH 43210, USA}}

\newcommand{\Alberta}{\affiliation{Department of Physics, University of Alberta, Edmonton, AB T6G 2E1, Canada}}

\newcommand{\ANU}{\affiliation{Research School of Astronomy and Astrophysics, Australian National University, Canberra, ACT 2611, Australia}}

\newcommand{\ASIAA}{\affiliation{Institute of Astronomy and Astrophysics, Academia Sinica, No. 1, Sec. 4, Roosevelt Road, Taipei 10617, Taiwan}}

\newcommand{\ASTROThreeD}{\affiliation{ARC Centre of Excellence for All Sky Astrophysics in 3 Dimensions (ASTRO 3D), Australia}}

\newcommand{\Carnegie}{\affiliation{Observatories of the Carnegie Institution for Science, 813 Santa Barbara Street, Pasadena, CA 91101, USA}}

\newcommand{\CfA}{\affiliation{Center for Astrophysics $\mid$ Harvard \& Smithsonian, 60 Garden Street, Cambridge, MA 02138, USA}}

\newcommand{\CITEVA}{\affiliation{Centro de Astronomía (CITEVA), Universidad de Antofagasta, Avenida Angamos 601, Antofagasta, Chile}}

\newcommand{\CNRS}{\affiliation{CNRS, IRAP, 9 Av. du Colonel Roche, BP 44346, F-31028 Toulouse cedex 4, France}}

\newcommand{\ESO}{\affiliation{European Southern Observatory, Karl-Schwarzschild Stra{\ss}e 2, D-85748 Garching bei M\"{u}nchen, Germany}}

\newcommand{\Heidelberg}{\affiliation{Astronomisches Rechen-Institut, Zentrum f\"{u}r Astronomie der Universit\"{a}t Heidelberg, M\"{o}nchhofstra\ss e 12-14, D-69120 Heidelberg, Germany}}

\newcommand{\ICRAR}{\affiliation{International Centre for Radio Astronomy Research, University of Western Australia, 35 Stirling Highway, Crawley, WA 6009, Australia}}

\newcommand{\INAF}{\affiliation{INAF -- Osservatorio Astrofisico di Arcetri, Largo E. Fermi 5, I-50157, Firenze, Italy}}

\newcommand{\IPAC}{\affiliation{Caltech-IPAC, 1200 E. California Blvd. Pasadena, CA 91125, USA}}

\newcommand{\IPARC}{\affiliation{Instituto de F\'{\i}sica de Part\'{\i}culas y del Cosmos IPARCOS, Facultad de Ciencias F\'{\i}sicas, Universidad Complutense de Madrid, E-28040, Spain}}

\newcommand{\IRAM}{\affiliation{Institut de Radioastronomie Millim\'{e}trique (IRAM), 300 Rue de la Piscine, F-38406 Saint Martin d'H\`{e}res, France}}

\newcommand{\ITA}{\affiliation{Universit\"{a}t Heidelberg, Zentrum f\"{u}r Astronomie, Institut f\"{u}r Theoretische Astrophysik, Albert-Ueberle-Str 2, D-69120 Heidelberg, Germany}}

\newcommand{\IWR}{\affiliation{Universit\"{a}t Heidelberg, Interdisziplin\"{a}res Zentrum f\"{u}r Wissenschaftliches Rechnen, Im Neuenheimer Feld 205, D-69120 Heidelberg, Germany}}

\newcommand{\JHU}{\affiliation{Department of Physics and Astronomy, The Johns Hopkins University, Baltimore, MD 21218, USA}}

\newcommand{\LAM}{\affiliation{Aix Marseille Univ, CNRS, CNES, LAM (Laboratoire d’Astrophysique de Marseille), Marseille, France}}

\newcommand{\Leiden}{\affiliation{Leiden Observatory, Leiden University, P.O. Box 9513, 2300 RA Leiden, The Netherlands}}

\newcommand{\Liverpool}{\affiliation{Astrophysics Research Institute, Liverpool John Moores University, IC2, Liverpool Science Park, 146 Brownlow Hill, Liverpool L3 5RF, UK}}

\newcommand{\Maryland}{\affiliation{Department of Astronomy, University of Maryland, College Park, MD 20742, USA}}

\newcommand{\MPE}{\affiliation{Max-Planck-Institut f\"{u}r extraterrestrische Physik, Giessenbachstra{\ss}e 1, D-85748 Garching, Germany}}

\newcommand{\MPIA}{\affiliation{Max-Planck-Institut f\"{u}r Astronomie, K\"{o}nigstuhl 17, D-69117, Heidelberg, Germany}}

\newcommand{\Nagoya}{\affiliation{Department of Physics, Nagoya University, Furo-cho, Chikusa-ku, Nagoya, Aichi 464-8602, Japan}}

\newcommand{\NAOJ}{\affil{National Astronomical Observatory of Japan, 2-21-1 Osawa, Mitaka, Tokyo, 181-8588, Japan}}

\newcommand{\Nichidai}{\affil{Department of Physics, General Studies, College of Engineering, Nihon University, 1 Nakagawara, Tokusada, Tamuramachi, Koriyama, Fukushima, 963-8642, Japan}}

\newcommand{\NRAO}{\affiliation{National Radio Astronomy Observatory, 520 Edgemont Road, Charlottesville, VA 22903-2475, USA}}

\newcommand{\OAN}{\affiliation{Observatorio Astron\'{o}mico Nacional (IGN), C/Alfonso XII, 3, E-28014 Madrid, Spain}}

\newcommand{\ObsParis}{\affiliation{Sorbonne Universit\'{e}, Observatoire de Paris, Universit\'{e} PSL, CNRS, LERMA, F-75014, Paris, France}}

\newcommand{\Princeton}{\affiliation{Department of Astrophysical Sciences, Princeton University, Princeton, NJ 08544 USA}}

\newcommand{\STScI}{\affiliation{Space Telescope Science Institute, 3700 San Martin Drive, Baltimore, MD 21218, USA}}

\newcommand{\Sydney}{\affiliation{Sydney Institute for Astronomy, School of Physics A28, The University of Sydney, NSW 2006, Australia}}

\newcommand{\TAPIR}{\affil{TAPIR, California Institute of Technology, Pasadena, CA 91125, USA}}

\newcommand{\TKU}{\affiliation{Department of Physics, Tamkang University, No.151, Yingzhuan Rd., Tamsui Dist., New Taipei City 251301, Taiwan}}

\newcommand{\Toulouse}{\affiliation{Universit\'{e} de Toulouse, UPS-OMP, IRAP, F-31028 Toulouse cedex 4, France}}

\newcommand{\Toledo}{\affiliation{University of Toledo, 2801 W. Bancroft St., Mail Stop 111, Toledo, OH 43606, USA}}

\newcommand{\UBonn}{\affiliation{Argelander-Institut f\"ur Astronomie, Universit\"at Bonn, Auf dem H\"ugel 71, 53121 Bonn, Germany}}

\newcommand{\UChile}{\affiliation{Departamento de Astronom\'{i}a, Universidad de Chile, Camino del Observatorio 1515, Las Condes, Santiago, Chile}}

\newcommand{\UCM}{\affiliation{Departamento de F\'{\i}sica de la Tierra y Astrof\'{\i}sica, Universidad Complutense de Madrid, E-28040, Spain}}

\newcommand{\UCSD}{\affiliation{Center for Astrophysics and Space Sciences, Department of Physics,  University of California, San Diego, 9500 Gilman Drive, La Jolla, CA 92093, USA}}

\newcommand{\UGent}{\affiliation{Sterrenkundig Observatorium, Universiteit Gent, Krijgslaan 281 S9, B-9000 Gent, Belgium}}

\newcommand{\UHawaii}{\affiliation{Institute for Astronomy, University of Hawaii, 2680 Woodlawn Drive, Honolulu, HI 96822, USA}}

\newcommand{\ULyon}{\affiliation{Univ Lyon, Univ Lyon 1, ENS de Lyon, CNRS, Centre de Recherche Astrophysique de Lyon UMR5574, F-69230 Saint-Genis-Laval, France}}

\newcommand{\UMass}{\affiliation{University of Massachusetts—Amherst, 710 N. Pleasant Street, Amherst, MA 01003, USA}}

\newcommand{\UWyoming}{\affiliation{Department of Physics and Astronomy, University of Wyoming, Laramie, WY 82071, USA}}


\correspondingauthor{Jiayi~Sun}
\email{sun208@mcmaster.ca}


\suppressAffiliations

\author[0000-0003-0378-4667]{Jiayi~Sun \begin{CJK*}{UTF8}{gbsn}(孙嘉懿)\end{CJK*}}
\altaffiliation{CITA National Fellow}
\McMaster
\CITA
\OSU

\author[0000-0002-2545-1700]{Adam~K.~Leroy}
\OSU
\CCAPP

\author[0000-0002-5204-2259]{Erik~Rosolowsky}
\Alberta

\author[0000-0002-9181-1161]{Annie~Hughes}
\CNRS
\Toulouse

\author[0000-0002-3933-7677]{Eva~Schinnerer}
\MPIA

\author{Andreas~Schruba}
\MPE

\author[0000-0001-9605-780X]{Eric~W.~Koch}
\CfA

\author[0000-0003-4218-3944]{Guillermo~A.~Blanc}
\Carnegie
\UChile

\author[0000-0003-2551-7148]{I-Da~Chiang \begin{CJK*}{UTF8}{bkai}(江宜達)\end{CJK*}}
\ASIAA
\UCSD

\author[0000-0002-9768-0246]{Brent~Groves}
\ANU

\author[0000-0001-9773-7479]{Daizhong~Liu}
\MPIA

\author[0000-0002-6118-4048]{Sharon~Meidt}
\UGent

\author[0000-0002-1370-6964]{Hsi-An~Pan}
\TKU

\author[0000-0003-3061-6546]{J\'er\^ome~Pety}
\IRAM
\ObsParis

\author[0000-0002-0472-1011]{Miguel~Querejeta}
\OAN

\author[0000-0002-2501-9328]{Toshiki~Saito}
\Nichidai
\NAOJ

\author[0000-0002-4378-8534]{Karin~Sandstrom}
\UCSD

\author[0000-0002-5783-145X]{Amy~Sardone}
\OSU
\CCAPP

\author[0000-0003-1242-505X]{Antonio~Usero}
\OAN

\author[0000-0003-4161-2639]{Dyas~Utomo}
\NRAO
\OSU

\author[0000-0002-0012-2142]{Thomas~G.~Williams}
\MPIA

\author[0000-0003-0410-4504]{Ashley~T.~Barnes}
\UBonn

\author[0000-0003-4826-9079]{Samantha~M.~Benincasa}
\OSU
\CCAPP

\author[0000-0003-0166-9745]{Frank~Bigiel}
\UBonn

\author[0000-0002-5480-5686]{Alberto~D.~Bolatto}
\Maryland

\author[0000-0003-0946-6176]{M\'ed\'eric~Boquien}
\CITEVA

\author[0000-0002-5635-5180]{M\'elanie~Chevance}
\Heidelberg
\ITA

\author[0000-0002-5782-9093]{Daniel~A.~Dale}
\UWyoming

\author[0000-0003-1943-723X]{Sinan~Deger}
\TAPIR

\author[0000-0002-6155-7166]{Eric~Emsellem}
\ESO
\ULyon

\author[0000-0001-6708-1317]{Simon~C.~O.~Glover}
\ITA

\author[0000-0002-3247-5321]{Kathryn~Grasha}
\altaffiliation{ARC DECRA Fellow}
\ANU
\ASTROThreeD

\author[0000-0001-9656-7682]{Jonathan~D.~Henshaw}
\MPIA
\Liverpool

\author[0000-0002-0560-3172]{Ralf~S.~Klessen}
\ITA
\IWR

\author[0000-0001-6551-3091]{Kathryn~Kreckel}
\Heidelberg

\author[0000-0002-8804-0212]{J.~M.~Diederik~Kruijssen}
\Heidelberg

\author[0000-0002-0509-9113]{Eve C.~Ostriker}
\Princeton

\author[0000-0002-8528-7340]{David~A.~Thilker}
\JHU

\collaboration{38}{as part of the PHANGS collaboration}


\begin{abstract}
We present a rich, multiwavelength, multiscale database built around the PHANGS--ALMA $\mathrm{CO}\,(2\text{--}1)$ survey and ancillary data.
We use this database to present the distributions of molecular cloud populations and sub-galactic environments in 80 PHANGS galaxies, to characterize the relationship between population-averaged cloud properties and host galaxy properties, and to assess key timescales relevant to molecular cloud evolution and star formation.
We show that PHANGS probes a wide range of kpc-scale gas, stellar, and star formation rate (SFR) surface densities, as well as orbital velocities and shear.
The population-averaged cloud properties in each aperture correlate strongly with both local environmental properties and host galaxy global properties.
Leveraging a variable selection analysis, we find that the kpc-scale surface densities of molecular gas and SFR tend to possess the most predictive power for the population-averaged cloud properties.
Once their variations are controlled for, galaxy global properties contain little additional information, which implies that the apparent galaxy-to-galaxy variations in cloud populations are likely mediated by kpc-scale environmental conditions.
We further estimate a suite of important timescales from our multiwavelength measurements.
The cloud-scale free-fall time and turbulence crossing time are ${\sim}5{-}20$~Myr, comparable to previous cloud lifetime estimates.
The timescales for orbital motion, shearing, and cloud--cloud collisions are longer, ${\sim}100$~Myr.
The molecular gas depletion time is $1{-}3$~Gyr and shows weak to no correlations with the other timescales in our data.
We publish our measurements online and expect them to have broad utility to future studies of molecular clouds and star formation.
\newpage
\end{abstract}




\section{Introduction} \label{sec:intro}

Molecular clouds are deeply integrated with their host galaxies by a number of intertwined physical processes.
The gas distribution, gravitational potential, radiation field, and feedback-driven flows in the host galaxy regulate molecular cloud formation and evolution \citep{Dobbs_etal_2014,Ballesteros-Paredes_etal_2020}.
The internal structure and dynamical properties of these clouds in turn set the initial conditions for star formation, which over time reshapes the matter and radiation distribution in the galaxy \citep{McKee_Ostriker_2007,Padoan_etal_2014,Klessen_Glover_2016,Girichidis_etal_2020}.
These complex interactions lead to strong, observable correlations between the properties of molecular clouds and the local and global properties of the host galaxy.
Characterizing these cloud--environment correlations is thus a promising avenue for understanding the physics governing molecular cloud evolution and, consequently, star formation and galaxy evolution.

Observations of molecular clouds in our Galaxy and a number of nearby galaxies have identified various empirical trends manifesting such cloud--environment correlations.
Within a galaxy, molecular clouds located closer to the galaxy center appear denser, more massive, and more turbulent \citep[e.g.,][also see \citealt{Heyer_Dame_2015}]{Oka_etal_2001,Colombo_etal_2014a,Freeman_etal_2017,Hirota_etal_2018,Miura_etal_2018,Brunetti_etal_2021}.
Similar trends have been found in galaxy-scale numerical simulations \citep[e.g.,][]{Pan_etal_2015a,Jeffreson_etal_2020,Tress_etal_2020b}.
Recent observational works also report that more massive and actively star-forming galaxies tend to host clouds with typically larger sizes, masses, surface densities, and velocity dispersions \citep[][but see \citealt{Bolatto_etal_2008,Fukui_Kawamura_2010,DonovanMeyer_etal_2013}]{Hughes_etal_2013a,Leroy_etal_2015a,Leroy_etal_2016,Schruba_etal_2019,Sun_etal_2020a}.

To proceed from the existing empirical knowledge to more concrete understandings of the cloud--environment correlations, major advances on two issues are necessary.
First, the characterization of environmental dependence often stops at a qualitative level in the molecular cloud literature: the ``environments'' are commonly defined in crude, categorical ways (e.g., galaxy centers, stellar bars, spiral arms), and they are merely considered as a secondary, moderating factor on the scaling relations followed by molecular clouds.
To better understand the underlying physics, a more direct approach would be to quantify the dependence of molecular cloud properties on a set of quantitative ``environmental metrics,'' such as the local gas and stellar mass surface density, star formation rate (SFR) surface density, and orbital shear \citep[e.g.,][]{Hughes_etal_2013a,Schruba_etal_2019}.
Second, many previous works (especially earlier ones) had to rely on observations in a small number of galaxies or sub-galactic regions, and thus only probed a limited range of host galaxy properties.
While such case studies could yield unique insights for specific targets, only systematic surveys across large galaxy samples can cover a wide, continuous range of host galaxy properties, produce representative population statistics, and make meaningful connections to galaxy evolution models.

The PHANGS--ALMA survey \citep{Leroy_etal_2021a} was designed to address both of these issues.
This survey provides sensitive, high resolution, wide field-of-view \CO21 imaging data for ${\sim}90$ nearby, high-mass, star-forming galaxies.
With these galaxies sampled uniformly along the star-forming main sequence, PHANGS--ALMA enables systematic studies of giant molecular clouds (GMCs; $M \gtrsim 10^5~\uM$) across an array of environments where most stars form in the local universe.
Furthermore, a rich set of multiwavelength ancillary data furnishes a multifaceted depiction of these host galaxies, making it possible to study their molecular cloud populations in full environmental context.

Indeed, one of the core science goals that motivated the PHANGS--ALMA survey was to characterize the dependence of molecular cloud populations on global and local galaxy properties.
Studies on this data set have presented population statistics for key molecular cloud properties such as mass, size, surface density, velocity dispersion, and virial parameter \citep[A.~Hughes et al.\ in preparation]{Sun_etal_2018,Sun_etal_2020a,Rosolowsky_etal_2021}.
They also noted significant variations among galaxies and across morphological regions within galaxies (e.g., centers, bars).
In the direct predecessor of this paper, \citet{Sun_etal_2020b} conducted a joint analysis on the PHANGS--ALMA CO data and multiwavelength ancillary data.
They showed that the variations in molecular gas turbulent pressure can be attributed to the dynamical balance between gravity and internal/\linebreak[0]{}external pressure in the gas, as previously argued in Galactic and extragalactic molecular cloud studies \citep[e.g.,][]{Field_etal_2011,Hughes_etal_2013a,Schruba_etal_2019}.

In this paper, we directly address this core science goal of the PHANGS--ALMA survey.
We build on a cross-spatial-scale analysis framework used by \citet{Sun_etal_2020b} and calculate population statistics for the molecular cloud properties measured in \citet{Sun_etal_2018,Sun_etal_2020a,Rosolowsky_etal_2021}; and A.~Hughes et al.\ (in preparation).
We further cross-match them with a large suite of environmental metrics depicting the local gas and stellar mass distribution, orbital kinematics, morphological structures, and star formation activities in the host galaxy.
This allows us to (1) present the full range of cloud populations and host galaxy environments captured by PHANGS--ALMA, (2) delineate the quantitative relationships between cloud characteristics (e.g., mass, surface density, velocity dispersion) and environmental metrics (e.g., gas, star, and SFR surface densities), and (3) identify a subset of relationships that carry unique explanatory/predictive power among all the observed cloud--environment correlations.

Another goal of this paper is to present a set of machine-readable data tables that consolidate all the aforementioned measurements from PHANGS-ALMA and ancillary surveys.
These high-level data products have already been used in a number of studies.
\citet{Herrera_etal_2020} and \citet{Barnes_etal_2021} have utilized previous versions of these tables to quantitatively assess several physical mechanisms relevant to molecular cloud and \HII\ region evolution, including stellar feedback and pressure balance.
Several ongoing works also rely on these tables to calculate the star formation efficiency and its link to small-scale turbulence, orbital shear, and disk instabilities (E.~Rosolowsky et al.\ in preparation; J.~Sun et al.\ in preparation; T.~Williams et al.\ in preparation).
In this paper, we also utilize the measurements in these tables to calculate a suite of characteristic timescales related to the gravity, turbulent motions, orbital motions, and star formation rate of molecular clouds.
The ratios among these timescales provide unique constraints on the viable mechanisms regulating molecular cloud evolution and star formation \citep[also see e.g.,][]{Wong_2009,Jeffreson_Kruijssen_2018,Kruijssen_etal_2019b,Chevance_etal_2020a,Chevance_etal_2020b,KimJY_etal_2020}.

The structure of this paper is as follows.
Section~\ref{sec:data} describes our galaxy sample and the sources of all data we use.
Section~\ref{sec:analysis} elaborates the cross-spatial-scale analysis framework we use to assemble the multiwavelength measurements into a coherent data structure.
Section~\ref{sec:stats} presents the distribution of various molecular cloud population properties and sub-galactic environmental properties measured in this work, and Section~\ref{sec:corr} characterizes the correlations between these two types of measurements.
To demonstrate an application scenario of our rich multiwavelength measurements, Section~\ref{sec:timescales} presents a set of characteristic timescales relevant to molecular cloud evolution and star formation.
Finally, Section~\ref{sec:summary} summarizes all our findings.


\section{Data} \label{sec:data}

\begin{figure*}[htbp]
\gridline{\fig{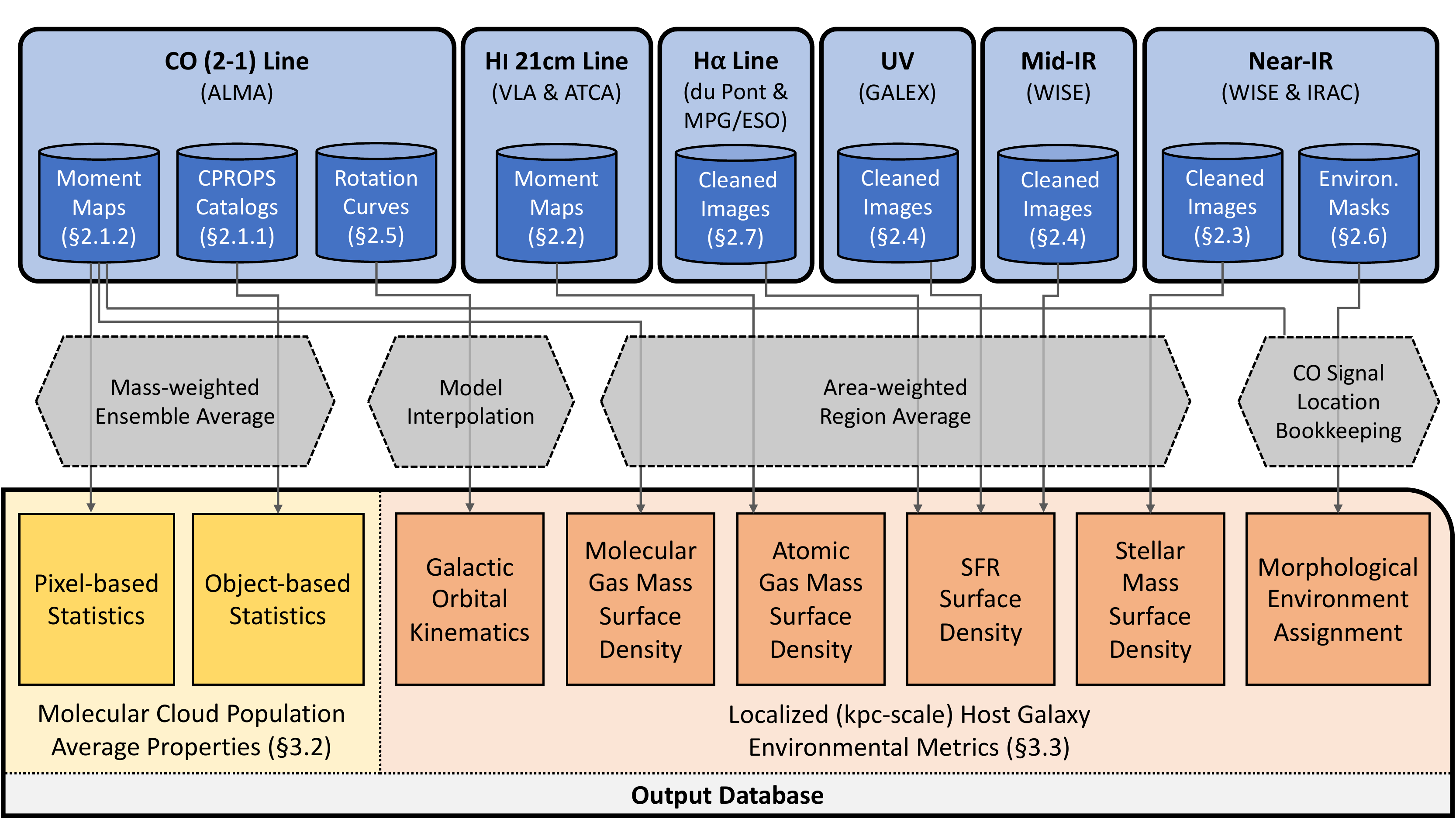}{0.99\textwidth}{}}
\vspace{-2\baselineskip}
\caption{
Schematics of the data sources, aggregation methods, and the derived physical quantities.
For easier navigation within this paper, we also note the section number relevant to each input data product and output quantities.}
\vspace{0.5\baselineskip}
\label{fig:schematic}
\end{figure*}

In this paper, we focus on a sample of \galnumtot\ galaxies (see Table~\ref{tab:sample}) selected from the full PHANGS--ALMA survey sample \citep[90 galaxies; see][]{Leroy_etal_2021a}.
We select these galaxies according to two criteria:
(1) their PHANGS--ALMA \CO21 observations have beam full-width-half-maximum (FWHM) sizes corresponding to physical scales of $150$~pc or smaller, so that each beam roughly probes a GMC-sized molecular gas structure; and
(2) they are not too heavily inclined ($i \lesssim 75^\circ$), so that we can unambiguously determine the locations of molecular clouds in the host galaxy.
A subset of 28 galaxies in this sample already appeared in \citet{Sun_etal_2020b}, where they utilized earlier versions of the same observational data sets and data analysis infrastructure.

In addition to the PHANGS--ALMA CO data, our target galaxies have abundant multiwavelength coverage, including radio, \mbox{mid-/}\linebreak[0]{}near-infrared (MIR/NIR), optical, and \mbox{near-/}\linebreak[0]{}far-ultraviolet (NUV/FUV) data (Sections~\ref{sec:data:HI}, \ref{sec:data:NIR}, \ref{sec:data:z0MGS}, and \ref{sec:data:other}).
High-level measurements such as CPROPS object catalogs created from CO data cubes \citep{Rosolowsky_etal_2021}, rotation curves derived from CO line kinematics \citep{Lang_etal_2020}, and morphological feature masks constructed from near-IR images \citep{Querejeta_etal_2021} are also available for most targets (Sections \ref{sec:data:CO:obj}, \ref{sec:data:rotcurve}, and \ref{sec:data:environ}).
These rich ancillary data provide us with comprehensive information about the multiphase ISM, stellar disk structures, star formation, and galactic dynamical properties on $\lesssim\,$kpc scales for our target galaxies.

In the following subsections, we detail the sources and characteristics of all raw data and high-level data products used in this study.
We provide a schematic summary of these input data at the top of Figure~\ref{fig:schematic}.

\subsection{PHANGS--ALMA CO Data} \label{sec:data:CO}

We use the PHANGS--ALMA \CO21 imaging data\footnote{We use PHANGS--ALMA internal data release v4, which corresponds to the first full public release.} \citep{Leroy_etal_2021a} to probe molecular gas properties on ${\lesssim}150$~pc scales.
These data cover the actively star-forming area in each galaxy (${\sim}100$~kpc$^2$ on average) and have sufficient depth and resolving power to detect and isolate the CO emission from individual GMCs (with a typical mass of $\gtrsim 10^5~\mathrm{M}_\odot$).
They include both interferometric and single dish observations and thus provide sensitivity to emission on all spatial scales.
We refer interested readers to \citet{Leroy_etal_2021a} for more details regarding sample selection, observational setup, and data product characteristics, and \citet{Leroy_etal_2021b} for an in-depth description of data calibration, imaging, and product creation procedures.

In this study, we measure properties of molecular cloud populations from the PHANGS--ALMA CO data using two different approaches.
The first approach measures molecular gas properties ``object-by-object.''
In this case, the objects of interest are identified by applying the cloud segmentation algorithm CPROPS to the PHANGS--ALMA CO data cubes \citep[A.~Hughes et al.\ in preparation]{Rosolowsky_Leroy_2006,Rosolowsky_etal_2021}.
The second approach treats the molecular gas as a spatially continuous medium and extracts measurement in a ``pixel-by-pixel'' fashion directly from the PHANGS--ALMA CO line moment maps, where the beam size corresponds roughly to the typical size of an individual GMC or giant molecular association \citep{Leroy_etal_2016,Sun_etal_2018,Sun_etal_2020a}.

While the two approaches access similar physical properties and often lead to similar results \citep[see][]{Sun_etal_2020a,Rosolowsky_etal_2021}, they complement each other in important ways.
The object-by-object approach treats each identified object as a fundamental structural unit, and by providing size estimates for these objects it probes the spatial organization of molecular gas.
The pixel-by-pixel method instead treats each resolution element as a fundamental unit, which preserves information from the smallest recoverable scale while remaining agnostic about the organization of the gas on larger scales.
When presenting our measurements in Section~\ref{sec:stats}, we compare the two approaches to illustrate how methodological choices could influence the main results.

\subsubsection{Object-by-object Measurements}
\label{sec:data:CO:obj}

We extract a set of molecular gas measurements for each object tabulated in the PHANGS--ALMA CPROPS catalogs\footnote{We use the v4 resolution-matched catalogs as described in A.~Hughes et al.\ (in preparation).} \citep[A.~Hughes et al.\ in preparation]{Rosolowsky_etal_2021}.
These catalogs are constructed by running CPROPS \citep{Rosolowsky_Leroy_2006} on data cubes at a set of common spatial resolutions (60, 90, 120, and 150~pc whenever available).
Note that A.~Hughes et al. (in preparation) present two versions of CPROPS catalogs for the PHANGS-ALMA sample: one is constructed from data cubes whose noise levels are homogenized among all galaxies, and the other from data cubes with the native noise.
We use the latter version in this paper, because (a) we would like to compare the object-by-object measurements to the corresponding pixel-by-pixel ones, which were derived from the original data cubes without noise homogenization; and (b) we would like to recover as much CO emission above the noise floor as possible.

For each object identified by the CPROPS algorithm, the catalog records its integrated CO line luminosity, $\LCOsub{obj}$, CO line width, $\sigCOsub{obj}$, and the two-dimensional projected radius on the sky\footnote{This radius is defined as the geometrical mean of the fitted semi-major and semi-minor axes for each identified object \citep[see][]{Rosolowsky_etal_2021}.}, $\Rsub{2D,\,obj}$.
These numbers are calculated after radially extrapolating each object to a hypothetical boundary at 0~K brightness temperature and then deconvolving the beam size and channel width.
From these basic observables, we estimate the following physical properties for each object:

\begin{itemize}[itemsep=0.5em,leftmargin=1em,parsep=0em,partopsep=0em]

\item[$\sbullet$] \emph{Molecular gas mass, $\Msub{obj}$}.
This is derived from the integrated \CO21 line luminosity $\LCOsub{obj}$ (in units of \uLco) via
\begin{equation}
\Msub{obj} = \alphaCOline{1}{0}\Rsub{21}^{-1}\LCOsub{obj}~.
\label{eq:M_obj}
\end{equation}
\noindent Here $\Rsub{21}=0.65$ is the adopted \CO21 to \CO10 line ratio \citep{denBrok_etal_2021,Leroy_etal_2021c}, and $\alphaCOline{1}{0}$ is a varying CO-to-\Htwo\ conversion factor for the \CO10 line.
By default, we use a metallicity-dependent $\alphaCO$ prescription as described in \citet{Sun_etal_2020b}:
\begin{equation}
\alphaCOline{1}{0} = 4.35~Z\spsc{\prime-1.6}~\mathrm{M_\odot~(\uIco~pc^2)^{-1}}~.
\label{eq:alphaCO}
\end{equation}
\noindent Here, $Z'$ is the inferred local gas phase abundance normalized to the solar value (see Section~\ref{sec:analysis:env}).
While we use Equation~\ref{eq:alphaCO} as our fiducial prescription, we also calculate $\alphaCO$ using several alternative prescriptions and include them in the data products (see Appendix~\ref{apdx:alphaCO}).

\item[$\sbullet$] \emph{Molecular gas surface density, $\Sigsub{obj}$}.
This is derived from $\Msub{obj}$ and $\Rsub{2D,\,obj}$ via
\begin{equation}
\Sigsub{obj}
= \frac{\onehalf\Msub{obj}}{\pi\Rsub{2D,\,obj}^2}\cos{i}
= \frac{\Msub{obj}}{2\pi\Rsub{2D,\,obj}^2}\cos{i}~.
\label{eq:Sigma_obj}
\end{equation}
\noindent Following \citet{Rosolowsky_etal_2021}, this estimate assumes a two-dimensional Gaussian profile for the projected gas mass distribution and focuses on the area within a FWHM, which includes half of the total gas mass.
The $\cos{i}$ term accounts for galaxy inclination (see Table~\ref{tab:sample}) by correcting the derived surface densities to face-on projection.
This correction was not present in the formulae used by \citet{Rosolowsky_etal_2021} and previous similar work.
We motivate this correction in Appendix~\ref{apdx:inclination}.

\item[$\sbullet$] \emph{One-dimensional velocity dispersion, $\sigsub{obj}$}. 
This is derived from the measured CO line width $\sigCOsub{obj}$ via
\begin{equation}
\sigsub{obj}
= \sigCOsub{obj}(\cos{i})^{0.5}~.
\label{eq:vdisp_obj}
\end{equation}
\noindent Here $\sigCOsub{obj}$ comes from the second moment (moment-2) of the CO line profile and is corrected for broadening due to the line spread function \citep[LSF; see][]{Rosolowsky_etal_2021}.
The extra $(\cos{i})^{0.5}$ term is an empirically determined correction that accounts for the dependence of the observed CO line width on galaxy inclination.
Appendix~\ref{apdx:inclination} details the origin of this term and the rationale for including it here.

\quad In the following discussions, we will assume that $\sigsub{obj}$ is dominated by turbulent motions, though this measurement will include additional contributions from thermal or ordered, streaming motions.
Fundamentally, it reflects the velocity dispersion along the line of sight direction within each object.

\item[$\sbullet$] \emph{Three-dimensional mean radius, $\Rsub{obj}$}.
This quantity is inferred from $\Rsub{2D,\,obj}$ via
\begin{equation}
\Rsub{obj} = 
\mathrm{min}\!\left[ \Rsub{2D,\,obj}, \sqrt[3]{\Rsub{2D,\,obj}^2\frac{H}{2\cos{i}}}\,\right]~.
\label{eq:R_obj}
\end{equation}
\noindent Here $H=100$~pc is an assumed molecular gas disk thickness perpendicular to the galaxy plane \citep{Heyer_Dame_2015}, and $\frac{H}{\cos{i}}$ would be the expected line-of-sight depth given the disk inclination.
Equation~\ref{eq:R_obj} assumes a spheroidal geometry when the object diameter on the sky exceeds this line-of-sight depth, and a spherical geometry otherwise.
This is similar to the treatments in \citet{Rosolowsky_etal_2021}, except that here we also correct for galaxy inclination (also see Appendix~\ref{apdx:inclination}).

\quad Our adopted value for $H$ is likely uncertain by a factor of ${\sim}\,2$ due to variations within a galaxy and among galaxies \citep[e.g.,][]{Yim_etal_2014,Bacchini_etal_2019a}. Systematic trends with galactocentric radius and global galaxy mass are also expected. A fixed value of $H=100$~pc cannot capture these variations, which means that our inferred $R_\mathrm{obj}$ values (and any measurements that rely on them, see bullet points below) are affected accordingly. Nevertheless, the functional form of Equation~\ref{eq:R_obj} suggests that at most $1/3$ of the fractional uncertainty on $H$ will propagate to $R_\mathrm{obj}$, which would only be marginally significant in comparison to other sources of systematic uncertainties (see discussions in Section~\ref{sec:stats:MC} and Appendix~\ref{apdx:scale}).

\item[$\sbullet$] \emph{Turbulent pressure, $\Pturbsub{obj}$}.
This is derived from $\Msub{obj}$, $\sigsub{obj}$, and $\Rsub{obj}$ via
\begin{equation}
\Pturbsub{obj}
\equiv \rhosub{obj}\sigsub{obj}^2
= \frac{\onehalf\Msub{obj}}{\frac{4}{3}\pi\Rsub{obj}^3}\sigsub{obj}^2
= \frac{3\Msub{obj}\sigsub{obj}^2}{8\pi\Rsub{obj}^3}~.
\label{eq:Pturb_obj}
\end{equation}
\noindent Here the mean density $\rhosub{obj}$ is derived from $\Rsub{obj}$ and the mass within the FWHM of a two-dimensional Gaussian profile \citep[Equation~16 in][]{Rosolowsky_etal_2021}.

\item[$\sbullet$] \emph{Virial parameter, $\alphavirsub{obj}$}.
This is derived from $\Msub{obj}$, $\sigsub{obj}$, and $\Rsub{obj}$ via
\begin{equation}
\alphavirsub{obj}
\equiv \frac{2E\sbsc{kin}}{|E\sbsc{grav}|}
= \frac{5\,\sigsub{obj}^2\Rsub{obj}}{G(\onehalf\Msub{obj})}
= \frac{10\,\sigsub{obj}^2\Rsub{obj}}{G\Msub{obj}}~.
\label{eq:alphavir_obj}
\end{equation}
\noindent This formula is derived by calculating the kinetic energy ($E\sbsc{kin}$) and gravitational potential energy ($E\sbsc{grav}$) for the gas within the two-dimensional FWHM size, assuming a uniform density distribution \citep[consistent with][]{Rosolowsky_etal_2021}.
With this definition, a virialized object would have $\alphavirsub{obj}=1$, whereas an object in energy equipartition would have $\alphavirsub{obj}=2$.
But we note that the virial parameter estimated in this way might not be a complete description of cloud dynamical states if there are strong magnetic field, surface pressure, or external tidal forces \citep[see discussions in, e.g.,][]{Ballesteros-Paredes_2006,Sun_etal_2020b,KimJG_etal_2020,LiuLJ_etal_2021}.

\end{itemize}

\subsubsection{Pixel-by-pixel Measurements}
\label{sec:data:CO:pix}

As an alternative to the object-by-object approach, we also derive molecular gas properties at fixed spatial resolutions (60, 90, 120, and 150~pc whenever available) measured pixel-by-pixel from the PHANGS--ALMA CO moment maps \citep[see][]{Sun_etal_2018,Sun_etal_2020a}.
The PHANGS-ALMA data reduction pipeline \citep{Leroy_etal_2021b} produces two versions of moment maps: a ``broad'' version that prioritizes high CO flux completeness through highly inclusive signal masking, and a ``strict'' version that features high signal-to-noise (S/N) CO line moment measurements thanks to more restrictive masking.
For our pixel-by-pixel analysis, we primarily use the ``strict'' moment maps so that only pixels with reliable measurements are included in our calculation.
To account for the lower CO flux completeness of the ``strict'' maps, we later estimate their flux completeness by comparing the ``strict'' and ``broad'' maps, and further correct for sensitivity-induced biases (see Section~\ref{sec:analysis:cloud:completeness}).

The ``strict,'' beam-matched moment maps provide CO line integrated intensity (moment-0), $\ICOsub{pix}$, CO line effective width \citep[see][]{Heyer_etal_2001}, $\sigCOsub{pix}$, and their associated uncertainties for every pixel with detected CO emission.
From these basic observables, we derive a set of molecular gas physical properties mirroring those from the object-by-object approach, and estimate their statistical uncertainties through Gaussian error propagation:

\begin{itemize}[itemsep=0.5em,leftmargin=1em,parsep=0em,partopsep=0em]

\item[$\sbullet$] \emph{Molecular gas surface density, $\Sigsub{pix}$}.
This quantity is derived from the integrated \CO21 line intensity $\ICOsub{pix}$ (in units of \uIco) for each pixel, via
\begin{equation}
\Sigsub{pix} = \alphaCOline{1}{0}\Rsub{21}^{-1}\ICOsub{pix}\cos{i}~.
\label{eq:Sigma_pix}
\end{equation}
\noindent Here $\Rsub{21}$ and $\alphaCOline{1}{0}$ represent the adopted CO line ratio and \CO10-to-\Htwo\ conversion factor as in Equation~\ref{eq:M_obj}.
The same $\cos{i}$ inclination correction from Equation~\ref{eq:Sigma_obj} also applies here.

\item[$\sbullet$] \emph{Molecular gas mass, $\Msub{pix}$}.
We also record the total molecular gas mass captured in each beam\footnote{Since the beam is usually over-sampled by the pixel grid in observational data, in theory $\Msub{pix}$ should only be derived for each \emph{independent} beam (rather than for each pixel) in order to conserve the total molecular gas mass budget. In this work, we only use $\Msub{pix}$ as an intermediate quantity to derive other pixel-based measurements, and none of these measurements requires an accurate gas mass accounting. Therefore, we do not explicitly distinguish between measurements per beam versus per pixel.} via
\begin{align}
\Msub{pix}
&= \alphaCOline{1}{0}\Rsub{21}^{-1}\ICOsub{pix}\Abeam~.
\label{eq:M_pix}
\end{align}
\noindent Here $\Abeam=(\pi/4\ln{2})\Dbeam^2$ is the effective area of the beam with a FWHM of $\Dbeam$ (i.e., 60, 90, 120, or 150~pc).
No inclination correction is required here since both $\ICOsub{pix}$ and $\Abeam$ are measured/defined in the projected plane of the sky.

\item[$\sbullet$] \emph{One-dimensional velocity dispersion, $\sigsub{pix}$}.
This quantity is derived from the LSF-corrected CO line width $\sigCOsub{pix}$ in each pixel and uses the same inclination correction as Equation~\ref{eq:vdisp_obj}:
\begin{equation}
\sigsub{pix}
= \sigCOsub{pix}(\cos{i})^{0.5}~.
\label{eq:vdisp_pix}
\end{equation}
\noindent Here $\sigCOsub{pix}$ represents the CO line \textit{effective width}, which is a different line width metric than the one based on the second moment used in Equation~\ref{eq:vdisp_obj}.
The effective width is a more robust line width metric than moment-2 at low S/N, but it could give biased results when there are multiple velocity components along the line of sight \citep{Henshaw_etal_2020}.

\item[$\sbullet$] \emph{Three-dimensional mean radius, $\Rsub{pix}$}.
We adopt the following three-dimensional size for the gas structure captured in each beam, mirroring Equation~\ref{eq:R_obj}:
\begin{equation}
\Rsub{pix} = \mathrm{min}\!\left[\frac{\Dbeam}{2},\,\sqrt[3]{\frac{\Dbeam^2H}{8\cos{i}}}\,\right]~.
\label{eq:R_pix}
\end{equation}
\noindent Again, the $\cos{i}$ term accounts for galaxy inclination by converting the molecular gas disk thickness (perpendicular to the galaxy plane) to the depth along the line of sight.
Note that Equation~\ref{eq:R_pix} yields a single $\Rsub{pix}$ value for each given beam size.

\item[$\sbullet$] \emph{Turbulent pressure, $\Pturbsub{pix}$}.
This is derived from $\Sigsub{pix}$ and $\sigsub{pix}$ via
\begin{equation}
\Pturbsub{pix}
\equiv \rhosub{pix}\sigsub{pix}^2
= \frac{3\Msub{pix}\sigsub{pix}^2}{4\pi\Rsub{pix}^3}~.
\label{eq:Pturb_pix}
\end{equation}
\noindent This assumes that the gas mass captured in each beam is uniformly distributed within a radius of $\Rsub{pix}$.
This is consistent with the geometrical assumptions adopted in previous studies \citep[e.g.,][]{Sun_etal_2020b}, yet it leads to an inconsistency with the object-based approach (Equation~\ref{eq:Pturb_obj}).
We comment on this issue in Section~\ref{sec:data:CO:notes}.

\item[$\sbullet$] \emph{Virial parameter, $\alphavirsub{pix}$}.
This is derived from $\Sigsub{pix}$ and $\sigsub{pix}$ via
\begin{equation}
\alphavirsub{pix}
\equiv \frac{2E\sbsc{kin}}{|E\sbsc{grav}|}
= \frac{5\,\sigsub{pix}^2\Rsub{pix}}{G\Msub{pix}}~.
\label{eq:alphavir_pix}
\end{equation}
\noindent This also assumes a spherical geometry and a uniform density distribution within $\Rsub{pix}$.
Similar to the situation with our turbulent pressure estimates, the geometrical assumptions here are not fully consistent with those adopted for the object-based analysis (see Section~\ref{sec:data:CO:notes} for further comments).

\end{itemize}

\subsubsection{Notes on the Common Grounds and Differences between the Object-based and Pixel-base Approaches}
\label{sec:data:CO:notes}

The object-based and pixel-based approaches show an apparent symmetry, in the sense that they have many measured quantities in common, such as molecular gas surface density, velocity dispersion, turbulent pressure, and virial parameter.
This allows us to make direct comparisons between the two approaches and assess how our methodological choices might influence the quantitative results.
However, it is worth emphasizing that, for several reasons, we do not necessarily expect the two approaches to yield exactly the same quantitative results.

First and foremost, the two approaches are motivated by two slightly different views for the structure and geometry of the molecular ISM in galaxies.
The object-based approach views the molecular ISM as a collection of dense, centrally concentrated structures, and the central goal of the CPROPS algorithm is to segment the observed CO emission distribution such that each identified CO-emitting object corresponds to a coherent structure like a GMC or a giant molecular association.
The pixel-based approach instead views the molecular ISM as a continuous distribution of gas while being agnostic about its spatial clustering, and the measurement process simply characterizes the gas captured in each beam.
In a sense, the two approaches see the same observational data through different lenses, and each attempts to extract measurable properties in a way that is most consistent with its adopted view.

Reflecting these different views, there are also important, practical differences in the methodologies between these two approaches, which make it non-trivial to draw direct comparisons between them.
In particular, the object-based approach aims to measure the true size and mass of each identified object by deconvolving the beam and extrapolating the detected part of each object to a hypothetical boundary at 0~K brightness temperature.
Such operations could in principle account for biases due to the finite resolution and sensitivity of the observations, but they are implicitly model-dependent and not easily adaptable to fit the pixel-based approach.

The distinct physical models underlying these two approaches are also reflected in the different auxiliary assumptions they adopt when calculating physical quantities.
The object-based approach assumes compact, Gaussian-shaped gas distributions and calculate gas surface density, turbulent pressure, and virial parameter for only the half of the gas located within the Gaussian FWHM (Equations~\ref{eq:Sigma_obj}--\ref{eq:alphavir_obj}, consistent with \citealt{Rosolowsky_etal_2021}).
In contrary, the pixel-based approach considers all the gas mass detected in each beam and assumes it is uniformly distributed within the beam area (Equations~\ref{eq:Pturb_pix}--\ref{eq:alphavir_pix}, in line with \citealt{Sun_etal_2018,Sun_etal_2020a}).

Considering these complications, we do not necessarily expect the two approaches to agree in their quantitative results, even though we start from the same CO data cubes and attempt to define measurable properties in a symmetric way.

\subsection{\texorpdfstring{\HI}{HI} Data} \label{sec:data:HI}

We use interferometric \HI\ 21~cm line data to trace the distribution of neutral atomic gas in each galaxy.
These include both new and archival observations taken by the Karl~G.~Jansky Very Large Array (VLA) and the Australia Telescope Compact Array (ATCA).

Among the \galnumhi\ galaxies with \HI\ data (see Table~\ref{tab:sample}), 20 have been observed as part of the PHANGS--VLA survey (A.~Sardone et al.\ in preparation).
The other galaxies have archival data from either large nearby galaxy surveys such as 
THINGS \citep[nine galaxies;][]{Walter_etal_2008},
VIVA \citep[six galaxies;][]{Chung_etal_2009},
HERACLES \citep[four galaxies;][]{Leroy_etal_2009},
LVHIS \citep[three galaxies;][]{Koribalski_etal_2018},
EveryTHINGS (two galaxies; I.~Chiang et al.\ in preparation),
or individual case studies with the VLA (seven galaxies) and the ATCA \citep[two galaxies;][]{Murugeshan_etal_2019}.
These \HI\ data sets have typical angular resolution of $15\arcsec{-}35\arcsec$ ($16{-}84$ percentile), which corresponds to linear scales of $0.7{-}2.8$~kpc (see Section~\ref{sec:analysis:env} for further discussions about \HI\ data resolution).
The $3\sigma$ sensitivity limit ranges $10{-}100\;\uIhi$ for the \HI\ line intensity.

Assuming optically thin 21~cm emission, we convert 21~cm line intensity $\Ihi$ to atomic gas surface density $\Sigatom$ via
\begin{equation}
    \frac{\Sigatom}{\uSig} = 2.0\times10^{-2} \left(\frac{\Ihi}{\uIhi}\right) \cos{i}~.
    \label{eq:SigHI}
\end{equation}
\noindent Here $\Sigatom$ includes the (extra 35\%) mass of helium and heavier elements. The $\cos{i}$ term accounts for galaxy inclination.

\subsection{Near-IR Data} \label{sec:data:NIR}

We use near-IR imaging data from the \textit{Spitzer Space Telescope} and the \textit{Wide-field Infrared Survey Explorer (WISE)} to trace the old stellar mass distribution (see Table~\ref{tab:sample}, column~8).
For \galnumirac\ galaxies in our sample, we use \textit{Spitzer} IRAC $3.6$~\micron\ images from the S$^4$G survey \citep[]{Sheth_etal_2010}.
For those without S$^4$G data, we instead use \textit{WISE} W1 band ($3.4$~\micron) images compiled by the $z0$MGS project \citep{Leroy_etal_2019}.
All these data are postprocessed by subtracting background emission, masking foreground stars in the field of view, and convolving the non-Gaussian point spread function (PSF) to a $7.5\arcsec$ Gaussian PSF using appropriate convolution kernels \citep{Aniano_etal_2011}.

We convert the stellar continuum intensity at $3.4$~\micron\ and $3.6$~\micron\ to stellar mass surface density, $\Sigstar$, location-by-location via
\begin{align}
    \frac{\Sigstar}{\uSig} &= 350 \left(\frac{\MtoLwiseone}{0.5}\right)\left(\frac{\Iiracone}{\uI}\right) \cos{i}~, \label{eq:Sigstar_3p6um}\\
    \frac{\Sigstar}{\uSig} &= 330 \left(\frac{\MtoLwiseone}{0.5}\right)\left(\frac{\Iwiseone}{\uI}\right) \cos{i}~.
    \label{eq:Sigstar_3p4um}
\end{align}
Here, $\MtoLwiseone$ is the stellar mass-to-light (M/L) ratio at $3.4$~\micron, which should be nearly identical to that at $3.6$~\micron.
We adopt a \emph{spatially varying} M/L ratio, which was estimated by \citet{Leroy_etal_2021a} for all PHANGS-ALMA targets based on an empirical relation between $\MtoLwiseone$ and the local SFR surface density to $3.4$~\micron\ surface brightness ratio.

\subsection{Mid-IR and UV Data}
\label{sec:data:z0MGS}

We use mid-IR images from \textit{WISE} and \mbox{far-/}\linebreak[0]{}near-UV images from the \textit{Galaxy Evolution Explorer (GALEX)} to trace the distribution of obscured and unobscured star formation.
These data are also compiled by the $z0$MGS project \citep{Leroy_etal_2019} and have been postprocessed by subtracting background emission, masking foreground stars, reprojecting to a shared astrometry, and then convolving to a $15\arcsec$ Gaussian PSF.

We combine the mid-IR and UV data and calculate the local star formation rate (SFR) surface density following the prescriptions described in \citet{Leroy_etal_2021a}.
By default we use the combination of \textit{GALEX} FUV (154~nm) and \textit{WISE} 22~\micron\ data to calculate the local SFR surface density
\begin{align}
\frac{\SigSFR}{\uSigSFR}
&= \left(8.9\times10^{-2}\,\frac{I\sbsc{154\,nm}}{\uI}\right. \nonumber\\
&\quad \left.+\,3.0\times10^{-3}\,\frac{I\sbsc{22\,\mu m}}{\uI}\right) \cos{i}~.
\label{eq:SigSFR_FUVW4}
\end{align}
\noindent For galaxies that do not have FUV data (see Table~\ref{tab:sample}, column~9), we instead combine \textit{GALEX} NUV (231~nm) and \textit{WISE} 22~\micron\ data (when NUV is available) or use the \textit{WISE} data alone (when NUV is not available either) to calculate SFR surface density
\begin{align}
\frac{\SigSFR}{\uSigSFR}
&= \left(8.9\times10^{-2}\,\frac{I\sbsc{231\,nm}}{\uI}\right. \nonumber\\
&\quad \left.+\,2.6\times10^{-3}\,\frac{I\sbsc{22\,\mu m}}{\uI}\right) \cos{i}~,
\label{eq:SigSFR_NUVW4}\\
\frac{\SigSFR}{\uSigSFR}
&= 3.8\times10^{-3}\,\left(\frac{I\sbsc{22\,\mu m}}{\uI}\right) \cos{i}~.
\label{eq:SigSFR_W4ONLY}
\end{align}
\noindent These prescriptions assume a Chabrier initial mass function \citep[IMF;][]{Chabrier_2003} via their calibration against \citet{Salim_etal_2016}, which is also consistent within $\approx 5\%$ with calibrations using a Kroupa IMF \citep{Kroupa_2001}.
The quantitative results agree with extinction-corrected H$\alpha$-based SFR estimates from the PHANGS--MUSE survey \citep[F.~Belfiore et al.\ in preparation]{Emsellem_etal_2021} at a  ${\sim}20{-}30\%$ level overall, but there is divergence in low SFR regions due to contributions from IR cirrus and/or old stellar populations \citep{Boquien_etal_2016}.
We refer the reader to \citet{Leroy_etal_2021a} for more details on the calibration of these SFR prescriptions.

\subsection{Rotation Curves} \label{sec:data:rotcurve}

We use rotation curves derived from CO line kinematics by \citet{Lang_etal_2020} to characterize galactic orbital kinematics (e.g., orbital period and shear) locally within each galaxy.
These rotation curves are measured from the same PHANGS--ALMA CO data set, and therefore cover roughly the same galactocentric radius range as the CO maps themselves.
They are available for \galnumrc\ out of the \galnumtot\ galaxies. 

The rotation curves in \citet{Lang_etal_2020} are measured and recorded with finite radial bin sizes (${\sim}150$~pc).
Due to the sparse distribution of CO detections across the field of view and the likely presence of unaccounted local streaming motions in the gas, the measured circular velocity sometimes fluctuates considerably between adjacent radial bins.
These bin-to-bin fluctuations make it challenging to reliably estimate any parameter that depends on the derivative of the rotation curves.

To address this issue, we use a set of functional fitting models constructed from the measured rotation curves (J.~Nofech et al.\ in preparation) rather than the raw measurements themselves.
These fitting models adopt the ``universal rotation curve'' functional form suggested by \citet{Persic_etal_1996}.
The fitting process effectively forces the rotation curve models to be smooth and have physically sensible slopes (i.e., with its logarithmic derivative between -0.5 and 1), while still matching the actual measurements as closely as possible.
We visually inspect all fitting results and conclude that the models represent the raw measurements reasonably well.

Based on these best-fit analytical models of the CO rotation curves and the estimated uncertainties on the model parameters, we extract at each radius the circular velocity, $\Vcirc$, the corresponding angular velocity, $\Omegacirc$, the logarithmic derivative of the rotation curve
\begin{equation}
\beta = \frac{\mathrm{d}\ln \Vcirc}{\mathrm{d}\ln \rgal}~,
\end{equation}
\noindent and Oort's $A$ parameter
\begin{equation}
\OortA = \frac{1}{2}\Omegacirc (1-\beta)~.
\end{equation}
\noindent These parameters (and their associated uncertainties) will be used to describe the local galactic dynamical properties at various locations within the target galaxies.

\subsection{Morphological Environment Masks} \label{sec:data:environ}

We use the environment masks presented in \citet{Querejeta_etal_2021} to distinguish different morphological regions in each galaxy.
These masks are constructed based on structural decomposition analysis and visual inspection of the IRAC $3.6$~\micron\ data \citep[also see][]{Herrera-Endoqui_etal_2015,Salo_etal_2015}.
The full set of environment masks mark the area covered by morphological features such as galaxy centers, stellar bars, spiral arms, rings, and lenses (see the last panel in Figure~\ref{fig:image}).
{The typical width of these environmental masks are set by the physical extent of the corresponding morphological features, which are often ${\gtrsim}\,1$~kpc wide for stellar bars and spiral arms but can be much narrower for the other features.}

In this work, we primarily use these masks to divide each galaxy into two types of environment: the area that falls into galaxy centers and stellar bars (referred to as ``center/bar'' hereafter), and the remaining outer disk area (``disk'' hereafter).
We make this distinction because we expect the physical conditions influencing GMCs to be different between these two regimes: the ``center/bar'' environment often sees galactic dynamics (i.e., gravitational torque and shear) playing a more prominent role, and in some galaxies AGN feedback can significantly impact the molecular gas in its central region.

\subsection{Other Data} \label{sec:data:other}

In addition to what has been described above, we also include measurements derived from other data sets in the analysis.
These measurements are not presented among the main scientific results in this paper, but they are part of our final data products and they have appeared in publications that used our data products \citep[e.g.,][]{Querejeta_etal_2021}.

We use continuum-subtracted, narrow-band H$\alpha$ imaging data to provide alternative estimates of star formation rate in \galnumha\ out of our \galnumtot\ targets.
These observations were obtained as part of the PHANGS--H$\alpha$ survey\footnote{We use PHANGS--H$\alpha$ internal data release version 2.3.} by either the Wide Field Imager (WFI) on the ESO/MPG \mbox{2.2-m} Telescope or the Direct CCD on the CIS \mbox{2.5-m} Ir\'en\'ee du~Pont Telescope (A.~Razza et al.\ in preparation).
The narrow-band H$\alpha$ data have been calibrated astrometrically and photometrically, corrected for sky emission, and masked for foreground stars; the continuum contribution was removed based on the associated $R$-band observations; and the continuum-subtracted data were further corrected for filter transmission and [\ion{N}{ii}] contamination.

We combine H$\alpha$ data with \textit{WISE} 22~\micron\ data to derive an attenuation-corrected SFR surface density following \citet{Calzetti_etal_2007} and \citet{Murphy_etal_2011}:
\begin{align}
\frac{\SigSFR}{\uSigSFR}
&= \left(2.7\times10^{13}\,\frac{I\sbsc{H\alpha}}{\uIha}\right. \nonumber\\
&\quad \left.+\,2.7\times10^{-3}\,\frac{I\sbsc{22\,\mu m}}{\uI}\right) \cos{i}~,
\label{eq:SigSFR_HaW4}
\end{align}
\noindent This prescription assumes constant star formation over 100~Myr and a Kroupa IMF \citep{Kroupa_2001}.
Given that the H$\alpha$ data and \textit{WISE} data have very different angular resolution (${\sim}1\arcsec$ versus $15\arcsec$), we first convolve the H$\alpha$ images to the \textit{WISE} resolution and estimate $\SigSFR$ via Equation~\ref{eq:SigSFR_HaW4}.
Then we determine the spatially varying $\SigSFR$-to-H$\alpha$ ratio at that coarser resolution and multiply it to the native resolution H$\alpha$ images to get the final, high resolution $\SigSFR$ maps.
This approach ensures that the average $\SigSFR$ value over a large area converges to the expectation from Equation~\ref{eq:SigSFR_HaW4}.


\section{Cross-spatial-scale Analysis} \label{sec:analysis}

\begin{figure*}[htbp]
\gridline{
\fig{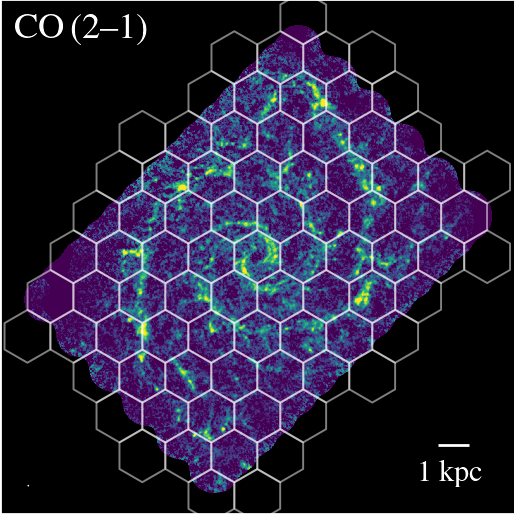}{0.325\textwidth}{}
\fig{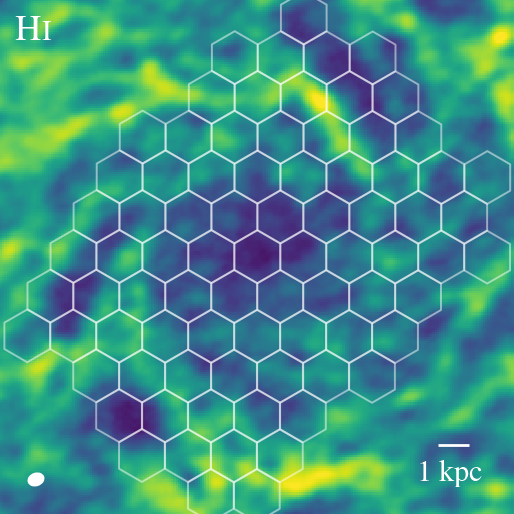}{0.325\textwidth}{}
\fig{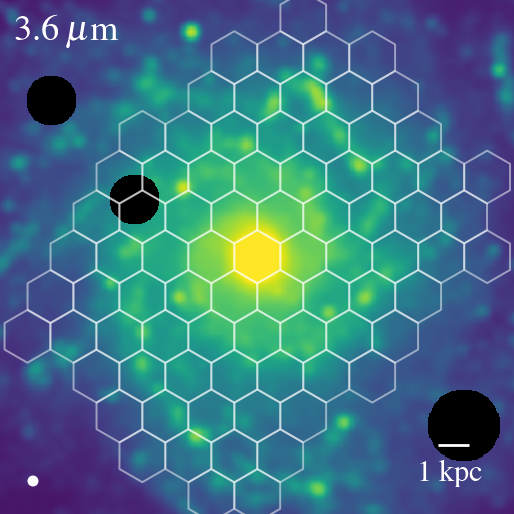}{0.325\textwidth}{}
}
\vspace{-2.1\baselineskip}
\gridline{
\fig{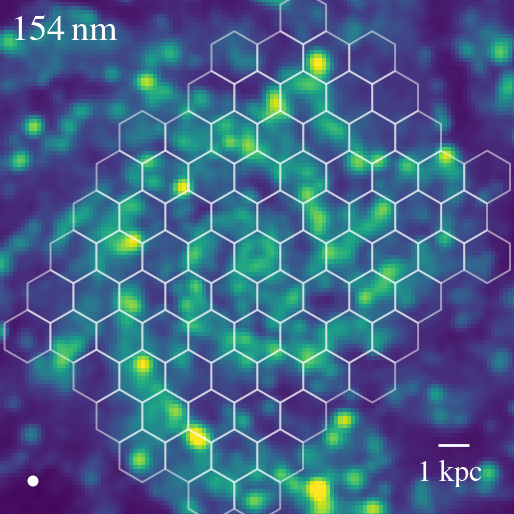}{0.325\textwidth}{}
\fig{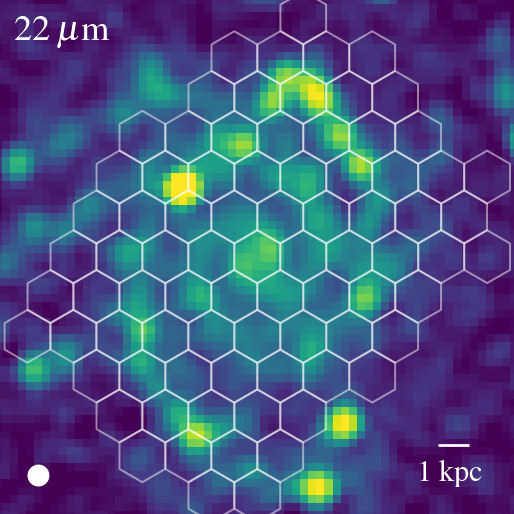}{0.325\textwidth}{}
\fig{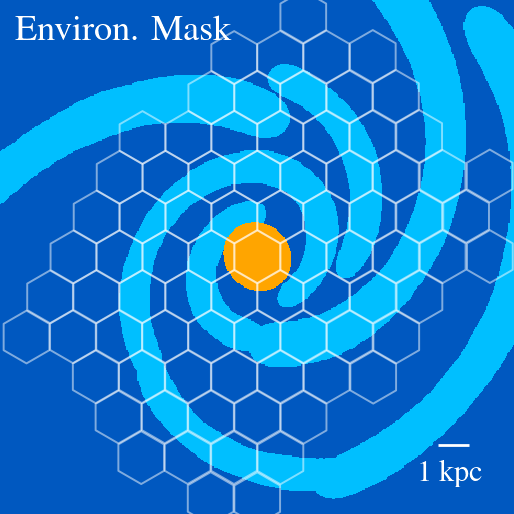}{0.325\textwidth}{}
}
\vspace{-2\baselineskip}
\caption{
This figure showcases a subset of the multiwavelength data that we assemble for the galaxy NGC~628.
The top row displays the PHANGS--ALMA \CO21\ line intensity map (tracing molecular gas), 
the THINGS VLA \HI\ 21-cm line intensity map (tracing neutral atomic gas), 
and the S$^4$G \textit{Spitzer} IRAC 3.6~\micron\ image (tracing stellar mass). 
The bottom row displays the \textit{GALEX} 154~nm image (tracing unobscured star formation), 
the \textit{WISE} 22~\micron\ image (tracing obscured star formation), 
and the PHANGS environment mask. 
In each panel, the scale bar at the lower-right corner shows the spatial extent of 1~kpc, whereas the white ellipse at the lower-left gives the beam size (except for the last panel).
The white grids demarcate the hexagonal apertures (\apersize\ in size) in which we extract molecular cloud population statistics and build a comprehensive inventory of host galaxy structural, kinematic, and star formation properties.
\vspace{1.0\baselineskip}
}
\label{fig:image}
\end{figure*}

We adopt a ``cross-spatial-scale'' analysis framework to connect molecular cloud properties (measured on $60{-}150$~pc scales) to galactic environmental properties (mostly measured on $\sim$kpc scales).
This analysis framework is inspired by a number of previous works \citep[e.g.,][]{Sandstrom_etal_2013,Leroy_etal_2016}.
Briefly, we divide the sky footprint of each galaxy into a set of averaging apertures, within which we aggregate high resolution molecular gas measurements to characterize the underlying cloud population.
We also attempt to build a full inventory of ancillary measurements to characterize various aspects of the host galaxy itself.
In this way, we assemble the diverse set of observational data described in Section~\ref{sec:data} into a coherent, multiwavelength database.
An early version of this database was constructed by \citet{Sun_etal_2020b}, with its subsequently improved versions used in several publications \citep[e.g.,][]{Herrera_etal_2020,Jeffreson_etal_2020,Barnes_etal_2021,Querejeta_etal_2021,Stuber_etal_2021}.
The source code for database construction, including generic tools for aggregating measurements from maps and catalogs into the existing database, is available on GitHub\footnote{\url{https://github.com/PhangsTeam/MegaTable}}, and a copy of the version used in this article is published on Zenodo \citep{MegaTable_3.0}.

\subsection{Defining Averaging Apertures} \label{sec:analysis:aperture}

We divide the sky footprint of each galaxy into a set of hexagonal apertures, as illustrated in Figure~\ref{fig:image}.
These apertures form a regular tiling in the plane of the sky, with a ``central'' aperture positioned right at the galaxy center.
Adjacent apertures are separated by a linear distance of \apersize, which implies that each aperture has a projected area of \aperarea\ on the sky.
The configuration of the hexagonal apertures here is analogous to the ``solution pixels'' used in \citet{Sandstrom_etal_2013}, except that the apertures in the current work do not overlap with each other.

For a complete coverage of the galaxy footprint, we include all apertures covering out to $\rgal = 1.5\Riso$ in each galaxy, where $\Riso$ is the galaxy radius defined by its 25~mag/arcsec$^2$ isophote (in $B$~band; see Table~\ref{tab:sample}).
This way, the constructed database for each galaxy includes almost all valid measurements from all data sets described in Section~\ref{sec:data}.
However, this work focuses on the correlation of molecular clouds and their galactic environments, and thus we will only present results from a subset of apertures that enclose non-zero signals in the PHANGS-ALMA CO moment maps.

In addition to the hexagonal tiling method described above, we also run a parallel line of analysis with a different binning scheme.
Specifically, we define a series of radial bins that are \annuluswidth\ in width and again cover out to $\rgal = 1.5\,\Riso$ in each galaxy.
Assembling measurements in these radial bins allows us to rigorously calculate their radial profiles, but at the expense of losing all non-axisymmetric information.
We publish the data products from these radial profile calculations together with those from the hexagonal aperture analysis (see Appendix~\ref{apdx:mrt}).
We do not present the results of this parallel line of analysis in this paper, but we expect the distributions of most measurements to be consistent with the hexagonal aperture averages once we use consistent weighting schemes (e.g., by the enclosed area or molecular gas mass; see Section~\ref{sec:stats}) for each aperture/ring.

\subsection{Aggregating Molecular Cloud Measurements}
\label{sec:analysis:cloud}

Within each aperture, we calculate the ensemble average of molecular cloud measurements using a molecular gas mass-weighted averaging scheme.
This is equivalent to a CO intensity-weighted averaging, because the $\alphaCO$ value is calculated per aperture rather than per object/pixel in this study (see Appendix~\ref{apdx:alphaCO}).
We use a ``$\brkt{}$'' symbol to denote this averaging operation:
\begin{equation}
\brkt{X\sbsc{\theta\,pc}} = 
\frac{\sum\limits_i M_i\, X\sbsc{\textit{i},\,\theta\,pc}}{\sum\limits_i M_i}~.
\label{eq:average}
\end{equation}
\noindent Here, $X\sbsc{\textit{i},\,\theta\,pc}$ represents a molecular gas property measured for the $i$-th object or pixel at $\theta$~pc resolution ($\theta=60,\,90,\,120,\,150$); it can be any of the object- or pixel-based measurements defined in Sections~\ref{sec:data:CO:obj} and~\ref{sec:data:CO:pix}.
$M_i$ is the molecular gas mass associated with the object or pixel for which $X\sbsc{\textit{i},\,\theta\,pc}$ is measured.
The summation in Equation~\ref{eq:average} includes all detected objects/\linebreak[0]{}pixels with their center coordinates located inside the sharp boundary of the averaging aperture.
In this case, each object/\linebreak[0]{}pixel belongs to a unique averaging aperture, and thus the averaging results in adjacent apertures are independent by construction.

Based on Equation~\ref{eq:average}, we can also estimate statistical uncertainties for the population-averaged cloud properties through Gaussian error propagation.
We take into account the uncertainties on both the quantity to be averaged, $X\sbsc{i,\,\theta\,pc}$, and the weight, $M\sbsc{i}$.
When aggregating the pixel-by-pixel measurements, we further consider the built-in correlation between adjacent pixels and scale the estimated uncertainty of the population average according to the oversampling factor.

Our aperture averaging scheme resembles the one adopted by \citet{Leroy_etal_2016} but differs from that approach in important ways.
In that work, the averaging is performed via a Gaussian kernel convolution, in which case the averaging result at any given location has a nonzero response to molecular clouds far away from that location.
This response pattern is designed to replicate the Gaussian beam of low-resolution data sets, and thus it may be preferable for rigorous calculations combining cloud-scale and kpc-scale measurements \citep[e.g.,][see also L.~Neumann\ in preparation]{Leroy_etal_2017a,Utomo_etal_2018}.
However, such an extended response pattern can lead to built-in correlations between averaging results at adjacent locations.
More importantly, it can yield biased population statistics when, for example, studying a region with little molecular gas next to a very gas-rich region (such as a galaxy center).
Since one of the main goals of this work is to derive reliable molecular cloud population statistics, we deem the ``sharp boundary'' scheme more appropriate here and will use it consistently for calculating both cloud population statistics and host galaxy properties (also see Section~\ref{sec:analysis:env}).

\subsubsection{Molecular Gas Clumping Factor}
\label{sec:analysis:cloud:clumping}

The averaging operation described above essentially extracts the (mass-weighted) expectation value of a molecular gas property from its probability distribution within each averaging aperture.
But one can also extract other types of statistics from the same distribution, such as the standard deviation of the gas surface density distribution (which quantifies the inhomogeneity of the medium), or the slope of the GMC mass function.
These other types of statistics can also provide unique observational constraints on the physical processes driving molecular cloud formation and evolution.

As part of the analysis done for this work, we calculate the molecular gas ``clumping factor,'' which is a dimensionless characterization of the surface density inhomogeneity in each aperture \citep{Leroy_etal_2013b}:
\begin{equation}
c\sbsc{pix,\,\theta\,pc} =
\frac{\left(\sum\limits_i \Sigsub{\textit{i},\,\theta\,pc}^2\right)N\sbsc{pix}}{\left(\sum\limits_i \Sigsub{\textit{i},\,\theta\,pc}\right)^2}~.
\label{eq:clumping}
\end{equation}
\noindent Here $\Sigsub{\textit{i},\,\theta\,pc}$ is the molecular gas surface density measured in the $i$-th pixel at $\theta\,pc$ resolution.
Similar to Equation~\ref{eq:average}, the summation includes all pixels with CO detections within the averaging aperture, and $N\sbsc{pix}$ is the total number of such pixels.
The right hand side of Equation~\ref{eq:clumping} can be interpreted as the ratio between the mass-weighted mean and the area-weighted mean of molecular gas surface density in the limit of infinite sensitivity \citep[see][]{Leroy_etal_2013b}.

We note that $c\sbsc{pix}$ is a measure of the width (i.e., second moment) of the surface density distribution among many similar parametrizations in the literature \citep[e.g., the smoothness index and the Gini coefficient; see][and references therein]{Davis_etal_2022}.
To measure this type of parameter reliably, a careful treatment of non-detections is particularly important.
We describe our strategy to handle non-detections in Section~\ref{sec:analysis:cloud:completeness}, and we illustrate the amplitude of the necessary corrections in Section~\ref{sec:stats:MC} and Appendix~\ref{apdx:completeness}.

\subsubsection{CO Flux Completeness and Corrections}
\label{sec:analysis:cloud:completeness}

The ensemble-average molecular cloud properties (Equation~\ref{eq:average}) and the molecular gas clumping factor (Equation~\ref{eq:clumping}) are both calculated based on pixels/\linebreak[0]{}objects that are \emph{detected} in the PHANGS--ALMA CO data.
For these calculations to reflect the true statistics of the \emph{entire} molecular cloud population in each region, the CO detections need to be reasonably complete, such that they represent a significant portion of the underlying cloud population.

Our object- and pixel-based measurements come from the PHANGS--ALMA CPROPS catalogs and the ``strict'' moment maps (Sections~\ref{sec:data:CO:obj} and~\ref{sec:data:CO:pix}), thus the completeness of our analysis is determined by the completeness of these data products.
Both data products adopt similar signal identification criteria to extract high-confidence CO detections in the original data cubes \citep{Leroy_etal_2021b,Rosolowsky_etal_2021}, which ensure reliable CO line measurements for the detected pixels/\linebreak[0]{}objects.
However, this comes at the price of excluding faint CO emission, which renders these data products incomplete in terms of both flux coverage and area coverage.

The extent of this effect can be quantified by the CO flux completeness, $\Fflux$, and area coverage fraction, $\Farea$, of the CPROPS catalog or the strict moment maps \citep[see tables~15 and~16 in][]{Leroy_etal_2021a}.
Here, we calculate $\Fflux$ and $\Farea$ for each averaging aperture and report these values along with the ensemble-average molecular cloud properties.
Specifically, within the footprint of each aperture, we calculate $\Farea$ by comparing the total area covered by CO detections in the ``strict'' moment-0 map to the total area of the aperture.
We calculate $\Fflux$ by comparing the total CO flux included in the ``strict'' moment-0 map to that in the corresponding ``broad'' moment-0 map.
The latter map is constructed with much more inclusive signal identification criteria than the strict map and has nearly 100\% flux completeness \citep[for more details, see][]{Leroy_etal_2021b}.

The incomplete CO flux and area coverage of the ``strict'' moment maps and CPROPS catalogs introduces a selection bias in our analysis.
The sense of this bias is that we miss places where CO emission is too faint to meet the masking criteria (e.g., areas occupied by small, low-mass molecular clouds or a diffuse gas component).
This selection bias affects many of the ensemble-average cloud properties calculated in this study, and is particularly severe for the clumping factor (see Section~\ref{sec:stats:MC} and Appendix~\ref{apdx:completeness}).

To account for this systematic bias, we introduce a correction factor for our measurements in each aperture based on the $\Fflux$ and $\Farea$ values in that aperture (see Appendix~\ref{apdx:completeness} for detailed derivations).
We assume that the CO intensity distribution (or equivalently, molecular gas surface density distribution) has a lognormal shape within each averaging aperture, and that the aforementioned selection bias prevents us from detecting CO emission below an intensity threshold.
Under these assumptions, we can solve for the width of the lognormal intensity distribution as well as its centroid (relative to the intensity threshold) from $\Fflux$ and $\Farea$.
This in turn allows us to calculate the appropriate correction factors to apply to the ensemble-average molecular cloud surface density, $\brkt{\Sigsub{obj}}$ and $\brkt{\Sigsub{pix}}$, and the clumping factor, $c\sbsc{pix}$.
We then assume that the correction factor calculated for $\brkt{\Sigsub{obj}}$ and $\brkt{\Sigsub{pix}}$ also applies to the ensemble-average cloud mass and turbulent pressure, but we leave molecular cloud size, velocity dispersion, and virial parameter uncorrected.
Though these latter quantities likely do suffer sensitivity-induced selection biases \citep[e.g., see the illustration of selection functions in][]{Sun_etal_2018}, the appropriate functional forms of their completeness corrections remain uncertain at present.
Finally, we scale the corresponding (statistical) uncertainty for each ensemble-average value by the same correction factor.

As illustrated in Section~\ref{sec:stats:MC} and Appendix~\ref{apdx:completeness}, thanks to the relatively high flux completeness of the PHANGS--ALMA CO data, the correction factors on the average cloud surface densities and the clumping factor are both moderate (${<}0.3$~dex for 90\% of the apertures with CO detections).
Nevertheless, we do expect our completeness correction scheme to be less reliable for apertures with low $\Fflux$ and/or $\Farea$, in which case the extrapolation is done based on very few measurements.
For this reason, we will exclude apertures with low $\Fflux$ or $\Farea$ when performing analyses that requires accurate cloud population statistics in individual apertures (see Section~\ref{sec:corr} and Appendix~\ref{apdx:completeness}).

\subsection{Aggregating Local Environmental Metrics}
\label{sec:analysis:env}

In addition to the compilation of ensemble-average molecular cloud properties described above, we assemble an inventory of ``environmental metrics'' that delineate various host galaxy local properties within each averaging aperture.
This inventory covers orbital kinematic properties (derived from rotation curves), gas-phase metallicity (predicted from scaling relations), surface densities of molecular gas, atomic gas, stellar mass, and SFR (estimated from multiwavelength imaging data), and morphological environment information (inherited from environmental masks).

We generally use two schemes to integrate these environmental metrics into the databases of aperture-wide statistics.
For those metrics that are calculated analytically (e.g., galactocentric coordinates, metallicity) or interpolated from analytical models (e.g., rotation curve-related properties), we directly record their values at the location of the aperture center.
For those metrics that rely on two-dimensional images, we use the native resolution images and calculate the \textit{unweighted} average among all pixels inside the sharp boundary of each aperture.
This latter scheme is consistent with the averaging scheme we used for aggregating molecular cloud properties (modulo the different weighting), and thus allows for direct comparisons between the averaging results.

We elaborate the specific treatment for each type of environmental metric below:

\begin{itemize}[itemsep=0.5em,leftmargin=1em,parsep=0em,partopsep=0em]

\item[$\sbullet$] \emph{Coordinates}.
For each hexagonal aperture, we record its central R.A.\ and Dec.\ coordinates.
Then based on the center coordinates, inclination angle, position angle, and the distance of the galaxy (see Table~\ref{tab:sample}), we calculate the deprojected galactocentric radius, $\rgal$, (in kpc units) at the aperture center and the deprojected azimuthal angle, $\phi\sbsc{gal}$, in the galaxy plane with respect to the major axis direction.
These coordinates uniquely determine the location of each aperture both on the sky and in the deprojected galaxy plane.

\item[$\sbullet$] \emph{Orbital kinematics}.
We report local orbital kinematic properties for apertures in the galaxy sample and galactocentric radius range covered by the rotation curve measurements from \citet{Lang_etal_2020}.
As detailed in Section~\ref{sec:data:rotcurve}, these orbital properties include the circular velocity, $\Vcirc$, angular velocity, $\Omegacirc$, logarithmic derivative of the rotation curve, $\beta$, and Oort's $A$ parameter, $\OortA$.
They are calculated by interpolating the functional fitting model of the rotation curves at the location of the aperture center.

\item[$\sbullet$] \emph{Metallicity}.
We report the predicted gas-phase metallicity in each aperture using a prescription similar to the one described in \citet{Sun_etal_2020b}, but with a few methodological improvements.
In short, we first infer the metallicity at $\rgal = 1.0\,\Reff$ in each galaxy based on a galaxy global mass--metallicity relationship \citep{Sanchez_etal_2019}, and then extrapolate to all $\rgal$ assuming a fixed radial metallicity gradient of $-0.1\,\mathrm{dex}/\Reff$ within each galaxy \citep{Sanchez_etal_2014}.
For better methodological consistency with the original references, here we approximate the galaxy effective radius as $\Reff \approx 1.68\,\Rdisk$, where $\Rdisk$ is the stellar disk scale length.
We also elevate the global stellar masses in Table~\ref{tab:sample} by $0.1$~dex before substituting their values into the mass--metallicity relationship.
We refer interested readers to Appendix~\ref{apdx:alphaCO} for more details about these adjustments.

\item[$\sbullet$] \emph{Molecular gas surface density (kpc-scale)}.
We report the \emph{area-weighted} mean molecular gas surface density, $\Sigmol$, in each kpc-scale aperture.
We emphasize the distinction between this measurement and the mass-weighted average of molecular cloud surface density, $\brkt{\Sigsub{[pix|obj]}}$, defined in Section~\ref{sec:analysis:cloud}.
The area-weighted mean $\Sigmol$ here is calculated from the total CO flux inside the hexagonal boundary of each kpc-sized aperture divided by its total deprojected area.
For this particular calculation, we use the native resolution ``broad'' moment-0 map to ensure a high flux completeness (see Section~\ref{sec:analysis:cloud:completeness}).
We then use the same metallicity-dependent CO-to-H$_2$ conversion factor to convert CO line intensity into mass surface density unit, as we do in Equation~\ref{eq:Sigma_pix}.

\quad We note that our methodology for calculating this kpc-scale aperture averaged $\Sigmol$ is different from the one used in \citet{Sun_etal_2020b}.
There, the kpc-scale $\Sigmol$ was derived via convolving the CO moment-0 maps to a fixed 1~kpc resolution and then sampling the convolved maps at the aperture centers.
As discussed above, the new averaging scheme in this paper leads to better methodological consistency with our calculation of molecular cloud population statistics.

\item[$\sbullet$] \emph{Atomic gas surface density}.
We report the area-weighted mean atomic gas surface density, $\Sigatom$, in all apertures for which we have \HI\ data (see Table~\ref{tab:sample}).
This is calculated in the same way as the area-weighted mean $\Sigmol$: we divide the total \HI\ 21~cm line flux inside the hexagonal aperture by the aperture area, and then convert it to mass surface density unit via Equation~\ref{eq:SigHI}.

\quad Since the \HI\ data resolution is typically comparable to or coarser than our adopted aperture size, our calculated $\Sigatom$ might not reflect the true atomic gas surface density inside the sharp aperture boundaries, but rather a slightly ``smoothed'' version of it.
However, the atomic gas distribution is usually much smoother than the molecular gas \citep[e.g., see][]{Leroy_etal_2013b}, and $\Sigatom$ only plays a minor role throughout this paper.
The resolution degradation is thus not a serious concern for the following analysis.

\item[$\sbullet$] \emph{Stellar mass surface density}.
We report the area-weighted mean stellar mass surface density, $\Sigstar$, in each aperture.
We calculate $\Sigstar$ via Equations~\ref{eq:Sigstar_3p6um} or \ref{eq:Sigstar_3p4um} based on the mean \textit{WISE} 3.4~\micron\ or IRAC 3.6~\micron\ surface brightness at $7.5\arcsec$ resolution within sharp aperture boundaries.
We determine the stellar M/L ratio, $\MtoLwiseone$, for each aperture by sampling the M/L ratio maps from \citet{Leroy_etal_2021a} at the location of the aperture center.

\item[$\sbullet$] \emph{SFR surface density}.
We report the area-weighted mean SFR surface density, $\SigSFR$, in each aperture.
This is primarily calculated via Equations~\ref{eq:SigSFR_FUVW4}--\ref{eq:SigSFR_W4ONLY} based on the best available UV/IR data combination (see Table~\ref{tab:sample}) at $15\arcsec$ resolution.
We note that this resolution could approach the averaging aperture size in the more distant targets in our sample, in which case concerns about correlated measurements could again arise.
To evaluate these concerns, we compare the UV/IR-based $\SigSFR$ measurements with H$\alpha$-based measurements (the latter includes data at much higher angular resolution; see Section~\ref{sec:data:other}).
We find quantitatively consistent results at $\SigSFR\gtrsim10^{-3}\;\uSigSFR$, which is the range of interest in this paper (see Figure~\ref{fig:hist_env} below).

\item[$\sbullet$] \emph{Morphological environment}.
We keep track of the morphological regions each averaging aperture inhabits in the host galaxy (see Section~\ref{sec:data:environ}).
Because of the kpc-scale sizes of these apertures, some of them could stretch across multiple morphological regions.
To deal with this ambiguity, we calculate the fraction of CO flux originating from each morphological region (especially galaxy centers and stellar bars) relative to the sum over the entire aperture.
We then classify all apertures that have a non-zero\footnote{Given the large aperture size, choosing a different threshold (e.g., 10\%) would make a negligible difference in the classification results.} CO flux contribution from galaxy centers or stellar bars as ``center/bar'' apertures, and all the remainder as ``disk'' apertures.

\end{itemize}

\subsection{Outcome of the Cross-spatial-scale Analysis}
\label{sec:analysis:outcome}

Our analysis yields a rich value-added database for each of the \galnumtot\ galaxies listed in Table~\ref{tab:sample}.
These databases present the molecular cloud populations residing in each galaxy, along with the large-scale gas and stellar mass distribution, kinematic information, morphological structures, and star formation activities of the galaxy disk itself.
Together, these high-level measurements have a broad range of applications (see Section~\ref{sec:summary}).
They are published in the form of machine-readable tables online (see Appendix~\ref{apdx:mrt}).

Our databases include \apernumtot\ apertures in total.
These apertures collectively cover the footprint of every target galaxy out to a galactocentric radius limit of $1.5\,\Riso$.
The majority of these apertures have local environmental measurements derived from multiwavelength data (such as UV, and IR), yet only a smaller subset of them have valid molecular gas measurements from PHANGS-ALMA CO data.
This is because the footprint of the PHANGS--ALMA survey is often more confined and covers only the inner, molecular gas-rich part of the galaxy disk \citep[see][]{Leroy_etal_2021a}.
Since this paper focuses on linking the molecular cloud population to their local environment, in the following sections, we restrict ourselves to a subset of \apernumwpix\ apertures that are inside the PHANGS--ALMA survey footprint and show detectable CO emission in the 60--150~pc scale ``strict'' moment maps.
Nonetheless, the full set of \apernumtot\ apertures will be included in the public data release given the rich information provided by the multiwavelength ancillary data alone.


\section{Distributions of Average Molecular Cloud Properties and Sub-galactic Environments in PHANGS--ALMA} \label{sec:stats}

In this section, we characterize the distributions of region-averaged molecular cloud properties and host galaxy local properties across the full PHANGS--ALMA data set.
To do this, we use the databases constructed in Sections~\ref{sec:analysis} and focus on \apernumwpix\ apertures with CO measurements from PHANGS-ALMA (including \apernumwpixdisk\ apertures classified as ``disk'' and the remainder as ``center/bar'').
In the main text, we will only present the statistics of molecular cloud measurements at 150~pc scales, which is the best common resolution achievable for all galaxies.
Quantitative comparisons across different resolutions are shown in Appendix~\ref{apdx:scale}.


\begin{figure*}
\gridline{\fig{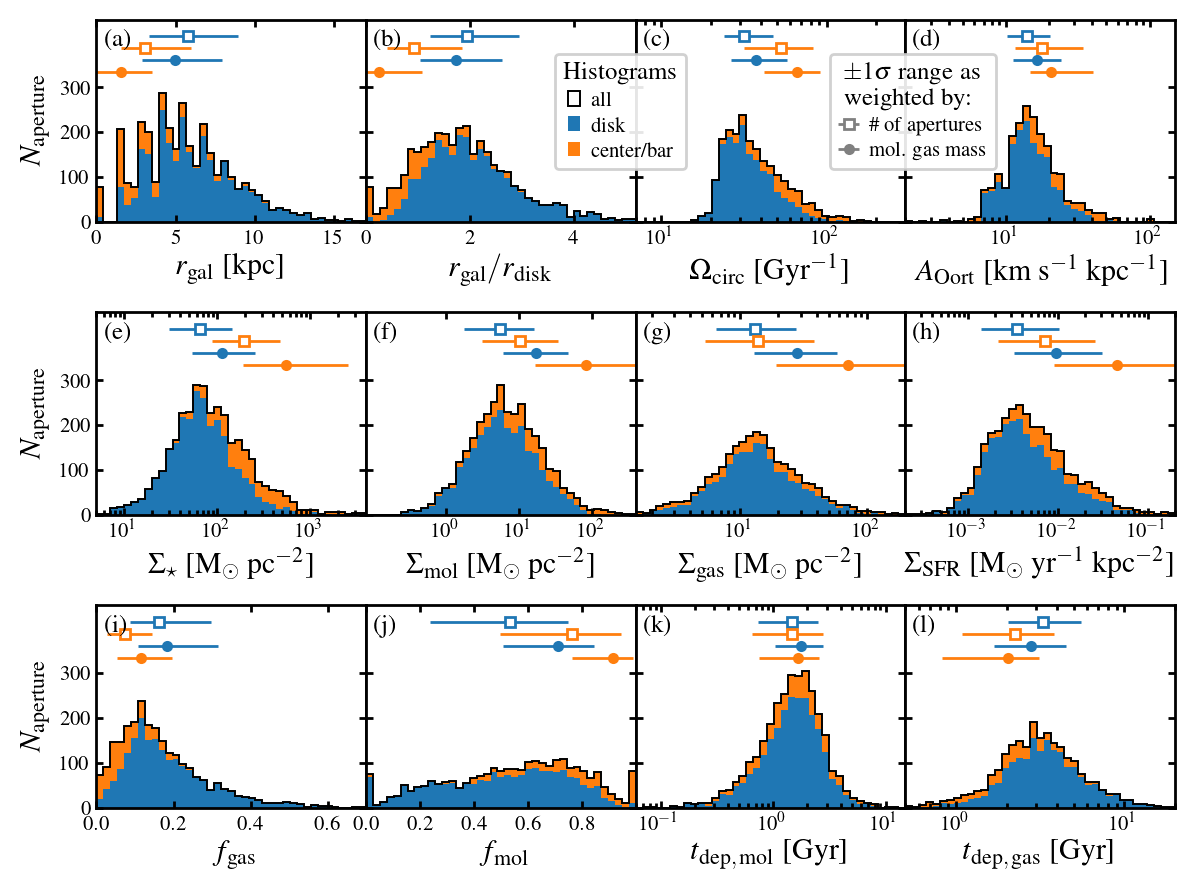}{0.99\textwidth}{}}
\vspace{-2.5\baselineskip}
\caption{
The full range of host galaxy local properties sampled by PHANGS--ALMA, outlined here by stacked histograms of 12 ``local environmental metrics'' across \apernumwpix\ apertures.
The panels show: (a) galactocentric radius, (b) galactocentric radius normalized by the disk scale length, (c) orbital angular velocity, (d) Oort's $A$ parameter, (e) stellar surface density, (f) molecular gas surface density, (g) total gas surface density, (h) SFR surface density, (i) gas fraction, (j) molecular fraction of the gas, (k) molecular gas depletion time, and (l) total gas depletion time.
We use blue and orange colors to distinguish the contributions from the ``disk'' and the ``center/bar'' subsamples in the summed histogram (black solid outline).
The symbols and horizontal error bars at the top show the median value and $\pm1\sigma$ range (i.e., $16{-}84$ percentile range) within each subsample.
The two symbol types correspond to two different weighting schemes for calculating the median values and percentiles: weighting by number of apertures (open squares) versus weighting by molecular gas mass (solid circles).
\vspace{1.0\baselineskip}
}
\label{fig:hist_env}
\end{figure*}

\subsection{Sub-galactic Environments Probed by PHANGS--ALMA}
\label{sec:stats:env}

Our multiwavelength measurements provide a multifaceted depiction of the range of local galactic environments probed by the PHANGS-ALMA survey.
To this end, Figure~\ref{fig:hist_env} shows the histograms of 12 local environmental metrics across \apernumwpix\ apertures.
We also calculate statistics such as the median value and $16{-}84$ percentile range for each environmental metric and tabulate them in Table~\ref{tab:stats_env}.
These statistics are calculated from the histogram using two different weighting schemes: simple counting of the number of apertures or weighting each aperture by the molecular gas mass it encloses.
The first scheme treats all apertures equally, and the calculated statistics reflect \emph{a typical kpc-sized area} covered by PHANGS-ALMA; the latter scheme instead treats each unit of gas mass equally, and the calculated statistics reflect the local environment in which \emph{most molecular gas resides}.

Below we split the 12 environmental metrics into four topical groups and comment on the corresponding histograms and statistics.

\begin{itemize}[itemsep=0.5em,leftmargin=1em,parsep=0em,partopsep=0em]

\item[$\sbullet$] \emph{Galactocentric radii.}
The PHANGS--ALMA CO measurements cover a wide radial range in terms of both absolute and normalized $\rgal$ (panels~\textit{a} and \textit{b}).
When weighting by the number of apertures, we find median values and $\pm1\sigma$ ranges of $\rgal=5.4^{+3.1}_{-2.7}$~kpc and $\rgal/\Rdisk=1.8^{+1.0}_{-0.8}$ across all apertures.
We find smaller values when weighting each aperture by its encircled molecular gas mass.
This reflects that the molecular gas distribution typically peaks toward the galaxy center, and thus apertures at smaller radii often enclose more molecular gas mass.

\quad We note that the $\rgal$ histogram appears ``quantized'' simply due to the fixed \apersize\ linear size of the hexagonal apertures and their tiling pattern on the sky.
This behavior is not obvious in the $\rgal/\Rdisk$ histogram because the normalization factor $\Rdisk$ varies among galaxies, which effectively ``smooth'' the histogram.

\item[$\sbullet$] \emph{Kinematic properties.}
For the subset of apertures located in the \galnumrc\ galaxies with CO kinematic measurements, we report the distributions of orbital angular velocity (panel~\textit{c}) and Oort's $A$ parameter (panel~\textit{d}).
Weighting all apertures equally, we find typical $\Omegacirc=32^{+16}_{-8}~\uOmega$ and $\OortA=14^{+6}_{-4}~\uOortA$ across galaxy disks, which translate to an orbital period of ${\sim}\,200~\ut$ and a local shearing timescale of ${\sim}\,70~\ut$ (i.e., the reciprocal of $\OortA$; also see Section~\ref{sec:timescales}).
These values suggest that the kinematic properties of a typical kpc-sized area probed by PHANGS-ALMA are very similar to those of the Solar Neighborhood \citep[$\Omegacirc=27.8~\uOmega$ and $\OortA=15.3~\uOortA$;][]{Bovy_2017a}.

\quad We also find that apertures located in galaxy centers and stellar bars show systematically higher $\Omegacirc$ and $\OortA$ values.
This is expected from their locations at smaller $\rgal$ and the stronger shear often observed in these environments.

\item[$\sbullet$] \emph{Galaxy disk mass components.}
Weighting all apertures in galaxy disks equally, we find typical surface densities of $\Sigstar=65^{+80}_{-34}~\uSig$ and $\Sigmol=5.4^{+10.8}_{-3.7}~\uSig$ (panels~\textit{e} and~\textit{f}).
Among the \galnumhi\ galaxies with \HI\ 21~cm line data (see Table~\ref{tab:sample}), we find a typical total gas surface density of $\Siggas=\Sigmol+\Sigatom=13^{+14}_{-7}~\uSig$ (panel~\textit{g}).
This gives a typical gas fraction of $\fgas=\Siggas/(\Sigstar+\Siggas)=0.16^{+0.14}_{-0.07}$ (panel~\textit{i}) and a molecular fraction of $\fmol=\Sigmol/\Siggas=0.53^{+0.22}_{-0.30}$ (panel~\textit{j}).
These values are modestly higher than the Solar Neighborhood values \citep[$\Sigstar=33.4~\uSig$, $\Sigatom=10.9~\uSig$, and $\Sigmol=1.0~\uSig$; see][and references therein]{McKee_etal_2015}.

\quad Examining the corresponding molecular gas mass-weighted statistics for galaxy disks, we find that most molecular gas mass resides in environments with even higher surface densities ($\Sigstar=110^{+140}_{-60}~\uSig$, $\Sigmol=17^{+29}_{-11}~\uSig$, $\Siggas=28^{+30}_{-15}~\uSig$) and molecular fraction ($\fmol=0.71^{+0.13}_{-0.20}$).
For comparison, these gas surface densities are likely higher than the averaged value across any kpc-sized neighborhood in our Galaxy \citep[e.g.,][though the central 1~kpc might be an exception given uncertainties in the conversion factor there]{Nakanishi_Sofue_2006,Spilker_etal_2021}.

\item[$\sbullet$] \emph{Star formation activity.}
The typical range of SFR surface density of ``disk'' apertures, when weighted by simple number counts, is $\SigSFR=3.5^{+6.8}_{-2.1}\times10^{-3}~\uSigSFR$ (panel~\textit{h}).
This is again comparable to the estimated Solar Neighborhood SFR surface density at the present day \citep[$\SigSFR=1.7\times10^{-3}~\uSigSFR$;][]{Bovy_2017b}.
Combined with the measured $\Sigmol$ and $\Siggas$ in our sample, this implies typical depletion times of $\tdepsub{mol}=\Sigmol/\SigSFR=1.5^{+1.0}_{-0.7}$~Gyr for the molecular gas (panel~\textit{k}) and $\tdepsub{gas}=\Siggas/\SigSFR=3.3^{+2.2}_{-1.3}$~Gyr for the total gas (panel~\textit{l}).

\quad In comparison, the molecular gas mass-weighted statistics reveal that most of the molecular gas mass resides in more actively star-forming environments with $\SigSFR=9.6^{+20.9}_{-6.4}\times10^{-3}~\uSigSFR$.
Yet associated gas depletion times appear similar to the aperture number-weighted values, with $\tdepsub{mol}=1.8^{+1.0}_{-0.7}$~Gyr and $\tdepsub{gas}=2.8^{+1.7}_{-1.1}$~Gyr.
In other words, the SFR surface density is proportionally higher in these environments as their gas surface densities are.

\end{itemize}

In summary, the PHANGS-ALMA survey covers an wide variety of host galaxy local environments.
When weighting all apertures equally, the most representative local environment in our sample closely resembles the Solar Neighborhood in many aspects.
Comparatively, most molecular gas mass is hosted in regions that are closer to the galaxy center, have higher surface densities of stars, gas, and SFR, and are possibly not matched by any kpc-scale regions in our Galaxy.


\begin{figure*}
\gridline{\fig{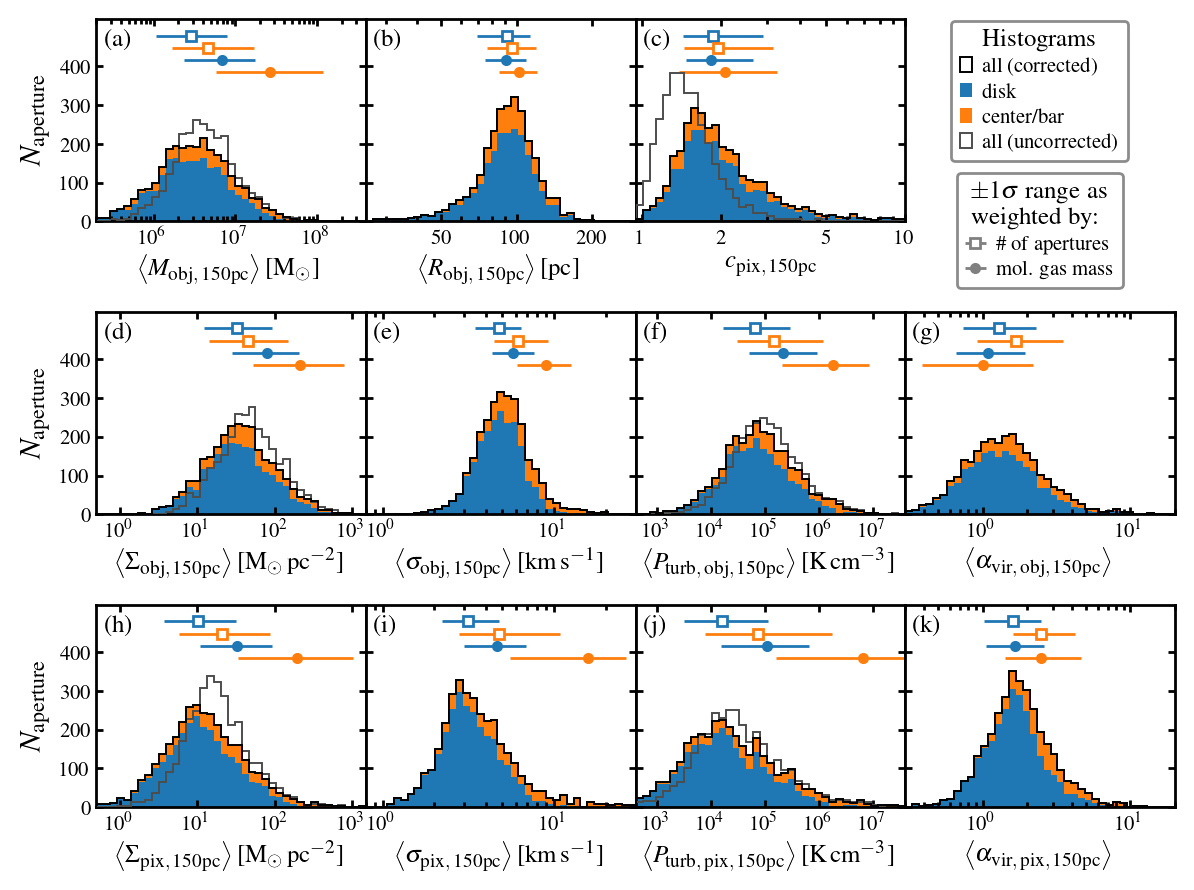}{0.99\textwidth}{}}
\vspace{-2.5\baselineskip}
\caption{
The ``demographic profile'' of molecular cloud populations captured in PHANGS--ALMA, illustrated here by stacked histograms of 11 population-averaged molecular cloud measurements derived from either object- or pixel-based approaches.
As described in Sections~\ref{sec:data:CO} and \ref{sec:analysis}, these quantities represent aperture-wise mass-weighted averages and their derivation accounts for the effect of galaxy inclination and finite data sensitivity.
The panels show: (a) object molecular gas mass, (b) object radius, (c) pixel-wise molecular gas clumping factor, (e--g) object-based molecular gas surface density, velocity dispersion, turbulent pressure, and virial parameter, (h--k) pixel-based molecular gas surface density, velocity dispersion, turbulent pressure, and virial parameter.
The derivation of these properties accounts for the effect of galaxy inclination (Section~\ref{sec:data:CO:obj}--\ref{sec:data:CO:pix}).
We have also applied completeness corrections on a subset of these quantities to offset sensitivity-related biases (Section~\ref{sec:analysis:cloud:completeness}).
The histograms of \emph{uncorrected} measurements (gray, unfilled) are shown in contrast to those of the \emph{corrected} measurements (black).
\vspace{1.0\baselineskip}
}
\label{fig:hist_MC}
\end{figure*}

\subsection{Molecular Cloud Populations in PHANGS-ALMA}
\label{sec:stats:MC}

Our calculations aggregate individual molecular cloud measurements to yield mass-weighted average properties for each aperture (Section~\ref{sec:analysis:cloud}).
The distributions of these population-averaged measurements offer a comprehensive portrait of how cloud populations vary across PHANGS--ALMA.
This is demonstrated by Figure~\ref{fig:hist_MC}, which shows histograms of population-averaged cloud properties measured from object/pixel-based approaches (originally measured at 150~pc).
These histograms include all \apernumwpix\ apertures with pixel-based data and \apernumwobj\ apertures with object-based data.
The latter number is smaller because CPROPS uses a slightly more stringent criterion for identifying objects in CO data cubes \citep[objects made of too few cube pixels are rejected even when they satisfy the S/N criteria;][]{Rosolowsky_etal_2021}.

Similar to Section~\ref{sec:stats:env}, here we calculate the median values and $16{-}84$ percentile ranges with two different weighting schemes (Table~\ref{tab:stats_gmc}).
We note that because most variables depicted in Figure~\ref{fig:hist_MC} (except $c\sbsc{pix}$) are already mass-weighted averages \emph{within individual apertures}, an additional mass-weighted averaging step \emph{across all apertures} can be interpreted as the mass-weighted average cloud properties combining all clouds in all galaxies in PHANGS--ALMA.
In comparison, the other weighting scheme (i.e., same weight for all apertures) gives us a view of the typical molecular cloud population likely to be found at a random location in a PHANGS--ALMA galaxy.
As results from the latter weighting method is less straightforward to interpret, below we focus mostly on the mass-weighted statistics in our discussion.

\begin{itemize}[itemsep=0.5em,leftmargin=1em,parsep=0em,partopsep=0em]

\item[$\sbullet$] \emph{Molecular cloud mass and size.}
The mass-weighted average molecular cloud mass at 150~pc resolution is $6.9^{+10.5}_{-4.5}\times10^6~\uM$ for all clouds in galaxy disks (panel~\textit{a}).
This value is high compared to the typical mass of molecular clouds in the Milky Way \citep[e.g.,][]{Rice_etal_2016,Miville-Deschenes_etal_2017,Colombo_etal_2019}, but is consistent with numbers measured in nearby galaxy studies \citep[e.g.,][]{Hughes_etal_2013a}.
As pointed out by \citet{Rosolowsky_etal_2021}, the finite resolution and sensitivity of the PHANGS--ALMA CO data limit our ability to identify molecular clouds with mass $\lesssim10^5~\uM$.
Specifically, the 60--150~pc resolution of the PHANGS--ALMA data would lead to individual, moderate size clouds being blended into a single object by CPROPS (also see discussions on the resolution-dependence of average cloud mass in Appendix~\ref{apdx:scale}).
That said, our completeness correction can partly remedy sensitivity-related biases by compensating for isolated, less massive clouds undetected in the high resolution CO observations (the histograms for the corrected measurements extend to lower values than those for the uncorrected measurements).

\quad In line with this consideration, the mass-weighted cloud effective radius measured at the same resolution spans $90^{+19}_{-15}$~pc, which slightly exceeds half the beam FWHM size (panel~\textit{b}).
This is consistent with a series of previous studies, all of which found that cloud segmentation algorithms tend to identify objects comparable to or slightly larger than the beam size \citep[e.g.,][]{Verschuur_1993,Pineda_etal_2009,Hughes_etal_2013a,Leroy_etal_2016}.

\item[$\sbullet$] \emph{Molecular cloud surface density.}
At 150~pc resolution, the mass-weighted molecular cloud surface density in galaxy disks is $78^{+124}_{-50}~\uSig$ from the object-based approach and $33^{+60}_{-22}~\uSig$ from the pixel-based approach (both weighted by gas mass; see panels~\textit{d} and \textit{h}).
These values are on the low end of the surface density distribution of Galactic molecular clouds \citep[e.g.,][]{Colombo_etal_2019}.
This likely reflects the coarser spatial resolution of our data compared to most Galactic studies, which means our measurements would be ``diluted'' by low column density sightlines within each beam.

\quad The quantitative differences between the object- and pixel-based approaches reflect that the they attempt to measure fundamentally different quantities (see Section~\ref{sec:data:CO:notes}).
In particular, the CPROPS algorithm attempts to measure the true surface density of an identified object.
Therefore, it includes additional de-convolution and extrapolation procedures, which lead to smaller cloud sizes and larger masses, and thus larger surface densities.
The pixel-by-pixel analysis instead measures the surface density point-by-point from a contiguous molecular gas distribution.
Without a priori expectation for the gas spatial distribution, it does not perform any de-convolution but simply extracts measurements at the resolution of the observations.
Given these differences, for marginally resolved clouds the pixel-based approach would simply yield the native, beam-averaged value at the data resolution, whereas the object-based approach would yield higher surface densities as a result of the de-convolution.

\quad Compared to the results for disk apertures, the cloud populations in galaxy centers and stellar bars have much higher mass-weighted mean surface densities of $210^{+560}_{-160}~\uSig$ (object-based) or $200^{+840}_{-160}~\uSig$ (pixel-based).
Such a trend has been highlighted in previous works on the same galaxies \citep{Sun_etal_2018,Sun_etal_2020a,Rosolowsky_etal_2021} and is also consistent with observations in our Galaxy \citep{Oka_etal_2001}.
We do caution that these results are more sensitive to the choice of $\alphaCO$ prescriptions (see Appendix~\ref{apdx:alphaCO}).
Several lines of evidence suggest lower $\alphaCO$ in galaxy centers \citep[e.g., see][]{Bolatto_etal_2013,Sandstrom_etal_2013,Israel_2020,Teng_etal_2021} and our fiducial prescription only mildly depresses $\alphaCO$ near galaxy centers.

\item[$\sbullet$] \emph{Molecular gas velocity dispersion.}
For molecular cloud populations located in galaxy disks, we find mass-weighted average velocity dispersions of $5.8^{+1.9}_{-1.5}~\uV$ (object-based; panel~\textit{e}) or $4.7^{+2.2}_{-1.7}~\uV$ (pixel-based; panel~\textit{i}).
The cloud populations in galaxy centers or stellar bars show systematically higher values ($9.0^{+3.6}_{-3.0}~\uV$ from object-based and $16^{+11}_{-10}~\uV$ from pixel-based statistics).
These typical values and their environmental dependence are broadly consistent with previous galactic and extragalactic studies \citep[e.g.,][]{Heyer_etal_2009,DonovanMeyer_etal_2013,Hughes_etal_2013a,Leroy_etal_2015a,Sun_etal_2020a,Rosolowsky_etal_2021}.

\quad The quantitative discrepancies between the two approaches here can also be explained by methodological differences.
As described above, the objects identified by CPROPS are often slightly larger than the beam size.
We would then expect larger CO line width measurements from the object-based approach due to both the size--line width relation in molecular clouds \citep[e.g.,][]{Larson_1981} and additional contributions from galaxy rotation and large-scale gas streaming motions.
This explains the sense of deviation for the measurements in galaxy disks.
However, in places where molecular clouds are spatially crowded (such as galaxy centers), the $ppv$ space segmentation in CPROPS helps to demarcate clouds that fall along the same line of sight but are separable in velocity space, whereas the pixel-based analysis simply measures the line effective width and thus cannot tell them apart \citep[see e.g.,][]{Henshaw_etal_2016}.
This explains why the pixel-based approach yields higher velocity dispersions with a wider spread in these environments.

\item[$\sbullet$] \emph{Molecular cloud turbulent pressure.}
Molecular clouds in galaxy disks have mass-weighted average turbulent pressure of $2.1^{+7.2}_{-1.6}\times10^5~\uP$ (object-based; panel~\textit{f}) or $1.1^{+5.4}_{-0.9}\times10^5~\uP$ (pixel-based; panel~\textit{j}).
These values are at least an order of magnitude higher in galaxy centers and stellar bars ($1.8^{+6.8}_{-1.6}\times10^6~\uP$ from object-based and $6.5^{+106.4}_{-6.4}\times10^6~\uP$ from pixel-based statistics), as anticipated from the high environmental pressure there \citep[see][for explicit comparisons between the two]{Schruba_etal_2019,Sun_etal_2020b}.

\quad The sense of deviation between the two approaches here is similar to that of the surface density and velocity dispersion measurements.
It is most apparent near the low pressure end, where the distribution almost always exceeds $10^4~\uP$ in the object-based statistics but extends to below $10^3~\uP$ in the pixel-based statistics.
This aligns with the intuition that the object-based calculations focus on the denser inner portion of molecular clouds (i.e., the half of gas within the FWHM of each object), whereas the pixel-based analysis treats every chunk of molecular gas equally at fixed resolution, and thus can reflect the behavior of the lower pressure, more diffuse gas when it dominates the mass budget.

\item[$\sbullet$] \emph{Molecular cloud virial parameter.}
Molecular cloud populations in galaxy disks exhibit a narrow range of mass-weighted average virial parameters: $1.1^{+0.8}_{-0.4}$ (object-based; panel~\textit{g}) or $1.6^{+0.9}_{-0.6}$ (pixel-based; panel~\textit{k}).
If taken at face value, these values would suggest that the dynamical state of molecular clouds are somewhere between virial equilibrium ($\alphavir=1$) and energy equipartition \citep[$\alphavir=2$; see also][]{Sun_etal_2018,Sun_etal_2020a,Rosolowsky_etal_2021}.
However, systematic uncertainties related to the sub-resolution gas distribution and the CO-to-\Htwo\ conversion factor are especially concerning for the $\alphavir$ measurements given the narrow dynamic range.
More, it can be challenging to determine the true dynamical state of the observed gas structures from a measured $\alphavir$ alone \citep[e.g.,][]{Ballesteros-Paredes_etal_2011a,Ibanez-Mejia_etal_2016,Lu_etal_2020}.
The most conservative conclusions from these data are that molecular cloud populations show a relatively narrow range of dynamical states and appear near energy equipartition across a wide range of environments.

\quad Comparing the mass-weighted average $\alphavir$ between cloud populations in galaxy disks versus those in galaxy centers and stellar bars, the pixel-based statistics indicate higher $\alphavir$ values for the latter, while the object-based statistics show essentially no difference.
This can be explained by the same line-of-sight blending effect that drive the pixel-based velocity dispersion measurements higher in centers and bars \citep[also see][]{Henshaw_etal_2016,Kruijssen_etal_2019a}.
We suggest to prefer the object-based results in this case, but again caution that we may be overestimating $\alphaCO$ in these regions.
If this is the case, the correct object-based values would also suggest higher $\alpha\sbsc{vir}$ in bars and galaxy centers.

\item[$\sbullet$] \emph{Molecular gas clumping factor.}
We find a completeness-corrected clumping factor of $1.9^{+1.1}_{-0.4}$ (uniform weight per aperture) or $1.9^{+1.0}_{-0.4}$ (weighted by molecular gas mass in each aperture) across our sample ({panel~\textit{c}}).
These values are markedly smaller than those reported in \citet[median value of $\sim$7 at 20--300~pc resolution]{Leroy_etal_2013b}, which were calculated from early CO observations targeting a handful of very nearby galaxies (including several CO-poor Local Group members).
This is partly due to the much higher sensitivity of the PHANGS--ALMA data set, and partly due to the improvement in the treatment of non-detections.
The small clumping factors we derive suggest that the molecular gas is much less clumpy than reported in previous studies on a smaller set of galaxies.

\end{itemize}

\input{stats_env}
\vspace{-2\baselineskip}
\input{stats_GMC}
\vspace{-2\baselineskip}

In summary, we observe substantial variations in the molecular cloud population-averaged properties across all apertures in our sample.
The mass-weighted average of all clouds in PHANGS-ALMA yields high masses, large sizes, and low surface densities compared to the cloud population in our Galaxy.
The contrast of cloud populations in galaxy disks and those in center/bar environments are qualitatively consistent with findings in previous extragalactic and Galactic studies.
While the object-based and pixel-based results display qualitatively similar trends, there exist important quantitative discrepancies, which reflect their different measurement approaches in expected ways.
Finally, after correcting for the completeness of CO detections, we find that the molecular gas in PHANGS-ALMA galaxies appears significantly less clumpy than previously determined from low sensitivity CO observations of a few very nearby galaxies.


\section{Environmental Dependence of Molecular Cloud Populations in PHANGS-ALMA}
\label{sec:corr}

In the previous section, we see strong variations in both the molecular cloud populations and the host galaxy local environments from aperture to aperture.
These variations are also known to correlate with each other --- numerous studies have shown that the physical properties of molecular clouds depend on their host galaxy environment in various ways (see discussions in Section~\ref{sec:intro}).
The rich set of measurements derived in this study allow for a systematic characterization of such cloud--environment connections across an unprecedented range of physical conditions.
In this section, we first summarize the basic, pairwise correlations between the population-averaged cloud properties and the local/global host galaxy properties (Section~\ref{sec:corr:pairwise}).
We then perform a variable selection method to identify a subset of host galaxy properties carrying the most predictive power, and to construct empirical predictive models for the molecular cloud properties (Section~\ref{sec:corr:model}).

Both the pairwise correlation analysis and the variable selection procedure requires high quality measurements for the cloud population statistics.
The latter also needs to be applied on a consistent sample of apertures that all have the relevant variables measured to good quality.
To meet these requirements, this part of the analysis works with a subsample of \apernumfin\ apertures from \galnumfin\ galaxies.
These apertures are selected because:
(1) they have the most complete multiwavelength data coverage, such that none of the cloud population statistics or host galaxy environmental metrics are missing; and
(2) the PHANGS--ALMA CO data have reasonably high flux completeness and area coverage fraction inside these apertures ($\Fflux>0.5$ and $\Farea>0.2$, see Appendix~\ref{apdx:completeness}), such that our measured cloud population statistics represent a significant portion of the molecular gas residing in that area.
This down-selection primarily restricts our sample to regions with higher $\Sigmol$ (with all apertures weighted equally, the median value and 16--84 percentile range is $14^{+17}_{-8}\;\uSig$ for the sub-sample, in comparison to $6.0^{+13.3}_{-4.1}\;\uSig$ for the parent sample).
Consequently, it also tends to select apertures in more massive, molecular gas-rich galaxies and at smaller $\rgal$ within each galaxy ($4.4^{+2.4}_{-1.8}\rm\;kpc$ for the sub-sample versus $5.4^{+3.1}_{-2.7}\rm\;kpc$ for the parent sample).
Nevertheless, the selected apertures still cover a considerable range of the relevant parameter space to allow for the following analyses.

\subsection{Pairwise Correlation}
\label{sec:corr:pairwise}

\begin{figure*}
\gridline{\fig{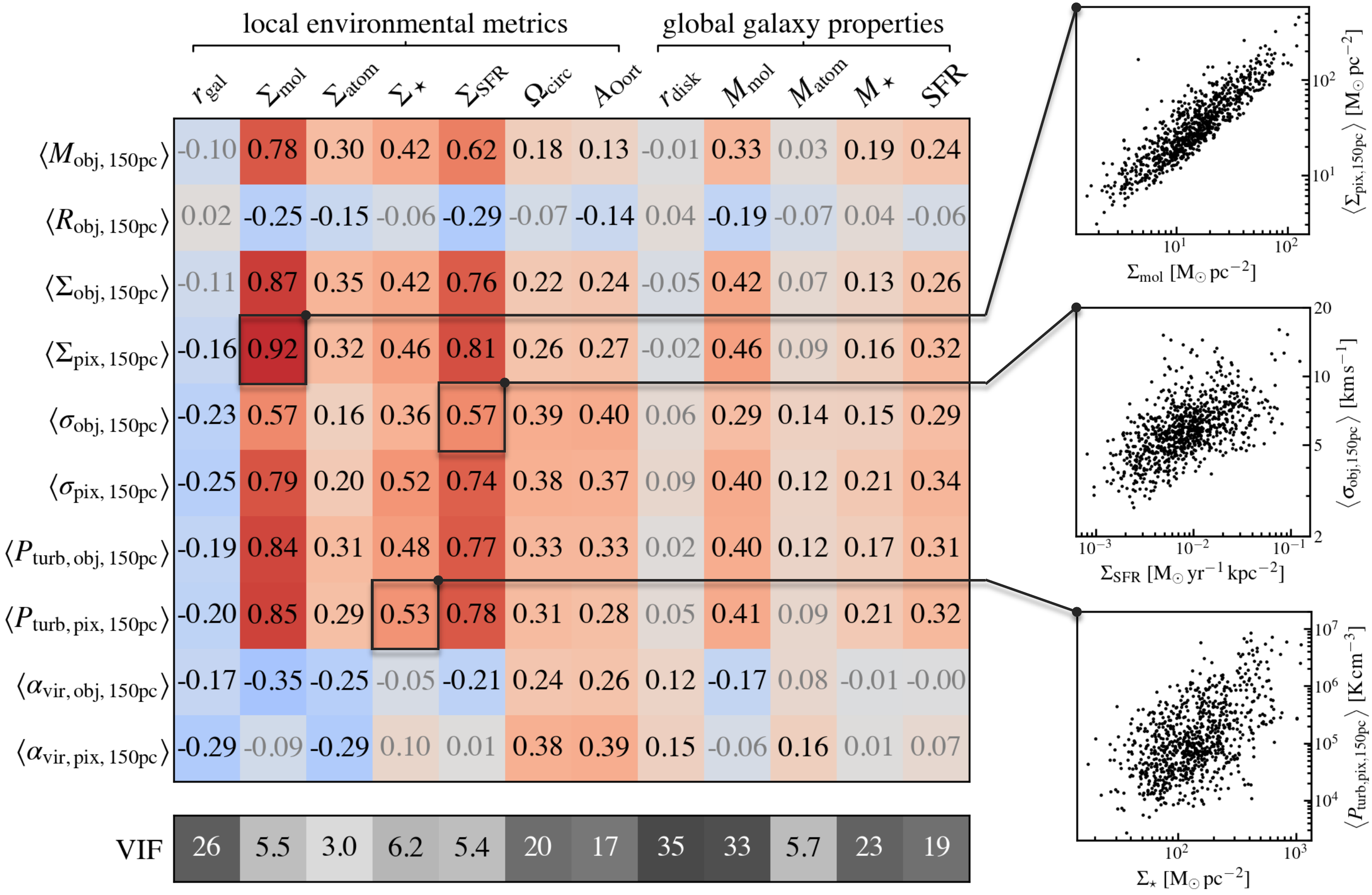}{0.99\textwidth}{}}
\vspace{-2.0\baselineskip}
\caption{
\textit{Top left:} Spearman's rank correlation coefficients between the population-average molecular cloud properties (dependent variables) and the host galactic properties (independent variables).
Darker red/\linebreak[0]{}blue colors indicate stronger positive/\linebreak[0]{}negative correlations.
The number in each entry is the corresponding correlation coefficient, with black/\linebreak[0]{}gray font colors indicating $p$-values smaller/\linebreak[0]{}larger than $0.001$.
\textit{Bottom left:} Variance inflation factors (VIFs) calculated for the independent variables.
Larger VIFs indicates higher multicollinearity, i.e., stronger mutual correlations among the independent variables.
\textit{Right:} Example scatter plots illustrating the correlations between three molecular cloud property--host galaxy property pairs.
}
\vspace{1.0\baselineskip}
\label{fig:corr_all}
\end{figure*}

To provide an empirical characterization of the observed cloud--environment connections, we calculate Spearman's rank correlation coefficients for each pair of population-averaged molecular cloud property and host galaxy environmental metric within the sub-selected \apernumfin\ apertures.
For this analysis, we consider seven local environmental metrics ($\rgal$, $\Sigmol$, $\Sigatom$, $\Sigstar$, $\SigSFR$, $\Omegacirc$, and $\OortA$), each of which holds a unique piece of information unavailable to all other variables.
We complement these local properties with five global galaxy properties ($\Rdisk$, $M\sbsc{mol}$, $M\sbsc{atom}$, $M_\star$, and SFR), so that our correlation analysis can also capture galaxy-to-galaxy trends.
These global properties are measured by \citep{Leroy_etal_2021a} for all PHANGS-ALMA galaxies.

The top left part of Figure~\ref{fig:corr_all} summarize the outcome of this pairwise analysis.
Broadly speaking, we find significant correlations between most pairs of population-averaged cloud properties and environmental metrics.
The signs of the correlation coefficients indicate that an average molecular cloud tends to be denser, more massive, more turbulent, and more strongly self-gravitating at places that are closer to the galaxy center, have higher gas and stellar content, show more active star formation, and feature shorter orbital period and stronger local shear.
The correlations of population-averaged cloud properties versus global galaxy properties are weaker, but most of them are statistically significant.
These findings agree well with previous observations targeting individual galaxy or smaller galaxy samples (see references in Section~\ref{sec:intro}).

Beyond these general trends, we highlight three interesting patterns in Figure~\ref{fig:corr_all}.
First, most molecular cloud properties show the strongest correlation with the (kpc-scale) aperture-averaged molecular gas surface density, $\Sigmol$.
While some systematic effects (e.g., uncertainties in the CO-to-\Htwo\ conversion factor or calibrations of the raw data) could influence both the independent and dependent variables here, these correlations, including the ones regarding cloud surface densities, still carry real information about multi-scale structures in the molecular gas.
Specifically, the correlation strength between the cloud-scale and kpc-scale surface densities partly reflects the inhomogeneity of the molecular gas on spatial scales between \apersize\ (aperture size) and $60{-}150$~pc (data resolution).
The limiting case of a perfect correlation appears only if the gas distribution is completely homogeneous, or if it is structured in a way that the clumping factor is the same in all apertures (which is not far from the reality given the narrow range of clumping factors observed across our sample; see Section~\ref{sec:stats:MC}).
On the contrary, we would expect no correlation if all molecular gas is concentrated into small, isolated clouds with fixed surface densities.

Second, for the cloud radius and virial parameter, the correlations with environmental metrics are weaker compared to the other cloud properties.
This is mainly due to the narrow dynamic range of these quantities in our data.
For the cloud radius, the narrow range is somewhat imposed by the limited data resolution and our adopted object identification algorithm (see Section~\ref{sec:data:CO:obj} and \ref{sec:stats:MC}).
For the virial parameter, the narrow range across our sample is more intrinsic and reflects the relative uniformity of the cloud dynamical state \citep[see][]{Sun_etal_2018,Sun_etal_2020b,Rosolowsky_etal_2021}.

Third, when comparing some of the local environmental metrics to their corresponding ``integrated'', galaxy global measurements (i.e., $\Sigmol$--$M\sbsc{mol}$, $\Sigatom$--$M\sbsc{atom}$, $\Sigstar$--$M_\star$, and $\SigSFR$--SFR), the correlation coefficients for the latter are always smaller.
One possible explanation is that the correlations between cloud populations and their local environment are more fundamental, to the extent that all galaxy-to-galaxy trends arise as their consequences.
In other words, the apparent relationships between cloud populations and global galaxy properties might be completely \emph{mediated} by the local properties.
This hypothesis is challenging to test based on the pairwise correlation coefficients alone, as the strengths of mutual correlations between the independent variables (a.k.a., multicollinearity) are not explicitly modeled.

The issue of multicollinearity is in fact a general concern that impacts more than the local--global quantity pairs identified above.
Many local environmental metrics considered here are known to follow scaling relations \citep[e.g., see reviews by][]{Kennicutt_Evans_2012,Sanchez_etal_2021}, and the same is true for the global galaxy properties \citep[e.g.,][]{Saintonge_Catinella_2022}.
This issue poses challenges to determining whether there are any secondary trends on top of the cloud--environment relationships with the largest correlation coefficients.
To quantify the severity of this issue, the bottom left part of Figure~\ref{fig:corr_all} shows the variance inflation factors (VIFs) of the same independent variable set.
A larger VIF means that a larger fraction of variations in that particular independent variable can be explained by the other independent variables.
The VIFs for many variables exceed commonly adopted cutoff values of $5{-}10$ \citep{James_etal_2013}, signaling strong multicollinearity.

In the following section, we address this issue of multicollinearity with an information criterion-based variable selection method.

\subsection{Variable Selection}
\label{sec:corr:model}

The goal of this section is to identify a subset of environmental metrics that are most directly relevant for setting each population-averaged cloud property.
We attempt to distinguish the most fundamental cloud--environment correlations from the ones that likely arise as indirect consequences of covariance among environment metrics.
Here we distinguish these underlying relations through variable selection.
For each cloud property (as a target variable), we compose an empirical predictive model using a minimal set of environmental metrics (feature variables) that carry most predictive power.
This approach has the advantage of removing irrelevant feature variables while optimizing prediction accuracy in the face of multicollinearity.
While this approach is still limited by the precision at which we can estimate each quantity of interest, it is an effective way to collapse a high-dimensional data set into concise and highly interpretable predictive models.

\subsubsection{Variable Selection Methodology}
\label{sec:corr:model:method}

The basis of this analysis is a multivariable linear regression in the logarithmic space.
That is, we restrict the model functional forms to simple linear combinations of logarithmic variables (including an intercept term), which are equivalent to products of power-laws of the original variables (with a normalization constant).
The regression is done independently for each population-averaged cloud property as a target variable, using all the environmental metrics in Figure~\ref{fig:corr_all} as available features.
Although we have applied inclination corrections to the measured cloud properties, we still include $\cos{i}$ as an extra feature to capture residual trends with inclination when they are present.
After converting our feature and target variables to their logarithms, we median-subtract all features to further reduce correlations between the fitting variables (i.e., power-law slopes versus the normalization constant).

With this regression setup, we perform a \lasso\ model fit \citep{Tibshirani_1996} and use the Bayesian Information Criterion \citep[BIC;][]{Schwarz_1978} for model selection.
This is implemented with the \texttt{LassoLarsIC} function in the \texttt{scikit-learn} Python library.
In detail, for a linear predictive model with the form $\hat{y_i} = \beta_0 + \sum_{j=1}^m\beta_j x_{ij}$, the \lasso\ regression minimizes the following objective function:
\begin{equation}
\frac{1}{2n} \sum_{i=1}^n(y_i-\hat{y_i})^2 +
\alpha \sum_{j=1}^m|\beta_j|~.
\label{eq:lasso}
\end{equation}
\noindent Here $i=1,2,...,n$ is the index of data (index of averaging aperture in our case) and $j=1,2,...,m$ is the index of features in the model.
The $\alpha$ parameter is a non-negative hyper-parameter, so that the second term in Equation~\ref{eq:lasso} adds a penalty for the use of any non-zero slope in the fitted model.
This particular ``regularization'' term is the reason that the \lasso\ as a regression method can also be used for variable selection.

The \lasso\ regression yields a best-fit model that minimizes Equation~\ref{eq:lasso} for each choice of the $\alpha$ parameter.
To guide subsequent model selection, we calculate the BIC value for each such model via
\begin{equation}
\mathrm{BIC} = n \ln(2 \pi \sigma^2) + \frac{\sum_{i=1}^{n} (y_i - \hat{y}_i)^2}{\sigma^2} + d\ln(n)~,
\end{equation}
\noindent where $d$ is the number of features with non-zero slope in that particular model, and $\sigma^2$ is the noise variance for the target variable.
{This expression of BIC is derived from its formal definition of $d\ln(n)-2\ln(\hat{L})$ assuming Gaussian error, with $\hat{L}$ being the maximum of the likelihood function.
It is consistent with other definitions in the literature \citep[e.g.,][]{James_etal_2013} up to irrelevant constants.}

{For the noise variance $\sigma^2$, the contribution from statistical uncertainties} is generally small for our measurements (typical fractional error ${\sim}$1--10\%).
We thus expect several sources of systematic uncertainties on the level of at least 0.1--0.3~dex to dominate the total noise variance.
These sources include (but are not limited to) the estimated $\alphaCO$ and the adopted $\Rsub{21}$ values, the unknown sub-resolution gas spatial and kinematic structures, and calibrations of the ALMA data.
Considering these uncertainties, we conservatively use a noise variance of $\sigma^2=0.1\;\mathrm{dex}^2$ (i.e., about a factor of 2) for the cloud masses, sizes, surface densities, and velocity dispersions.
We use a larger variance of $\sigma^2=0.25\;\mathrm{dex}^2$ (i.e., about a factor of 3) for the turbulent pressures and virial parameters, as they are particularly sensitive to the assumed geometry of the gas.

To complete the model selection procedure, we compare the BIC of all candidate models to their minimum ($\mathrm{BIC_{min}}$) and identify a subset of candidate models that satisfy $\Delta\mathrm{BIC}\equiv\mathrm{BIC}-\mathrm{BIC_{min}}\leq10$ \citep[see e.g.,][for justifications of this $\Delta\mathrm{BIC}=10$ threshold]{Kass_Raftery_1995}.
Among this subset, we select the one model that corresponds to the largest $\alpha$ value, which typically includes the fewest features.
This last step allows us to further suppress any less relevant features, as there is less strong evidence for their inclusion in the model.

\subsubsection{Variable Selection Outcomes}
\label{sec:corr:model:outcome}

\input{lasso_models}

\vspace{-2\baselineskip}

\begin{figure*}
\gridline{\fig{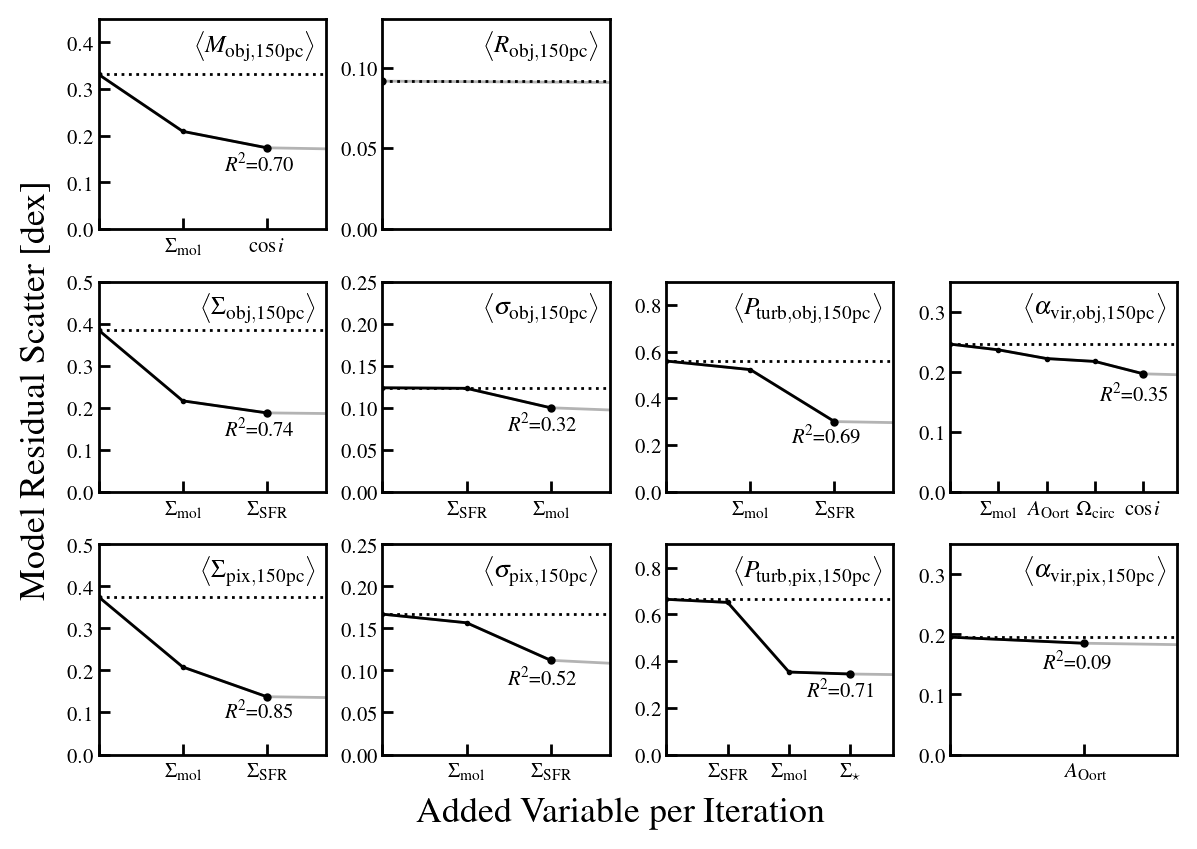}{0.99\textwidth}{}}
\vspace{-2.5\baselineskip}
\caption{
{The full path of the \lasso\ regression up to the ``preferred'' model for each population-averaged cloud property as a target variable.
The model residual scatter decreases as each new feature variable is added into the model.
The ``preferred'' model (large black dot) is selected by comparing the BIC of all models along the full regression path, as explained in Section~\ref{sec:corr:model:method}.
Its coefficient of determination ($R^2$, see text label) can be calculated from the model residual scatter and the total scatter in the target variable (horizontal dotted line).}
}
\vspace{0.5\baselineskip}
\label{fig:lasso_path}
\end{figure*}

With the \lasso\ regression and the BIC-based model selection, we find a set of ``preferred'' power-law predictive models, whose analytical forms are tabulated in Table~\ref{tab:models}.
We also report in Table~\ref{tab:models} the model residual scatter, the model coefficient of determination ($R^2$, which quantifies the model explanatory power), and the BIC difference between the preferred model and a null model with only the normalization term.
Figure~\ref{fig:lasso_path} illustrates the full path of the \lasso\ regression up to the ``preferred'' model.
The model residual scatter reduces as each new feature variable is added into the model.

We find that only a small number ($0{-}4$) of environmental metrics are included in the preferred model for each molecular cloud property, which means that most of the correlations in Figure~\ref{fig:corr_all} can be attributed to a more concise set of fundamental correlations.
This result is in sharp contrast to the impression one would get from the correlations in Figure~\ref{fig:corr_all}, which indicates ubiquitous, significant trends for nearly all local and global galaxy properties.
Evidently, for most galactic properties considered in this work, their apparent correlations with molecular cloud properties are potentially explicable via their covariance with other galaxy properties, such that the predictive models need only a few variables.
Once the modulating effect of those few variables are accounted for, we see no evidence that the remainder play a significant role at the current precision level of our measurements.

Our variable selection exercise allows us to draw some interesting conclusions based on the functional form of the power-law predictive models in Table~\ref{tab:models}.
First of all, the absence of global galaxy properties (except inclination angle, see below) in these models implies that their correlations with cloud properties are not fundamental: these correlations probably originate from the tighter connections between molecular clouds and their local (sub-galactic) environment.
As star-forming galaxies follow various scaling relations, galaxies with larger size, mass, and SFR would include more sub-galactic regions with higher mass and SFR surface densities, which subsequently entails cloud populations with higher average masses, surface densities, velocity dispersions, and turbulent pressures.
This is likely the driver of the observed systematic variations in molecular cloud properties from galaxy to galaxy \citep[e.g.,][]{Hughes_etal_2013a,Sun_etal_2020a,Rosolowsky_etal_2021}.

Furthermore, the galactocentric radius $\rgal$ does not appear in any of the predictive models either.
Given that most physical properties of the host galaxy (including many not considered in this work) are strong functions of galactocentric radius, its general absence in the predictive models is rather encouraging.
It suggests that those environmental metrics included in the models are doing a decent job in capturing most systematic trends: they likely make better proxies than $\rgal$ for many relevant physical quantities not considered in this work (e.g., radiation field, magnetic field, and cosmic ray strength).

The predictive models for specific molecular cloud properties also provide insights into various aspects of molecular cloud formation and evolution.
Below we comment on these models individually:

\begin{itemize}[itemsep=0.5em,leftmargin=1em,parsep=0em,partopsep=-0.5em]

\item[$\sbullet$] \emph{Molecular cloud mass.}
The average molecular cloud mass shows primary dependence on the kpc-scale molecular gas surface density and galaxy inclination: together, these two quantities can explain 70\% of all variations.
On the one hand, we can make sense of the former dependence in light of the theoretical expectation that gravitationally unstable gas disks tend to fragment into objects at a specific mass scale (i.e., the Toomre mass).
This mass scale is often linked to the local gas surface density and the disk vertical scale height $H$ via $M_T \approx \pi H^2 \Sigmol$ \citep[e.g.,][]{Murray_etal_2010}.
While $H$ is not available for variable selection, our derived predictive models for molecular cloud mass do exhibit slightly sub-linear dependencies on $\Sigmol$.
This is consistent with the expected anti-correlation between $\Sigmol$ and $H$, as $\Sigmol$ rapidly declines with galactocentric radius, while $H$ mildly increases with it in the inner part of nearby disk galaxies \citep[e.g.,][]{Yim_etal_2011,Yim_etal_2014}.

\quad On the other hand, a non-trivial portion of the trends with $\Sigmol$ and $\cos{i}$ could originate from observational and methodological limitations.
The nearly constant cloud radii given by CPROPS, in combination with small clumping factors (see Section~\ref{sec:stats:MC}), would naturally produce strong correlations between masses of the identified clouds and the large-scale $\Sigmol$.
The additional $\cos{i}$ dependence also signifies stronger source blending in more inclined galaxies, despite CPROPS's attempt to deblend based on information in the velocity space (see Section~\ref{sec:data:CO:obj}).
Therefore, any attempt to interpret the predictive model should also take these non-physical factors into consideration.

\item[$\sbullet$] \emph{Molecular cloud radius.}
The BIC-based variable selection favors the ``null'' model for the average molecular cloud radius, i.e., the one that includes only a normalization term and nothing else.
In other words, there is no strong evidence that any of the environmental metrics considered here can effectively predict the (small) variations in the cloud radius.
This agrees with the notion that the sizes of the CPROPS-identified objects are more influenced by algorithm-related factors (e.g., deblending criteria) and data characteristics (e.g., beam size; see Figure~\ref{fig:resolution}) than physical properties of the gas or the host galaxy environment \citep[also see][]{Hughes_etal_2013a,Rosolowsky_etal_2021}.

\item[$\sbullet$] \emph{Molecular cloud surface density.}
For this quantity (measured with either object-based or pixel-based approach), we find significant, secondary trends with $\SigSFR$ on top of the prominent correlations with the large-scale $\Sigmol$.
A possible explanation is that regions with more clumpy molecular gas (i.e., larger clumping factor, higher $\brkt{\Sigsub{150\,pc}}/\Sigmol$ ratio) are possibly more subject to gravitational instabilities and the gas there has more chance to form stars.
This points at a possibility that knowing the molecular gas clumping factor on 60--150~pc scales could allow for better prediction of $\SigSFR$ at a given $\Sigmol$ on kpc scales \citep[i.e., improving upon the Schmidt-Kennicutt relation;][]{Schmidt_1959,Kennicutt_1998}.

\quad It is also worth noting that the preferred models for the average cloud surface densities do not include a $\cos{i}$ dependent term.
Though partly by construction, this still affirms the effectiveness of the inclination correction on the cloud surface densities (see Section~\ref{sec:data:CO:obj} and \ref{sec:data:CO:pix} as well as Appendix~\ref{apdx:inclination}).

\item[$\sbullet$] \emph{Molecular gas clumping factor.} We do not separately construct a power-law predictive model for the clumping factor because this quantity can be well approximated by $\brkt{\Sigsub{pix}}/\Sigmol$ when the data sensitivity is sufficiently high (see Section~\ref{sec:analysis:cloud:clumping}). Instead, one can easily derive a predictive model for this quantity by dividing the model for $\brkt{\Sigsub{pix}}$ by $\Sigmol$.

\item[$\sbullet$] \emph{Molecular cloud velocity dispersion.}
The preferred models for this quantity include the same environmental metrics, $\Sigmol$ and $\SigSFR$, as the models for the cloud surface densities.
Since molecular gas surface density and velocity dispersion correlate strongly even on a cloud-to-cloud level (see discussions in Section~\ref{sec:intro}), it is not surprising that the same environmental metrics turn out to be most relevant for both molecular cloud properties after population averaging.
However, the preferred predictive models for the cloud velocity dispersion explain a much less fraction of its total observed scatter (32--52\%) compared to those for the cloud surface density (74-85\%).
At least part of this is attributable to the latter fraction being exaggerated, because the cloud-scale and kpc-scale molecular gas surface densities tend to co-vary for many non-physical reasons (e.g., relying on the same conversion factor).
It is also possible that the physical drivers of gas velocity dispersion variations are less well-captured by the set of environmental metrics included in this work \citep[e.g., gas inflow rate is another possible driver of velocity dispersion variations; see][]{Krumholz_Burkert_2010}.

\item[$\sbullet$] \emph{Molecular cloud turbulent pressure.}
The functional forms of the preferred models for this cloud property are broadly consistent with the expectation from the models for the average cloud surface density and velocity dispersion.
Interestingly, the preferred model for the pixel-based turbulent pressure also includes an extra term depending on the large-scale stellar mass surface density, $\Sigstar$, albeit with a small power-law index (0.08).
Adding this term improves the model $R^2$ by only a small amount (from $0.70$ to $0.71$; see Figure~\ref{fig:lasso_path}), but it lowers the model BIC by more than 10, which means that our data clearly favor the model with an extra dependence on $\Sigstar$.
This extra dependence is in line with theoretical models proposing that molecular clouds can be influenced by the external gravitational potential of the host galaxy stellar disk \citep[e.g.,][]{Meidt_etal_2018,Meidt_etal_2020,Sun_etal_2020b,LiuLJ_etal_2021}.

\item[$\sbullet$] \emph{Molecular cloud virial parameter.}
For this quantity, the preferred models for the object-based and pixel-based results show the largest deviation.
The preferred model for the object-based measurement includes four galaxy properties ($\Sigmol$, $\Omegacirc$, $\OortA$, and $\cos{i}$), whereas that for the pixel-based only includes $\OortA$.
This difference is probably related to the narrower dynamic range in $\brkt{\alphavirsub{pix,\,150pc}}$ than in $\brkt{\alphavirsub{obj,\,150pc}}$ (see Section~\ref{sec:stats:MC}).

\quad Moreover, the appearance of $\Omegacirc$, $\OortA$, and $\cos{i}$ in the models points at potential influence of galactic rotation on the inferred cloud dynamical state.
Particularly, if the measured velocity dispersion includes contributions from differential galactic rotation (i.e., beam smearing), it would lead to positive correlations with $\Omegacirc$ and $\OortA$ because they reflect the strength of the differential motion, and with $\cos{i}$ because the beam smearing effect is more prominent in more inclined galaxies.

\end{itemize}
\vspace{0.5\baselineskip}

In summary, through the \lasso\ regression and BIC-based model selection we compose power-law predictive models for all population averaged molecular cloud properties.
These models capture the primary cloud--environment correlations with at most four environmental metrics as independent variables.
The most commonly involved environmental metrics in these models are the large-scale molecular gas surface density, $\Sigmol$, and SFR surface density, $\SigSFR$.
Furthermore, the general absence of global galaxy properties in these models suggests that galaxy-to-galaxy variations in molecular cloud populations might be the mere consequences of their tighter connections with sub-galactic environments.


\section{Characteristic Timescales of Molecular Cloud Evolution in PHANGS--ALMA}
\label{sec:timescales}

Molecular cloud formation and evolution are influenced by a number of physical processes including turbulence driving and cascade, gravitational collapse, galactic rotation and shearing motions, cloud--cloud collisions, and gas depletion due to star formation.
These processes not only operate over a vast span of spatial scales, but also feature different characteristic timescales.
Estimating their timescales across diverse galactic environments allows us to demonstrate the balance (or not) between these processes and infer how star formation is regulated under distinct physical conditions \citep{Wong_2009,Jeffreson_Kruijssen_2018,Kruijssen_etal_2019b,Chevance_etal_2020a,Chevance_etal_2020b,KimJY_etal_2020}.

In this section, we estimate six different characteristic timescales as a use case demonstration for our rich multiwavelength measurements.
We detail the definition and derivation of each timescale in Section~\ref{sec:timescales:method}, and compare the quantitative results in Section~\ref{sec:timescales:results}.

\subsection{Timescale Definitions}
\label{sec:timescales:method}

\begin{itemize}[itemsep=0.5em,leftmargin=1em,parsep=0em,partopsep=-0.5em]

\item[$\sbullet$] \emph{Free-fall time, $t\sbsc{ff}$.}
This is the timescale for a molecular cloud to collapse in free-fall due to self-gravity, provided no pressure support to counterbalance it.
We estimate this timescale from the mean volume density of molecular clouds under the assumption of spherical symmetry.
The population-average free fall time, $\bar{t}\sbsc{ff}$, is subsequently calculated via
\begin{equation}
    \frac{1}{\bar{t}\sbsc{ff}}
    = \brkt{\frac{1}{t\sbsc{ff}}}
    \simeq \brkt{\sqrt{\frac{32G \rho\sbsc{mol}}{3\pi}}}
    = \brkt{\sqrt{\frac{8G M\sbsc{mol}}{\pi^2 R\sbsc{cloud}^3}}}~.
    \label{eq:t_ff}
\end{equation}
\noindent Here $M\sbsc{mol}$ and $R\sbsc{cloud}$ are the cloud mass and radius, estimated from either object- or pixel-based approaches\footnote{We use $M\sbsc{mol}=M\sbsc{obj}/2$ for the object-based measurements to be consistent with our calculations in Section~\ref{sec:data:CO:obj}.}.
The ``$\brkt{}$'' symbol denotes the same mass-weighted averaging scheme as defined in Section~\ref{sec:analysis:aperture}.
We define this population-averaged free fall time as a \emph{mass-weighted harmonic mean} so that it appropriately reflects the overall timescale for the whole cloud population\footnote{{This mass-weighted harmonic mean can be very convenient in the following scenario: if all clouds form stars on their corresponding free-fall timescale with the same efficiency per free-fall time ($\epsilon\sbsc{ff}$), one can easily derive the total SFR of a cloud population via $\epsilon\sbsc{ff}\,M\sbsc{tot}/\bar{t}\sbsc{ff}$, where $M\sbsc{tot}$ is the total gas mass held by the cloud population, and $\bar{t}\sbsc{ff}$ the population-averaged free fall time defined by the mass-weighted harmonic mean.}} \citep[also see][]{Jeffreson_Kruijssen_2018,Utomo_etal_2018}.
We apply a similar completeness correction to this measurement as we did for the other population-averaged cloud properties (see Appendix~\ref{apdx:completeness} for more detail).

\item[$\sbullet$] \emph{Turbulence crossing time, $t\sbsc{cr}$.}
This is the timescale for the turbulent flow to cross the span of a molecular cloud.
We drive it from the cloud radius and the (one-dimensional) turbulent velocity dispersion, and then calculate the mass-weighted harmonic mean as
\begin{equation}
    \frac{1}{\bar{t}\sbsc{cr}}
    = \brkt{\frac{1}{t\sbsc{cr}}}
    \simeq \brkt{\frac{\sigmol}{R\sbsc{cloud}}}.
    \label{eq:t_cr}
\end{equation}
\noindent Here $R\sbsc{cloud}$ and $\sigmol$ are the radius and one-dimensional velocity dispersion of individual molecular clouds, again derived from either object-based or pixel-based analyses (see Sections~\ref{sec:data:CO:obj} to~\ref{sec:data:CO:pix}).
Under this definition, the crossing time is related to the free-fall time and the virial parameter via $t\sbsc{ff}/t\sbsc{cr} \approx 0.50\,\alphavir^{0.5}$.
Therefore, the crossing time of a virialized molecular cloud would be roughly two times longer than its free-fall time.

\item[$\sbullet$] \emph{Orbital time, $t\sbsc{orb}$.}
This is the period of the orbital revolution around the galaxy center.
We derive it from the orbital angular velocity measured from the CO rotation curves (see Section~\ref{sec:data:rotcurve}):
\begin{equation}
    t\sbsc{orb} = 2\pi/\Omegacirc~.
    \label{eq:t_orb}
\end{equation}

\item[$\sbullet$] \emph{Shearing time, $t\sbsc{shear}$.}
This is the timescale for two objects to move closer/farther by a unit length azimuthally, given that they are on two circular orbits separated radially by the same unit length.
It equals the reciprocal of Oort's $A$ parameter (see Section~\ref{sec:data:rotcurve}):
\begin{equation}
    t\sbsc{shear} = \OortA^{-1} = \frac{2}{\Omegacirc\,(1-\beta)}~.
    \label{eq:t_shear}
\end{equation}

\item[$\sbullet$] \emph{Cloud--cloud collision time, $t\sbsc{coll}$.}
Most generally, this is the timescale for any \emph{particular} molecular cloud to collide with another cloud (i.e., it is \emph{not} the timescale for such collisions to happen within a given area).
We estimate this timescale following a simplified model of shear-induced collision \citep{Tan_2000}.
The key assumptions are that molecular clouds are randomly distributed in each aperture, and that cloud--cloud collision happens \textit{only} when clouds catch up with other clouds on adjacent circular orbits due to orbital shear.
In this scenario, we can estimate a population-averaged collision time, $\bar{t}\sbsc{coll}$, via:
\begin{align}
    \frac{1}{\bar{t}\sbsc{coll}}
    = \brkt{\frac{2v\sbsc{shear}}{\lambda\sbsc{mfp}}}
    &\simeq \brkt{\frac{2R\sbsc{cloud}/t\sbsc{shear}}{(2R\sbsc{cloud}N\sbsc{cloud})^{-1}}} \nonumber\\
    &= \frac{4}{t\sbsc{shear}}N\sbsc{cloud}\brkt{R\sbsc{cloud}^2}~.
    \label{eq:t_coll}
\end{align}
\noindent Here $v\sbsc{shear} \simeq R\sbsc{cloud}/t\sbsc{shear}$ is the shear velocity of two orbits separated radially by $R\sbsc{cloud}$, the average impact parameter among all collisions.
$\lambda\sbsc{mfp} \simeq (2R\sbsc{cloud}N\sbsc{cloud})^{-1}$ is the mean free path of cloud--cloud collisions given a linear cross section of $2R\sbsc{cloud}$ and an area number density of $N\sbsc{cloud}$.
The extra factor of 2 on the numerator accounts for the fact that the other cloud can be located on either an inner orbit (smaller $\rgal$) or an outer orbit (larger $\rgal$) relative to the cloud in question \citep{Tan_2000}.

\quad Equation~\ref{eq:t_coll} makes it straightforward to estimate $\bar{t}\sbsc{coll}$ from measurable quantities in the object-based approach (or specifically, $N\sbsc{cloud}$ as area density of identified objects and $\brkt{R\sbsc{cloud}^2}$ as mass-weighted average of object radius squared).
Yet it is not trivial to measure them with the pixel-based approach in a totally symmetric way.
Alternatively, we follow a similar line of reasoning as \citet{Tan_2000} to approximate $\bar{t}\sbsc{coll}$ from other pixel-based measurements:
\begin{align}
    \frac{1}{\bar{t}\sbsc{coll}}
    \simeq \frac{4}{t\sbsc{shear}}N\sbsc{cloud}\,R\sbsc{cloud}^2
    &\simeq \frac{4}{t\sbsc{shear}}\frac{\Sigmol}{\pi\Sigsub{cloud}} \nonumber\\
    &\simeq \frac{4}{\pi\,t\sbsc{shear}\,c\sbsc{pix}}~.
    \label{eq:t_coll_pix}
\end{align}
\noindent The second step assumes all molecular gas is concentrated into clouds with characteristic surface densities $\Sigma\sbsc{cloud}$ and radii $R\sbsc{cloud}$, such that $\Sigmol \approx N\sbsc{cloud}\, (\Sigma\sbsc{cloud}\,\pi R\sbsc{cloud}^2$).
The third step follows from the definition of the molecular gas clumping factor, $c\sbsc{pix}$, as the contrast between molecular gas surface densities on cloud scales and kpc scales \citep[see discussion in Section~\ref{sec:analysis:cloud:clumping} and][]{Leroy_etal_2013b}.

\quad We caution that the simplifying assumptions involved in Equations~\eqref{eq:t_coll} and~\eqref{eq:t_coll_pix} likely bias our $\bar{t}\sbsc{coll}$ estimates high.
In reality, molecular clouds are not evenly distributed and have random motions in addition to circular rotation \citep[also see][]{Dobbs_etal_2015}.
Possible blending of multiple clouds in a single beam or a single identified object can also lead to longer estimated collision timescales than reality.

\item[$\sbullet$] \emph{Molecular gas depletion time, $\tdepsub{mol}$.}
This is the timescale to convert all molecular gas into stars at the current SFR, provided no other sources or sinks for the gas:
\begin{equation}
    \tdepsub{mol} = \Sigmol/\SigSFR ~.
    \label{eq:t_dep}
\end{equation}

\end{itemize}

\subsection{Timescale Comparisons}
\label{sec:timescales:results}

\begin{figure}[!htb]
\gridline{\fig{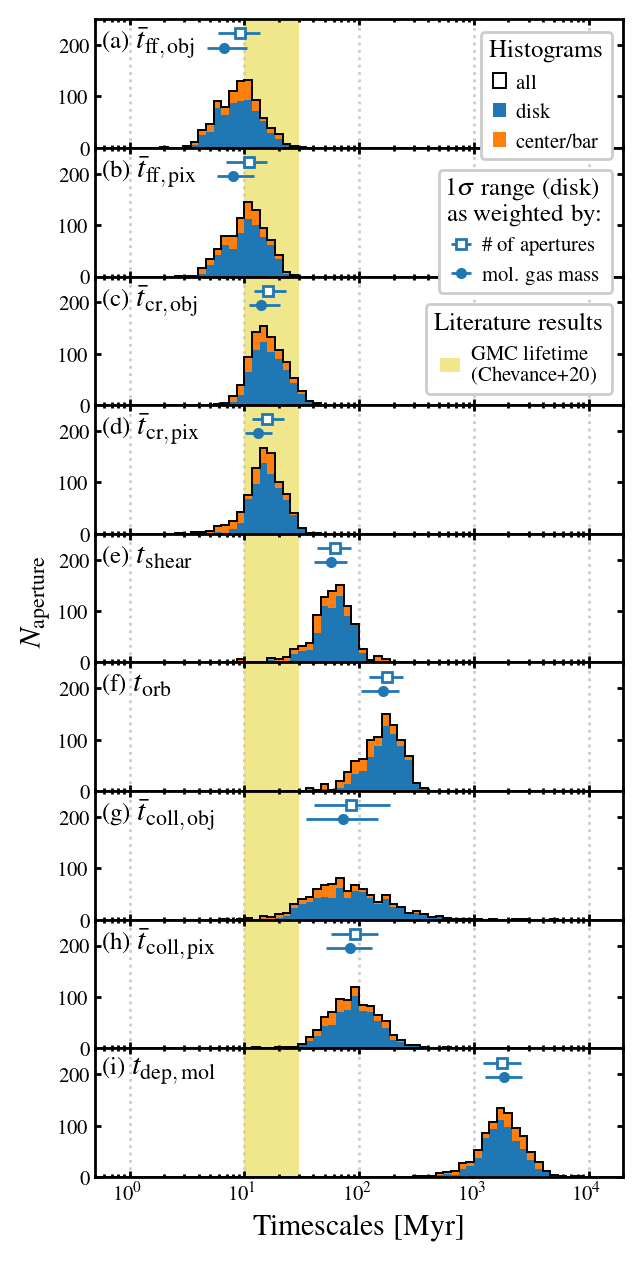}{0.46\textwidth}{}}
\vspace{-2.5\baselineskip}
\caption{
Stacked histograms of molecular cloud free-fall time (\textit{a} \& \textit{b}), turbulence crossing time (\textit{c} \& \textit{d}), shearing time  (\textit{e}), orbital time (\textit{f}), cloud--cloud collision time (\textit{g} \& \textit{h}), and molecular gas depletion time (\textit{i}).
Symbols and colors have the same meaning as in Figure~\ref{fig:hist_env}.
The shaded region in light brown marks the range of molecular cloud lifetimes measured in a subset of PHANGS--ALMA galaxies \citep[$10{-}30$~Myr;][]{Chevance_etal_2020a,Chevance_etal_2020b,KimJY_etal_2020}.
}
\label{fig:timescales}
\vspace{-0.5\baselineskip}
\end{figure}

\begin{figure*}
\gridline{\fig{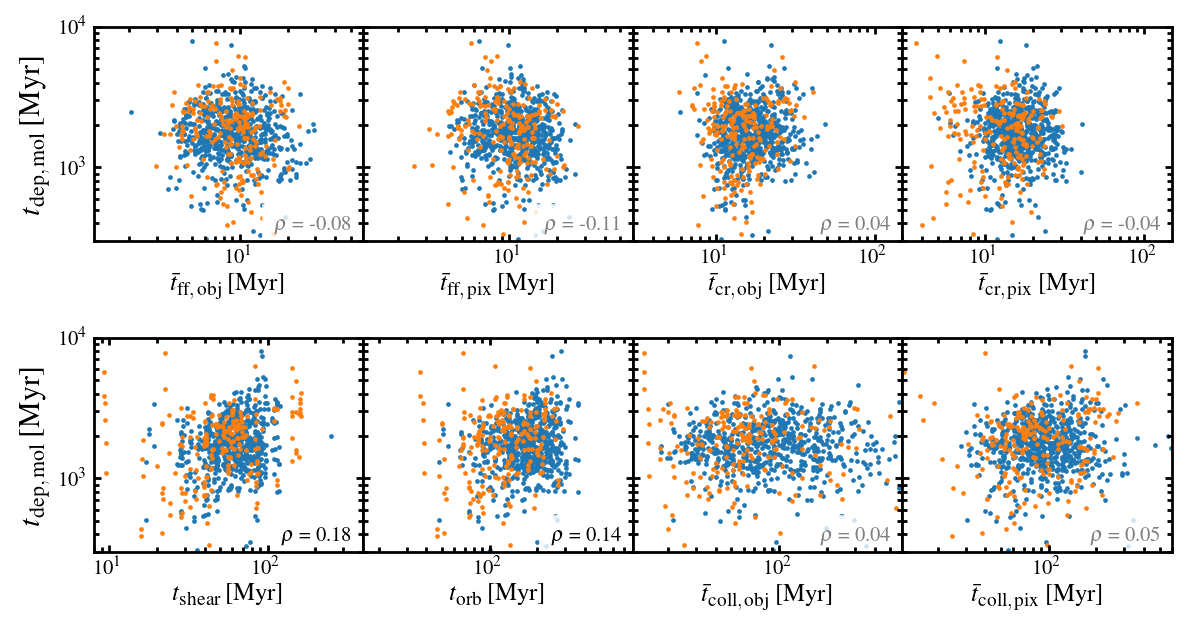}{0.99\textwidth}{}}
\vspace{-2.5\baselineskip}
\caption{
Correlations between molecular gas depletion time ($t\sbsc{dep\,mol}$) versus each of the other timescales shown in Figure~\ref{fig:timescales}.
The $x$-axes of all panels are scaled to show the same logarithmic range.
Measurements in the ``disk'' and ``center/bar'' samples are denoted by data points in blue and orange, respectively.
We display Spearman's rank correlation coefficient for all measurements at the lower right corner in each panel, with black (gray) font color indicating a $p$-value smaller (larger) than 0.001.
\vspace{0.5\baselineskip}
}
\label{fig:scatter_timescales}
\end{figure*}

Figure~\ref{fig:timescales} shows the statistical distributions of all six timescales, including their variants derived from object- or pixel-based measurements.
For reference, we also mark the range of molecular cloud lifetimes, $t\sbsc{life}$, measured from the spatial \mbox{(de-)}correlation of molecular gas and young star tracers in a subset of PHANGS--ALMA targets \citep{Chevance_etal_2020a,Chevance_etal_2020b,KimJY_etal_2020}.
Table~\ref{tab:stats_timescales} summarizes the median values and $1\sigma$ ranges of all our estimated timescales.

Based on Figure~\ref{fig:timescales} and Table~\ref{tab:stats_timescales}, we can identify three distinct groups of timescales separated by roughly an order of magnitude apart from each other.
The first group consists of $\bar{t}\sbsc{ff}$ and $\bar{t}\sbsc{cr}$, both around $5{-}20~\ut$.
These timescales correspond to physical processes taking place inside molecular clouds.
The median values of $t\sbsc{ff}$ and $t\sbsc{cr}$ differs by less than a factor of two, which is no more than a restatement that most molecular clouds have virial parameters of order unity.
Furthermore, they appear comparable to (or only slightly shorter than) $t\sbsc{life}$, as seen in previous observations and simulations\footnote{We note that some studies in the literature argue for a longer molecular cloud lifetime on the order of $100$~Myr \citep[e.g.,][]{Scoville_Hersh_1979,Koda_etal_2009}. These studies typically use the timescales of galactic dynamical processes (such as orbital time or spiral arm crossing time) as anchoring points to derive molecular cloud lifetimes, which might partly explain why their estimated cloud lifetimes are comparatively longer.} \citep{Fukui_etal_1999,Elmegreen_2000,Kawamura_etal_2009,Murray_2011,Grudic_etal_2018,Kim_etal_2018,Kruijssen_etal_2019b,Benincasa_etal_2020a,Chevance_etal_2020a}.
This is not inconsistent with our estimated $\alphavir=1{-}2$, as even clouds in free-fall collapse can yield apparent virial parameters of ${\sim}2$ \citep[e.g.,][]{Ballesteros-Paredes_etal_2011a,Camacho_etal_2016}.

The second group of timescales consists of $t\sbsc{shear}$, $t\sbsc{orb}$, and $\bar{t}\sbsc{coll}$, all of which are ${\sim}100~\ut$.
These timescales characterize dynamical processes taking place on kpc scales or even over entire galaxies.
The order of magnitude contrast between them and the cloud ``internal'' timescales discussed above implies that the effects of galactic-scale dynamics on individual molecular clouds are likely modest.
More specifically, $t\sbsc{shear}\gg\bar{t}\sbsc{cr}\;\text{or}\;\bar{t}\sbsc{ff}$ indicates that shearing motions are generally small on cloud scales relative to motions generated by turbulence or gravitational collapse \citep[at least in most regions targeted by PHANGS--ALMA; also see][]{Utreras_etal_2020};
$t\sbsc{orb}\gg t\sbsc{life}$ means that molecular clouds can only last a small fraction of a complete orbital revolution around the galaxy center \citep[see][]{Chevance_etal_2020a};
and $\bar{t}\sbsc{coll}\gg t\sbsc{life}$ suggests that cloud--cloud collisions do not happen to most molecular clouds throughout their lifetimes \citep[c.f.\ \citealt{Dobbs_etal_2015}]{Blitz_Shu_1980,Jeffreson_Kruijssen_2018}.

The last group consists of $\tdepsub{mol}$ by itself, which is about $1{-}3$~Gyr across our sample.
This range is consistent with many previous studies on the molecular dominated regions in nearby, massive, star-forming galaxies \citep[e.g.,][]{Bigiel_etal_2008,Leroy_etal_2008,Utomo_etal_2017,Muraoka_etal_2019,Ellison_etal_2020c}.
The large ratios between $\tdepsub{mol}$ and all other characteristic timescales reaffirm the notion that star formation is inefficient:
the implied star formation efficiency is $0.5{-}1\%$ per free-fall time or turbulence crossing time \citep[see][J.~Sun et al.\ in preparation]{Evans_etal_2014,Lee_etal_2016,Vutisalchavakul_etal_2016,Utomo_etal_2018}, ${\sim}1\%$ per cloud lifetime \citep{Kruijssen_etal_2019a,Chevance_etal_2020a,Chevance_etal_2021,KimJY_etal_2020}, and ${\sim}10\%$ per orbital revolution or cloud--cloud collision \citep{Silk_1997,Kennicutt_1998a}.

Beyond the typical values of the estimated timescales and their ratios, we also examine which timescales correlate the best with $\tdepsub{mol}$.
Figure~\ref{fig:scatter_timescales} shows that all the other timescales we considered show weak to no correlations with $\tdepsub{mol}$ (judging from the small correlation coefficients).
The only statistically significant trends are with $t\sbsc{shear}$ and $t\sbsc{orb}$.
They exhibit mild positive correlations with $\tdepsub{mol}$ with coefficients of $\rho=0.18$ and $0.14$, quantitatively consistent with the results in \citet{Wong_2009}.
Since these two timescales only differ by a factor of $\pi/(1-\beta)$, and the measured $1-\beta$ has a narrow dynamic range across our sample and a relatively large uncertainty, it is expected that $t\sbsc{shear}$ and $t\sbsc{orb}$ contain virtually the same amount of information and have similar predictive power for $\tdepsub{mol}$.

\input{stats_timescales}
\vspace{-2\baselineskip}


\section{Summary} \label{sec:summary}

This work examines the fundamental correlations between molecular clouds and their host galaxy environments in \galnumtot\ nearby, massive, star-forming galaxies targeted by the PHANGS--ALMA survey.
It directly addresses one of the core science questions that motivated the PHANGS--ALMA survey: \emph{how do molecular cloud populations depend on local and global properties of the host galaxy?}
Taking advantage of the large, representative galaxy sample and the homogeneous, high-quality data provided by PHANGS--ALMA, we provide a first systematic description of the environmental dependence of the cloud populations residing in typical star-forming environments across the local universe.

To achieve this overarching goal, we use PHANGS--ALMA \CO21 imaging data products \citep{Sun_etal_2018,Sun_etal_2020a,Leroy_etal_2021a,Leroy_etal_2021b} and CPROPS-based object catalogs \citep[A.~Hughes et al.\ in preparation]{Rosolowsky_etal_2021} to determine a rich set of molecular gas properties on $60{-}150$~pc scales.
We further complement these molecular cloud scale measurements with multiwavelength observations covering UV, optical, IR, and radio bands \citep[e.g.,][A.~Razza et al.\ in preparation, A.~Sardone et al.\ in preparation]{Leroy_etal_2019,Querejeta_etal_2021}, as well as high-level data products including rotation curves \citep{Lang_etal_2020} and morphological feature catalogs \citep{Querejeta_etal_2021}.
Together, these ancillary data present the kpc-scale gas and stellar mass distributions, star formation activities, kinematic properties, and morphological structures in the host galaxy (see Figure~\ref{fig:schematic} for a schematic summary).

Following the cross-spatial-scale analysis framework developed by \citet{Sun_etal_2020b}, we divide the sky footprint of each target galaxy into a series of hexagonal apertures, each \apersize\ in size (Figure~\ref{fig:image}).
We aggregate cloud-scale molecular gas measurements within each aperture, and then calculate the mass-weighted, population-averaged properties.
We also compile measurements of host galaxy properties as area-weighted averages across the $\sim$kpc-scale apertures.
Our analysis covers \apernumtot\ apertures in total, and \apernumwpix\ apertures with \emph{both} cloud population measurements \emph{and} host galaxy measurements.
We publish these rich multiwavelength measurements online in machine-readable formats (see Appendix~\ref{apdx:mrt}).

Utilizing these databases, we construct basic statistical profiles for both the molecular cloud populations probed by the PHANGS--ALMA survey and the kpc-scale sub-galactic environments they inhabit.
We quantify empirical correlations between cloud population properties and host galaxy environmental metrics.
We further perform a data-driven variable selection technique and identify a small subset of environmental metrics as primary predictors of the cloud population statistics.
Our main findings are as follows:

\begin{enumerate}[itemsep=0.5em,leftmargin=1em,parsep=0em,partopsep=0.0em]

\item The PHANGS--ALMA survey samples a wide range of host galaxy local properties.
This is illustrated by the broad distributions of galactocentric radii, orbital kinematic properties, as well as kpc-scale gas, stellar, and star formation rate surface densities among all the kpc-scale averaging apertures (Figure~\ref{fig:hist_env} and Table~\ref{tab:stats_env}).
Judged purely by aperture number counts, the most typical sub-galactic environment in our sample closely resembles the Solar Neighborhood.
In comparison, most molecular gas mass is hosted in apertures with higher gas, stellar, and SFR surface densities, which are likely not matched by any kpc-scale region in the Milky Way.

\item Molecular cloud populations vary substantially across kpc-scale regions.
This is seen in population-averaged cloud properties such as mass, surface density, velocity dispersion, and turbulent pressure from aperture to aperture (Figure~\ref{fig:hist_MC} and Table~\ref{tab:stats_gmc}).
These population-averaged measurements have been corrected for the effect of galaxy inclination and finite data sensitivity with novel methods.
We conclude that variations of cloud properties within and among galaxies are not merely random scatter from cloud to cloud, but reflect systematic change across sub-galactic environments.

\item Cloud population average properties appear significantly correlated with many local and global host galaxy properties (Figure~\ref{fig:corr_all}).
The sense of these correlations indicate that cloud populations with higher average mass, surface density, and turbulence strength prefer galactic environments at smaller galactocentric radii, higher gas, star, and SFR surface densities, shorter orbital period, and stronger shear.
Similar trends are also present with global galaxy properties.

\item 
Our BIC-based variable selection analysis yields a set of power-law predictive models that capture the most prominent trends for each cloud population-averaged property (Table~\ref{tab:models}).
The small number of independent variables appeared in these models suggests that most cloud--environment correlations can be reduced to the primary dependencies on a few local environmental metrics, especially on the kpc-scale molecular gas and SFR surface densities.
The absence of global galaxy properties in these predictive models suggests that the correlations between molecular clouds and their local kpc-scale environment are more fundamental, and that galaxy-to-galaxy variations might arise merely as their consequences.

\end{enumerate}
\vspace{0.1\baselineskip}

The rich multiwavelength measurements derived in this work have broad applications.
We demonstrate one application scenario by deriving and comparing a set of characteristic timescales relevant to molecular cloud evolution and star formation (Figure~\ref{fig:timescales} and Table~\ref{tab:stats_timescales}).
This further inquiry leads to the following findings:

\begin{enumerate}[itemsep=0.5em,leftmargin=1em,parsep=0em,partopsep=0.0em]
\setcounter{enumi}{4}

\item The molecular cloud population average free-fall time and turbulent crossing time are around $5{-}20~\ut$, comparable to typical molecular cloud lifetimes estimated in a subset of our target galaxies \citep{Kruijssen_etal_2019a,Chevance_etal_2020a,Chevance_etal_2021,KimJY_etal_2020}.
These results support the notion that, when averaged across co-spatial populations, typical molecular clouds have virial parameters of order unity and only live for a few ``internal'' dynamical timescales.

\item The characteristic timescales of galactic-scale dynamical processes (including orbital revolution, shearing, and cloud--cloud collision) are around $100~\ut$, or about an order of magnitude longer than the cloud ``internal'' timescales or their estimated lifetimes.
This contrast seems to suggest that galactic dynamical processes would have to be highly efficient to have a pronounced impact on molecular clouds throughout their short lifetime.

\item The molecular gas depletion time ranges $1{-}3$~Gyr across our sample, implying star formation efficiencies of $0.5{-}1\%$ per cloud free-fall time or crossing time, ${\sim}1\%$ per cloud lifetime, and ${\sim}10\%$ per orbital time or cloud--cloud collision time.

\item Among all molecular cloud internal timescales and galactic dynamical timescales we considered, only orbital time and shearing time show statistically significant (yet weak) correlations with the molecular gas depletion time (Figure~\ref{fig:scatter_timescales}).

\end{enumerate}
\vspace{0.1\baselineskip}

Our rich multiwavelength measurements have already supported multiple observational studies on PHANGS galaxies.
These studies cover a broad range of topics, including the dynamical equilibrium of the ISM \citep{Sun_etal_2020b}, pressures in \HII\ regions \citep{Barnes_etal_2021}, morphological features in the stellar disks \citep{Querejeta_etal_2021}, nuclear gas outflows \citep{Stuber_etal_2021}, and the molecular gas--star formation cycle \citep{Pan_etal_2022}.
We also expect similar applications in future studies on PHANGS targets examining molecular cloud lifecycle (J.~Kim et al.\ in preparation), molecular cloud star formation efficiency per free-fall time (J.~Sun et al.\ in preparation), and galaxy disk global instabilities (T.~Williams et al.\ in preparation).

Beyond the projects mentioned above, we expect these databases to be useful for many purposes, and we highlight as few of them here.
(1) Our cloud population measurements can be directly compared to similar measurements in other types of galaxies, such as dwarf galaxies \citep{Mizuno_etal_2001a,Leroy_etal_2006,Schruba_etal_2017,Imara_Faesi_2019}, starburst galaxies \citep{Ueda_etal_2012,Brunetti_etal_2021,Krieger_etal_2021}, bulge-dominated early-type galaxies \citep{Utomo_etal_2015,Espada_etal_2019,LiuLJ_etal_2021}, or even lensed, high-$z$ galaxies \citep{Dessauges-Zavadsky_etal_2019}.
Such comparisons could highlight commonalities and differences among star-forming environments across the universe.
(2) Both our database and our power-law models can be used to predict molecular cloud properties in other samples of star-forming galaxies with only kpc-resolution data \citep[e.g.,][]{Bolatto_etal_2017,Sorai_etal_2019,LinLH_etal_2020}.
(3) Our databases provide a comprehensive set of initial conditions and outcome properties for benchmarking numerical simulations of the cold interstellar gas at high spatial resolution \citep[e.g.,][]{Benincasa_etal_2013,Kim_Ostriker_2017,Dobbs_etal_2019,Jeffreson_etal_2020,LiH_etal_2020,Tress_etal_2020b}.
(4) Our measurements allow for crucial tests of analytical star formation theories \citep[e.g.,][]{Krumholz_McKee_2005,Hennebelle_Chabrier_2011,Padoan_Nordlund_2011,Federrath_Klessen_2012,Krumholz_etal_2018,Burkhart_Mocz_2019,Orr_etal_2021a}, as well as empirical calibrations of ``sub-grid star formation recipes'' in galaxy evolution models \citep[e.g.,][]{Olsen_etal_2017,Vallini_etal_2018,Popping_etal_2019a}.

We plan to keep maintaining and improving the databases, thereby making them a long-term reference for the community.
In particular, crucial next steps will come from incorporating measurements of ionised gas and stellar populations from the PHANGS--MUSE survey \citep{Emsellem_etal_2021}, as well as star clusters from the PHANGS--HST survey \citep{Lee_etal_2021}.
Future versions of these databases will be released at the same location online as indicated in Appendix~\ref{apdx:mrt}.


\vspace{\baselineskip}
{
This work was carried out as part of the PHANGS collaboration.
The work of JS is partially supported by the Natural Sciences and Engineering Research Council of Canada (NSERC) through the Canadian Institute for Theoretical Astrophysics (CITA) National Fellowship.
The work of JS, AKL, and DU is partially supported by the National Science Foundation (NSF) under Grants No.~1615105, 1615109, and 1653300.
The work of JS and AKL is partially supported by the National Aeronautics and Space Administration (NASA) under ADAP grants NNX16AF48G and NNX17AF39G.
EWR acknowledges the support of the NSERC, funding reference number RGPIN-2017-03987.
EWK acknowledges support from the Smithsonian Institution as a Submillimeter Array Fellow. 
GAB gratefully acknowledges support by the ANID BASAL project FB210003.
HAP acknowledges funding from the European Research Council (ERC) under the European Union’s Horizon 2020 research and innovation programme (grant agreement No.~694343) and the Ministry of Science and Technology of Taiwan under grant 110-2112-M-032-020-MY3.
JP acknowledges support by the Programme National ``Physique et Chimie du Milieu Interstellaire'' (PCMI) of CNRS/INSU with INC/INP, co-funded by CEA and CNES.
MQ acknowledges support from the Spanish grant PID2019-106027GA-C44, funded by MCIN/AEI/10.13039/501100011033.
AS is supported by an NSF Astronomy and Astrophysics Postdoctoral Fellowship under award AST-1903834.
TGW acknowledges funding from the ERC under the European Union’s Horizon 2020 research and innovation programme (grant agreement No.~694343).
ATB and FB would like to acknowledge funding from the ERC under the European Union’s Horizon 2020 research and innovation programme (grant agreement No.~726384/Empire).
MB gratefully acknowledges support by the ANID BASAL project FB210003.
MC gratefully acknowledges funding from the Deutsche Forschungsgemeinschaft (DFG) in the form of an Emmy Noether Research Group (grant number CH2137/1-1).
MC and JMDK gratefully acknowledge funding from the DFG in the form of an Emmy Noether Research Group (grant number KR4801/1-1) and the DFG Sachbeihilfe (grant number KR4801/2-1), and from the ERC under the European Union’s Horizon 2020 research and innovation programme via the ERC Starting Grant MUSTANG (grant agreement No.~714907).
SCOG and RSK acknowledge financial support from the DFG via the collaborative research center (SFB 881, Project-ID 138713538) ``The Milky Way System''  (subprojects A1, B1, B2, and B8). They also acknowledge funding from the Heidelberg Cluster of Excellence ``STRUCTURES'' in the framework of Germany’s Excellence Strategy (grant EXC-2181/1, Project-ID 390900948) and from the European Research Council via the ERC Synergy Grant ``ECOGAL'' (grant 855130).
KG is supported by the Australian Research Council through the Discovery Early Career Researcher Award (DECRA) Fellowship DE220100766 funded by the Australian Government.
KK gratefully acknowledges funding from the DFG in the form of an Emmy Noether Research Group (grant number KR4598/2-1, PI Kreckel).
The work of ECO is partially supported by NASA under ATP grant No.~NNX17AG26G.

This paper makes use of the following ALMA data, which have been processed as part of the PHANGS--ALMA \CO21 survey: \\
\noindent ADS/JAO.ALMA\#2012.1.00650.S, \linebreak 
ADS/JAO.ALMA\#2013.1.00803.S, \linebreak 
ADS/JAO.ALMA\#2013.1.01161.S, \linebreak 
ADS/JAO.ALMA\#2015.1.00121.S, \linebreak 
ADS/JAO.ALMA\#2015.1.00782.S, \linebreak 
ADS/JAO.ALMA\#2015.1.00925.S, \linebreak 
ADS/JAO.ALMA\#2015.1.00956.S, \linebreak 
ADS/JAO.ALMA\#2016.1.00386.S, \linebreak 
ADS/JAO.ALMA\#2017.1.00392.S, \linebreak 
ADS/JAO.ALMA\#2017.1.00766.S, \linebreak 
ADS/JAO.ALMA\#2017.1.00886.L, \linebreak 
ADS/JAO.ALMA\#2018.1.01321.S, \linebreak 
ADS/JAO.ALMA\#2018.1.01651.S, \linebreak 
ADS/JAO.ALMA\#2018.A.00062.S. \linebreak 
ALMA is a partnership of ESO (representing its member states), NSF (USA), and NINS (Japan), together with NRC (Canada), NSC and ASIAA (Taiwan), and KASI (Republic of Korea), in cooperation with the Republic of Chile. The Joint ALMA Observatory is operated by ESO, AUI/NRAO, and NAOJ. The National Radio Astronomy Observatory (NRAO) is a facility of NSF operated under cooperative agreement by Associated Universities, Inc (AUI).

This work is based in part on observations made with NSF's Karl~G.~Jansky Very Large Array 
(VLA; project code:
14A-468, 14B-396, 16A-275, 17A-073, 
18B-184). 
VLA is also operated by NRAO.

This work is based in part on observations made with the Australia Telescope Compact Array (ATCA). ATCA is part of the Australia Telescope National Facility, which is funded by the Australian Government for operation as a National Facility managed by CSIRO.

This work is based in part on observations made with the \textit{Spitzer Space Telescope}, which is operated by the Jet Propulsion Laboratory, California Institute of Technology under a contract with NASA.

This work makes use of data products from the \textit{Wide-field Infrared Survey Explorer (WISE)}, which is a joint project of the University of California, Los Angeles, and the Jet Propulsion Laboratory/California Institute of Technology, funded by NASA.

This work is based in part on observations made with the \textit{Galaxy Evolution Explorer (GALEX)}. \textit{GALEX} is a NASA Small Explorer, whose mission was developed in cooperation with the Centre National d'Etudes Spatiales (CNES) of France and the Korean Ministry of Science and Technology. \textit{GALEX} is operated for NASA by the California Institute of Technology under NASA contract NAS5-98034.

This work is based in part on data gathered with the CIS 2.5m Ir\'en\'ee du Pont Telescope and the ESO/MPG 2.2m Telescope at Las Campanas Observatory, Chile.

This work has made use of the NASA/IPAC Infrared Science Archive (IRSA) and the NASA/IPAC Extragalactic Database (NED), which are funded by NASA and operated by the California Institute of Technology. Relevant datasets to this work include \citet{S4G_2010} and \citet{z0MGS_2019}.

We acknowledge the usage of the HyperLeda database\footnote{\url{http://leda.univ-lyon1.fr}} \citep{Makarov_etal_2014} and the SAO/NASA Astrophysics Data System\footnote{\url{http://www.adsabs.harvard.edu}}.
}

\facilities{ALMA, VLA, ATCA, Spitzer, WISE, GALEX, Max Planck:2.2m, Du Pont, IRSA}

\software{
\texttt{NumPy} \citep{NumPy_2020},
\texttt{SciPy} \citep{SciPy_2020},
\texttt{Astropy} \citep{Astropy_2013,Astropy_2018},
\texttt{pandas} \citep{Pandas_1.3.4},
\texttt{scikit-learn} \citep{Scikit-learn_2011},
\texttt{Matplotlib} \citep{Matplotlib_2007},
\texttt{APLpy} \citep{APLpy_2012},
\texttt{MegaTable} \citep{MegaTable_3.0},
\texttt{adstex} (\url{https://github.com/yymao/adstex}).
}


\appendix

\section{Galaxy Sample}
\label{apdx:sample}

\setcounter{table}{0}
\renewcommand\thetable{\thesection\arabic{table}}

We list our galaxy sample in Table~\ref{tab:sample}.

\input{sample}
\vspace{-2\baselineskip}

\onecolumngrid

\section{Prescriptions for the Metallicity and the CO-to-H$_2$ Conversion Factor}
\label{apdx:alphaCO}

\setcounter{equation}{0}

In this work, we adopt empirical relation-based prescriptions to infer a local gas-phase metallicity (Section~\ref{sec:analysis:env}) and its associated CO-to-\Htwo\ conversion factor (Sections~\ref{sec:data:CO:obj} to~\ref{sec:data:CO:pix}) in each kpc-sized aperture.
Here we detail these prescriptions and the rationale behind them.

\subsection{Metallicity}

To account for possible variations of CO-to-\Htwo\ conversion factor across our galaxy sample, a key first step is to get homogeneous and reliable metallicity estimates.
Although extensive compilations of (global and resolved) metallicity measurements for nearby galaxies exist in the literature \citep[e.g.,][]{Pilyugin_etal_2014,DeVis_etal_2019}, we do not yet have a uniform sample of resolved metallicity measurements with the same calibration scheme for all PHANGS--ALMA targets.
In this work, we instead rely on two well-calibrated scaling relations to capture the general trends of metallicity variation across our sample.

We assume a global galaxy mass--metallicity relation \citep{Sanchez_etal_2019} and a fixed radial metallicity gradient within each galaxy \citep{Sanchez_etal_2014}, such that
\begin{align}
\log_{10}\!Z'(\Reff) &= 0.04 + 0.01\,\left(\log_{10}\!\frac{\Mstar}{M_\odot} - 11.5\right)\, \exp\left(-\log_{10}\!\frac{\Mstar}{M_\odot} + 11.5\right)\,, \label{eq:MZR}\\
\log_{10}\!Z'(\rgal) &= \log_{10}\!Z'(\Reff) - 0.1\frac{\rgal}{\Reff}~. \label{eq:Z_grad}
\end{align}
\noindent Here $Z'(\Reff)$ is the local gas-phase abundance at $\rgal=\Reff$ normalized by the solar value [$12+\log\mathrm{(O/H)}=8.69$],
$Z'(\rgal)$ is the normalized abundance at arbitrary $\rgal$,
and $\Mstar$ is the galaxy global stellar mass derived by assuming a Chabrier IMF \citep{Chabrier_2003}.
Note that these scaling relations are appropriate for abundance measurements adopting the O3N2 calibration in \citet{Pettini_Pagel_2004}.

While Equations~\ref{eq:MZR} and~\ref{eq:Z_grad} are identical to the formulae used in \citet{Sun_etal_2020b}, we make two methodological improvements when applying them in this work.
First, we elevate the $\Mstar$ values in Table~\ref{tab:sample} by 0.1~dex before inserting them into Equation~\ref{eq:MZR}.
According to \citet{Sanchez_etal_2019}, this 0.1~dex offset can largely correct for systematic effects due to (a) differences between Salpeter and Chabrier IMFs, and (b) the finite aperture size of their IFU data (see their Appendix~A and Figure~A1).
Second, we estimate $\Reff$ by multiplying the stellar disk scale length, $\Rdisk$, by a factor of $1.68$.
This step mirrors the procedure for deriving $\Reff$ in \citet{Sanchez_etal_2019}.
Overall, these two methodological changes improve the self-consistency of our metallicity prescription.

As a sanity check, we compare our predictions to the observed two-dimensional metallicity distributions in 18 galaxies in the PHANGS--MUSE survey \citep[also see \citealt{Emsellem_etal_2021}]{Williams_etal_2021a}.
Modulo the uncertain translation between different metallicity calibration schemes \citep[i.e., O3N2 versus \mbox{S-cal}; see][]{Pettini_Pagel_2004,Pilyugin_Grebel_2016}, the predictions and the actual measurements show similar median values (within 0.05~dex) across this subsample, although the dynamic range of our predictions appears narrower than the observed range (0.12~dex versus 0.21~dex).

\subsection{\texorpdfstring{CO-to-H\textsubscript{2} Conversion Factor}{CO-to-H2 Conversion Factor}}

The CO-to-\Htwo\ conversion factor, $\alphaCO$, is expected to depend strongly on metallicity \citep[e.g.,][]{Wilson_1995,Arimoto_etal_1996,Israel_1997,Wolfire_etal_2010,Glover_MacLow_2011,Feldmann_etal_2012,Narayanan_etal_2012,Schruba_etal_2012,Amorin_etal_2016,Accurso_etal_2017,Gong_etal_2020}.
Within a star-forming galaxy, $\alphaCO$ can vary by more than a factor of 2 \citep{Leroy_etal_2011,Blanc_etal_2013,Sandstrom_etal_2013},
with part of it attributable to metallicity variations (at least in the low temperature ``outer disk'' regime).
These considerations motivate us to use a metallicity-dependent $\alphaCO$ prescription as a fiducial choice in this work.
Nevertheless, we also consider a few other prescriptions and provide these alternative estimates in the published datasets.

Our fiducial estimate follows the same metallicity-dependent prescription as described in \citet{Sun_etal_2020b}:
\begin{equation}
\frac{\alphaCOline{1}{0}}{\ualphaCO} = 4.35\;Z\spsc{\prime-1.6}~,
\label{eq:alphaCO_S20}
\end{equation}
\noindent where $Z'$ is the predicted local metallicity from Equations~\ref{eq:MZR} and~\ref{eq:Z_grad}.
The adopted power-law slope in Equation~\ref{eq:alphaCO_S20} is motivated primarily by the metallicity-dependent part of the xCOLD~GASS calibration \citep{Accurso_etal_2017}, whereas the normalization is anchored to the Galactic value at solar metallicity \citep[including the gas mass contribution by helium; see][]{Bolatto_etal_2013}.
This prescription gives similar predictions to many other prescriptions in the literature \citep[e.g.,][]{Genzel_etal_2012,Schruba_etal_2012,Amorin_etal_2016} within the metallicity range probed in this work \citep[e.g., see Figure~6 in][]{Accurso_etal_2017}.

Beyond this fiducial $\alphaCO$ prescription, we calculate four alternative prescriptions, following and expanding on \citet{Sun_etal_2020b}.
The first is simply a constant value matching the Galactic average:
\begin{equation}
\frac{\alphaCOline{1}{0}}{\ualphaCO} = 4.35~.
\label{eq:alphaCO_MW}
\end{equation}

The second prescription follows \citet{Narayanan_etal_2012} and infers $\alphaCO$ from both metallicity ($Z'$) and the flux-weighted \CO21\ line intensity ($\brkt{I\sbsc{CO(2{-}1)}}$):
\begin{equation}
\frac{\alphaCOline{1}{0}}{\ualphaCO} 
= 8.5\;Z^{\prime-0.65}\;\min\!\!\left[1,\;1.5\times\left(\frac{\brkt{\I\sbsc{CO(2{-}1)}}}{\uIco}\right)^{-0.32}\right]~.
\label{eq:alphaCO_N12}
\end{equation}
\noindent The above formula is adapted from equation~(11) in \citet{Narayanan_etal_2012} with two notable distinctions.
First, the original formula depends on $\brkt{I\sbsc{CO(1{-}0)}}$, whereas our formula converts that dependence to $\brkt{I\sbsc{CO(2{-}1)}}$ assuming a line ratio of $\Rsub{21}=0.65$ \citep{denBrok_etal_2021,Leroy_etal_2021c}.
Second, we increase the normalization by a factor of 1.36 to correct for helium contribution.

The third prescription follows \citet{Bolatto_etal_2013} and infers $\alphaCO$ from metallicity ($Z'$), molecular cloud surface density (proxied by $\brkt{\Sigsub{mol,\,pix}}$), and kpc-scale total surface density including both gas and stellar mass ($\Sigsub{total}$):
\begin{align}
\frac{\alphaCOline{1}{0}}{\ualphaCO} 
&= 2.9\;\exp\!\left(\frac{40\;\uSig}{Z'\brkt{\Sigsub{mol,\,pix}}}\right)\;\left(\frac{\Sigsub{total}}{100\;\uSig}\right)^{-\gamma}~,\nonumber\\
&\text{with } \gamma=
\begin{cases}
0.5, &\text{if } \Sigsub{total}>100\;\uSig \\
0. &\text{otherwise}
\end{cases}
\label{eq:alphaCO_B13}
\end{align}
\noindent Since calculating $\brkt{\Sigsub{mol,\,pix}}$ and $\Sigsub{total}$ relies on knowing $\alphaCO$ in the first place, we solve for $\alphaCO$ iteratively until the output of Equation~\ref{eq:alphaCO_B13} converges to the assumed value for calculating $\brkt{\Sigsub{mol,\,pix}}$ and $\Sigsub{total}$ \citep[also see Equations~24--26 in][]{Sun_etal_2020b}.

We combine the above prescriptions for $\alphaCOline{1}{0}$ with the adopted \CO21-to-\CO10 line ratio $\Rsub{21}=0.65$ \citep{denBrok_etal_2021,Leroy_etal_2021c} to get the appropriate conversion factor for the \CO21\ line, $\alphaCOline{2}{1}$.
We then apply these $\alphaCOline{2}{1}$ values on a per aperture basis --- that is, we assume a constant conversion factor within the $\sim$kpc scale extent of each averaging aperture.
These treatments largely follow \citet{Sun_etal_2020b}.

As another improvement over \citet{Sun_etal_2020b}, here we add a fourth alternative prescription following \citet{Gong_etal_2020}.
This simulation-motivated prescription considers the dependence on metallicity ($Z'$), CO line integrated intensity ($\ICO$), and the physical beam size ($D\sbsc{beam}$).
It directly predicts the conversion factor for the \CO21\ line without relying on a separately assumed $\Rsub{21}$ value.
The original formula is expressed in number column density convention \citep[i.e., Equation 4b in Table~3 in][]{Gong_etal_2020}:
\begin{equation}
\frac{X\sbsc{CO(2\text{--}1)}}{10^{20}\;\mathrm{cm}^{-2}\,(\uIco)^{-1}}
= 21.1\,Z\spsc{\prime-0.50}\,\left(\frac{I\sbsc{CO(2\text{--}1)}}{\uIco}\right)^{-0.97+0.34\log_{10}\left(\frac{D\sbsc{beam}}{\mathrm{pc}}\right)}\,\left(\frac{D\sbsc{beam}}{\mathrm{pc}}\right)^{-0.41}~.
\label{eq:alphaCO_G20}
\end{equation}
\noindent We convert it into mass surface density units via
\begin{equation}
\frac{\alphaCOline{2}{1}}{\ualphaCO}
= 2.18\,\left(\frac{X\sbsc{CO(2\text{--}1)}}{10^{20}\;\mathrm{cm}^{-2}\,(\uIco)^{-1}}\right)~.
\end{equation}
\noindent Since this prescription is calibrated at $\lesssim100$~pc scales, we derive the $\alphaCOline{2}{1}$ values pixel-by-pixel at 60--150~pc scales and then calculate CO line intensity weighted mean values within the $\sim$kpc-scale averaging apertures.
This is done for all four resolution levels considered in this work (60, 90, 120, and 150~pc).

We include our estimated $\alphaCOline{2}{1}$ values from all aforementioned prescriptions in the published databases.
This allows for easy conversions if the reader wish to adopt an alternative prescription instead of our fiducial choice.
Concretely, our molecular cloud measurements scale with the adopted conversion factor as $\brkt{M}\propto\brkt{\Sigma}\propto\brkt{\Pturb}\propto\alphaCOline{2}{1}$ and
$\brkt{\alphavir}\propto\alphaCOline{2}{1}^{-1}$;
the kpc-scale molecular gas surface density scales as $\Sigmol\propto\alphaCOline{2}{1}$;
the timescale measurements scale as $\bar{t}\sbsc{ff}\propto\alphaCOline{2}{1}^{-0.5}$ and $t\sbsc{dep,\,mol}\propto\alphaCOline{2}{1}$.
The impact of different prescriptions on some of the molecular gas measurements is examined in detailed by \citet{Sun_etal_2020b}.

We expect future works to improve the handling of the CO-to-H$_2$ conversion factor even further.
Particularly, combining a varying $\Rsub{21}$ \citep[either observed directly or predicted based on similar observations;][]{Leroy_etal_2021c} with an $\Rsub{21}$-dependent $\alphaCO$ prescription \citep{Gong_etal_2020} would allow us to better capture the gas excitation temperature variations, especially in galaxy centers where this effect becomes very pronounced.
We also defer a more thorough comparisons between the different $\alphaCO$ prescriptions to a subsequent paper.

\section{Inclination Corrections}
\label{apdx:inclination}

\renewcommand\thefigure{\thesection\arabic{figure}}
\setcounter{figure}{0}

In Section~\ref{sec:data:CO:obj} and \ref{sec:data:CO:pix}, we introduce inclination corrections on several molecular cloud properties measured at 60--150~pc scales.
These corrections represent an important methodological change, as many previous works \citep[including][using the same data set]{Sun_etal_2018,Sun_etal_2020a,Rosolowsky_etal_2021} have assumed spherical geometry for observations on these spatial scales and, consequently, have not applied such corrections.
This spherical approximation assumes that at the resolution scale, the structure of the molecular ISM is isotropic.
While this assumption is common throughout the literature, it is not well tested.
Furthermore, the 60--150 pc resolutions are also comparable to the thickness of the molecular gas disk, which implies anisotropy.
In this appendix, we show that there is an inclination dependence in our measurements and motivate specific functional forms for empirical corrections.

\begin{figure}[htp]
\vspace{-1\baselineskip}
\gridline{
\fig{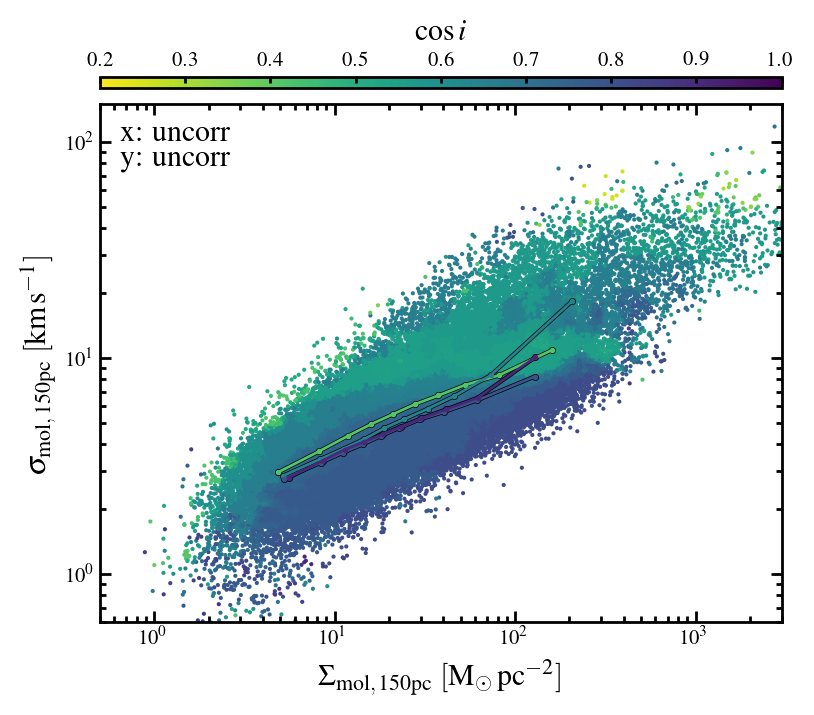}{0.49\textwidth}{}
\fig{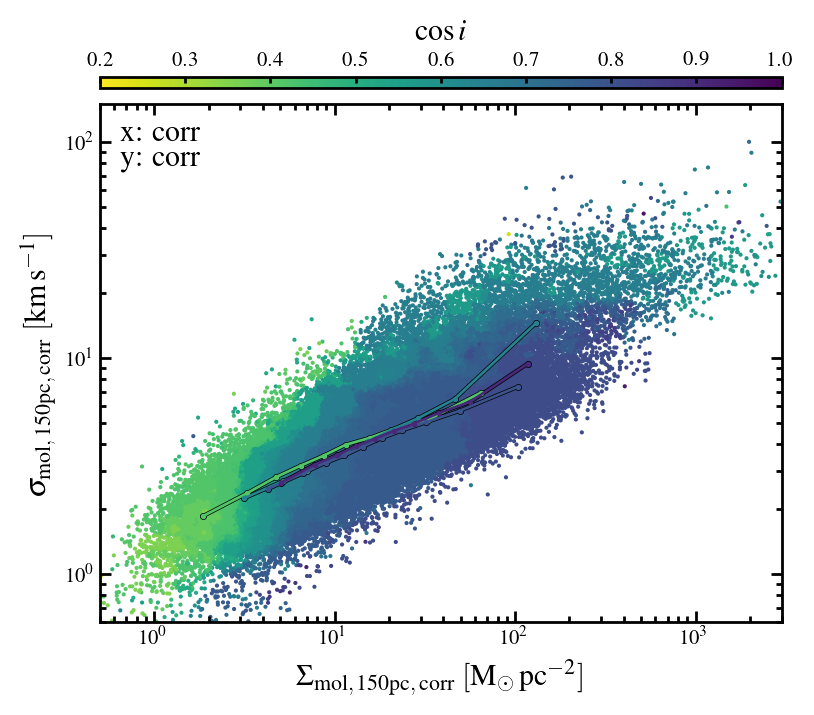}{0.49\textwidth}{}
}
\vspace{-2\baselineskip}
\caption{
The relationship between molecular gas surface density ($\Sigsub{mol}$) and velocity dispersion ($\sigsub{mol}$) measured pixel-by-pixel at 150~pc scales, without and with inclination corrections (\textit{left} and \textit{right}, respectively).
Both plots are made using the published data tables in \citet{Sun_etal_2020a}.
The colors of the data points represent the $\cos{i}$ values of the host galaxies (see colorbar), but have been median-filtered to bring out the overall trend across the parameter space.
The four colored lines in each panel show the running median of $\sigsub{mol}$ at fixed $\Sigsub{mol}$ for galaxies in four different inclination bins.
The left panel (without inclination corrections) reveals a mild but statistically significant trend of elevated $\sigsub{mol}$ at fixed $\Sigsub{mol}$ in galaxies with higher inclination (i.e., smaller $\cos{i}$).
This trend largely disappears in the right panel (with the inclination corrections applied).
}
\vspace{-0.5\baselineskip}
\label{fig:incl}
\end{figure}

The primary motivation for introducing these corrections is that the observed molecular gas properties (such as surface density and velocity dispersion) at $\sim\,$100~pc scales show apparent correlations with the host galaxy inclination angle.
For example, A.~Hughes et al.\ (in preparation) found such trends in the measured properties of CPROPS-identified objects among the PHANGS-ALMA galaxy sample.
A parallel examination of the pixel-by-pixel measurements derived from the same dataset \citep{Sun_etal_2020a} also reveals a similar trend (see Figure~\ref{fig:incl} left panel).
Since inclination angle is not an intrinsic property of galaxies, the presence of these inclination-related trends signifies systematic biases in the molecular gas measurements derived with both CPROPS-based and pixel-by-pixel approaches.

To quantify these biases, we carry out a modified version of the variable selection analysis in Section~\ref{sec:corr:model}.
We suppress the inclination corrections for all molecular cloud properties (i.e., the $\cos{i}$ terms in Equations~\ref{eq:M_obj}--\ref{eq:alphavir_pix}) while keeping $\cos{i}$ in the list of feature variables.
This re-analysis yields a new set of power-law predictive models similar to those in Table~\ref{tab:models}, but many of these models now carry an extra $(\cos{i})^\beta$ term.
Specifically, in the predictive models for $\Sigsub{obj}$ and $\Sigsub{pix}$, the power of the $\cos{i}$ terms are close to $\beta=-1.0$; while in the models for $\sigsub{obj}$ and $\sigsub{pix}$ they are close to $\beta=-0.5$.
In other words, to eliminate the apparent $\cos{i}$ dependence in these models (which should not be present were these models reflecting purely physical trends), one would have to multiply $\cos{i}$ to the measured surface densities and $(\cos{i})^{0.5}$ to the velocity dispersions.
This motivates the correction terms in Equations~\ref{eq:Sigma_obj}, \ref{eq:vdisp_obj}, \ref{eq:Sigma_pix}, and \ref{eq:vdisp_pix}.

We also inspect the effects of these inclination corrections on the original pixel-by-pixel measurements in \citet{Sun_etal_2020a} without doing any cloud population averaging.
Figure~\ref{fig:incl} compares the surface density--velocity dispersion relation before and after applying the inclination corrections.
Without these corrections, the median velocity dispersion at given surface densities appears tends to be higher in galaxies with high inclination angles (smaller $\cos{i}$).
This trend largely disappears when we apply the inclination corrections to both axes.
This result suggests that, despite being motivated by the observed cloud--environment correlations, our adopted inclination corrections can also remove the unphysical inclination dependence in the relationships among molecular cloud internal properties.

Our adopted inclination corrections are empirical and data-driven, but their functional forms have physical implications.
For the surface densities, a multiplicative term of $\cos{i}$ is the exact correction one would use for disk-like structures with their orientations aligned with the whole galaxy disk.
Our proposed interpretation is that the interstellar gas forms filamentary networks, which preferentially align with the large-scale galaxy disk even at $\sim\,$100~pc scales \citep{Zucker_etal_2018}.
This preference in orientation should eventually disappear at smaller scales, but evidently, the PHANGS-ALMA observations do not yet reach the transitional spatial scale.

For the velocity dispersions, there are at least two effects that can produce some inclination dependence:
(1) contributions from ordered, in-plane motions of the gas (e.g., beam smearing), and
(2) anisotropy of the gas velocity dispersion (usually with the in-plane components larger than the vertical component; see e.g., Jeffreson et al.\ submitted).
Both effects would predict higher velocity dispersions at higher inclination angles, which is consistent with the direction of the adopted correction factor, but neither would call for a specific functional form of $(\cos{i})^{0.5}$ for this correction.
Alternatively, if one assumes that the velocity structure of the turbulent gas can be described by a line width--size relation of $\sigmol(l) \propto l^{0.5}$ \citep[e.g.,][]{Solomon_etal_1987}, and that this relationship still holds at 60--150~pc scales, then the varying line-of-sight depth with inclination (i.e., $l \propto 1/\cos{i}$) could imply $\sigmol \propto (\cos{i})^{-0.5}$, which matches our empirical result.

We stress that all the inclination-dependent trends we identify above are real measurements that only emerge in statistical analysis of many galaxies at similarly high resolution.
They imply that the molecular gas structures are clearly anisotropic at 60--150~pc scales.
Existing and future surveys at even higher resolution (e.g., in very nearby galaxies or in CO-bright sub-regions of PHANGS galaxies) can help us extend this analysis to smaller spatial scales, where the transition from anisotropic to isotropic structures presumably occurs.
This transitional spatial scale can be further compared with estimates of the gas disk scale height from independent methods \citep[e.g.,][Jeffreson et al.\ submitted]{Koch_etal_2020}.

\section{Completeness Corrections} 
\label{apdx:completeness}

\setcounter{equation}{0}

\renewcommand\thefigure{\thesection\arabic{figure}}
\setcounter{figure}{0}

\newcommand{\sigmaln}{\sigma\sbsc{int}}
\newcommand{\ICOln}{\bar{I}\sbsc{CO,\,int}}
\newcommand{\ICOth}{I\sbsc{CO,\,th}}
\newcommand{\Fcompl}{f\sbsc{compl}}
\newcommand{\correctI}{F\sbsc{correct,\,\Sigma}}
\newcommand{\correctc}{F\sbsc{correct,\,c}}
\newcommand{\correctt}{F\sbsc{correct,\,t}}

In Section~\ref{sec:analysis:cloud:completeness}, we identify a systematic bias affecting the population-averaged molecular cloud properties.
This systematic bias originates from the incomplete CO flux recovery in the PHANGS--ALMA ``strict'' moment maps and the associated CPROPS catalogs.
We introduce a completeness correction to account for this bias, making use of the measured CO flux completeness, $\Fflux$, and area coverage fraction, $\Farea$, in each averaging aperture.
In this appendix, we present the mathematical derivation of this completeness correction method.

We assume that the intrinsic CO intensity probability distribution function (PDF) within each averaging aperture follows a lognormal distribution:
\begin{equation}
    P(\ICO)
    = \frac{1}{\sqrt{2\pi}\,\sigmaln\,\ICO}\,\exp\left[-\frac{(\ln\ICO-\ln\ICOln)^2}{2\sigmaln^2}\right]~.
    \label{eq:lognormal}
\end{equation}
\noindent This assumption is motivated by observational constraints on the CO intensity distribution in nearby galaxies \citep[e.g.,][]{Hughes_etal_2013b,Leroy_etal_2016,Egusa_etal_2018,Sun_etal_2018,Sun_etal_2020a}.
We further assume that the CO emission included in the ``strict'' moment maps constitutes everything above a threshold intensity $\ICOth$ (i.e., the maps have a sharp, well-defined sensitivity limit), and that the ``strict'' maps capture all emission along the sightlines with CO detections.
Under these assumptions, we can express $\Fflux$ and $\Farea$ in terms of $\sigmaln$, $\ICOln$, and $\ICOth$:
\begin{align}
    \Farea
    &= \int^{+\infty}_{\ICOth} P(\ICO)\,\mathrm{d}\ICO
    = \frac{1}{2}\left[1 - \mathrm{erf}\left(\frac{\ln(\ICOth/\ICOln)}{\sqrt{2}\sigmaln}\right) \right]~, \\
    \Fflux
    &= \frac{\int^{+\infty}_{\ICOth} \ICO\,P(\ICO)\,\mathrm{d}\ICO}{\int^{+\infty}_{0} \ICO\,P(\ICO)\,\mathrm{d}\ICO}
    = \frac{1}{2}\left[1 - \mathrm{erf}\left(\frac{\ln(\ICOth/\ICOln) - \sigmaln^2}{\sqrt{2}\sigmaln}\right) \right]~.
\end{align}
\noindent Notice that these two relations allow us to inversely solve for $\sigmaln$ and $\ln(\ICOth/\ICOln)$ and express their combinations in terms of the measurable quantities $\Farea$ and $\Fflux$:
\begin{align}
    \frac{\ln(\ICOth/\ICOln)}{\sigmaln}
    &= \sqrt{2}\;\mathrm{erf^{-1}}(1-2\Farea)~, \\
    \sigmaln
    &= \sqrt{2}\left[\mathrm{erf^{-1}}(1-2\Farea) - \mathrm{erf^{-1}}(1-2\Fflux)\right]~.
\end{align}
\noindent Here ``$\mathrm{erf^{-1}}$'' stands for the inverse error function.

We then calculate the appropriate correction factors for the population-averaged molecular cloud properties and express them in terms of $\Farea$ and $\Fflux$.
The three population-averaged properties we consider here are the the flux-weighted average cloud surface density, $\brkt{\Sigma}$ (Section~\ref{sec:data:CO:obj}--\ref{sec:data:CO:pix}), the molecular gas clumping factor, $c\sbsc{pix}$ (Section~\ref{sec:analysis:cloud:clumping}), and the flux-weighted average of the reciprocal of free-fall time, $\brkt{t\sbsc{ff}^{-1}}$ (Section~\ref{sec:timescales:method}).
\begin{align}
    \correctI
    \equiv \frac{\brkt{\Sigma}\sbsc{true}}{\brkt{\Sigma}\sbsc{obs}}
    &\propto \frac{\int^{+\infty}_{0} \ICO^2\,P(\ICO)\,\mathrm{d}\ICO}{\int^{+\infty}_{0} \ICO\,P(\ICO)\,\mathrm{d}\ICO}
    \left[\frac{\int^{+\infty}_{\ICOth} \ICO^2\,P(\ICO)\,\mathrm{d}\ICO}{\int^{+\infty}_{\ICOth} \ICO\,P(\ICO)\,\mathrm{d}\ICO}\right]^{-1}
    = \frac{\Fflux}{\Fcompl(2)}~, \label{eq:correction_Sigma}\\
    \correctc
    \equiv \frac{c\sbsc{pix,\,true}}{c\sbsc{pix,\,obs}}
    &\propto \frac{\int^{+\infty}_{0}\ICO^2\,P(\ICO)\,\mathrm{d}\ICO}{\left(\int^{+\infty}_{0}\ICO\,P(\ICO)\,\mathrm{d}\ICO\right)^2}
    \left[\frac{\left(\int^{+\infty}_{\ICOth}\ICO^2\,P(\ICO)\,\mathrm{d}\ICO\right)\left(\int^{+\infty}_{\ICOth}P(\ICO)\,\mathrm{d}\ICO\right)}{\left(\int^{+\infty}_{\ICOth} \ICO\,P(\ICO)\,\mathrm{d}\ICO\right)^2}\right]^{-1} \nonumber\\
    &= \frac{\Fflux^2}{\Farea\,\Fcompl(2)}~, \label{eq:correction_clumping}\\
    \correctt
    \equiv \frac{\brkt{t\sbsc{ff}^{-1}}\sbsc{true}}{\brkt{t\sbsc{ff}^{-1}}\sbsc{obs}}
    &\propto \frac{\int^{+\infty}_{0} \ICO^{3/2}\,P(\ICO)\,\mathrm{d}\ICO}{\int^{+\infty}_{0} \ICO\,P(\ICO)\,\mathrm{d}\ICO}
    \left[\frac{\int^{+\infty}_{\ICOth} \ICO^{3/2}\,P(\ICO)\,\mathrm{d}\ICO}{\int^{+\infty}_{\ICOth} \ICO\,P(\ICO)\,\mathrm{d}\ICO}\right]^{-1}
    = \frac{\Fflux}{\Fcompl(3/2)}~, \label{eq:correction_tff}\\
    \mathrm{where} \quad \Fcompl(\beta)
    &= \frac{\int^{+\infty}_{\ICOth} \ICO^\beta\,P(\ICO)\,\mathrm{d}\ICO}{\int^{+\infty}_{0} \ICO^\beta\,P(\ICO)\,\mathrm{d}\ICO}
    = \frac{1}{2}\left[1 - \mathrm{erf}\left(\frac{\ln(\ICOth/\ICOln) - \beta\sigmaln^2}{\sqrt{2}\sigmaln}\right) \right] \nonumber\\
    &= \frac{1}{2}\left[1 - \mathrm{erf}\Big(\beta\,\mathrm{erf^{-1}}(1-2\Fflux) - (\beta-1)\,\mathrm{erf^{-1}}(1-2\Farea) \Big) \right]~. \nonumber
\end{align}
\noindent Note that the second steps in Equations~\ref{eq:correction_Sigma}--\ref{eq:correction_tff} is valid because we adopt a constant $\alphaCO$ within each aperture.

\begin{figure}[htp]
\vspace{-0.5\baselineskip}
\gridline{
\fig{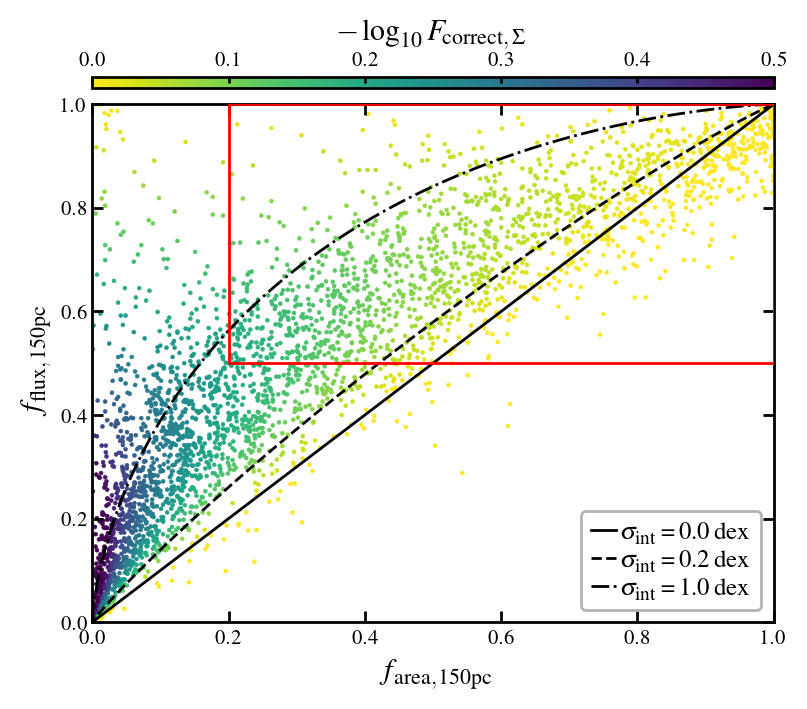}{0.49\textwidth}{}
\fig{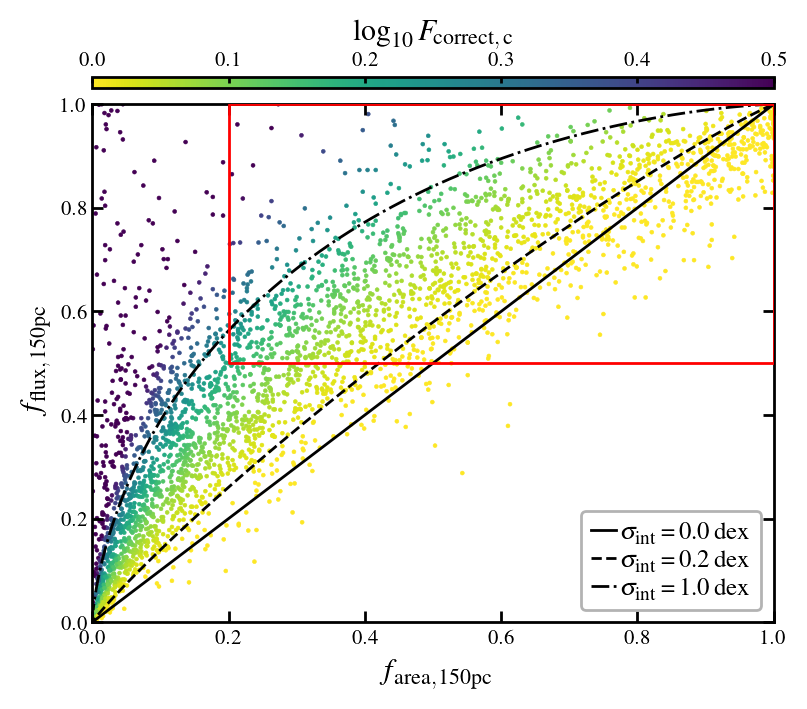}{0.49\textwidth}{}
}
\vspace{-2\baselineskip}
\caption{
Area coverage fraction, $\Farea$, versus CO flux recovery fraction, $\Fflux$, on 150~pc scales within each of the \apernumwpix\ hexagonal apertures.
Black lines show the expected $\Farea$ vs. $\Fflux$ relations for sensitivity-limited CO observations given lognormal-shaped intrinsic $\ICO$-PDFs (different line styles correspond to lognormal PDFs with 1$\sigma$ widths ranging from 0 to 1.0~dex).
The colors of the data points reflect the amplitude of the appropriate correction factors on $\brkt{\Sigma\sbsc{150pc}}$ (\textit{left}) and $c\sbsc{pix,\,150pc}$ (\textit{right}) according to Equations~\ref{eq:correction_Sigma} and~\ref{eq:correction_clumping}.
A red box highlights the parameter space with $\Fflux>0.5$ and $\Farea>0.2$, which are the criteria for selecting the subsample of \apernumfin\ apertures with high fidelity measurements in Section~\ref{sec:corr}.
}
\vspace{-0.5\baselineskip}
\label{fig:completeness}
\end{figure}

Figure~\ref{fig:completeness} shows the joint distribution of $\Farea$ and $\Fflux$ for all \apernumwpix\ hexagonal apertures, with color-codes reflecting the amplitude of the derived correction factors $\correctI$ and $\correctc$.
The $\Farea$ and $\Fflux$ values for $\sim$68\% of apertures are consistent with intrinsic lognormal $\ICO$-PDFs with $\sigmaln$ ranging from 0 to 1.0~dex.
This broadly matches the observed $\ICO$-PDF widths of 0.2 to 0.6~dex in sub-regions of nearby galaxies \citep[e.g.,][]{Hughes_etal_2013b,Sun_etal_2018,Sun_etal_2020a}
The implied $\ICO$ PDF widths for another $\sim$12\% of apertures exceed 1.0~dex, which seem too high to be physical.
These apertures tend to have low $\Farea$, possibly suggesting that these apertures include true ``empty'' areas devoid of CO emission, in which case the lognormal PDF assumption is no longer appropriate.
Finally, the remaining $\sim$20\% of apertures have $\Fflux < \Farea$.
Given the high $\Farea$ in most of these regions, it is likely that there is missing CO flux along sight lines with CO detections.
We calculate the completeness correction for these apertures by assuming an ad-hoc $\Fflux$ value equal to $\Farea$.

The color trends in Figure~\ref{fig:completeness} indicate that apertures with low $\Fflux$ or $\Farea$ would require very significant completeness corrections, which means that the CO detections in these apertures are too ``unrepresentative'' of the underlying cloud population for us to extract reliable statistics.
This motivates us to select a subsample of apertures with high $\Fflux$ \emph{and} high $\Farea$ for the more careful correlation analyses in Section~\ref{sec:corr}.
The selection criteria, $\Fflux > 50\%$ and $\Farea > 20\%$, are also illustrated in Figure~\ref{fig:completeness}.
Among all apertures that meet these criteria, the correction factors on $\brkt{\Sigma}$ have a median value of 0.03~dex and a maximum of 0.2~dex, which means that the uncorrected $\brkt{\Sigma}$ values are already close to the inferred true population average.
For the same set of apertures, the correction factors on $c\sbsc{pix}$ have a smaller median of 0.01~dex but a much larger maximum of 0.6~dex.
The few apertures with very large correction factors are those with high $\Fflux$ but low $\Farea$, where a simple lognormal PDF is likely not a good description of the underlying CO intensity distribution.

\section{Resolution Dependence}
\label{apdx:scale}

\renewcommand\thefigure{\thesection\arabic{figure}}
\setcounter{figure}{0}
\setcounter{equation}{0}

\newcommand{\galnumallscale}{15}
\newcommand{\apernumallscale}{328}

In this work, we derive molecular cloud measurements from the PHANGS--ALMA CO data at four common resolution levels: 60, 90, 120, and 150~pc.
The main text focuses on results derived at 150~pc scale so as to cover the full sample of apertures while keeping the presentation succinct.
In this appendix, we draw comparisons across all four spatial scales to illustrate the scale-dependence of the measured molecular cloud properties.

\begin{figure}[t!]
\gridline{\fig{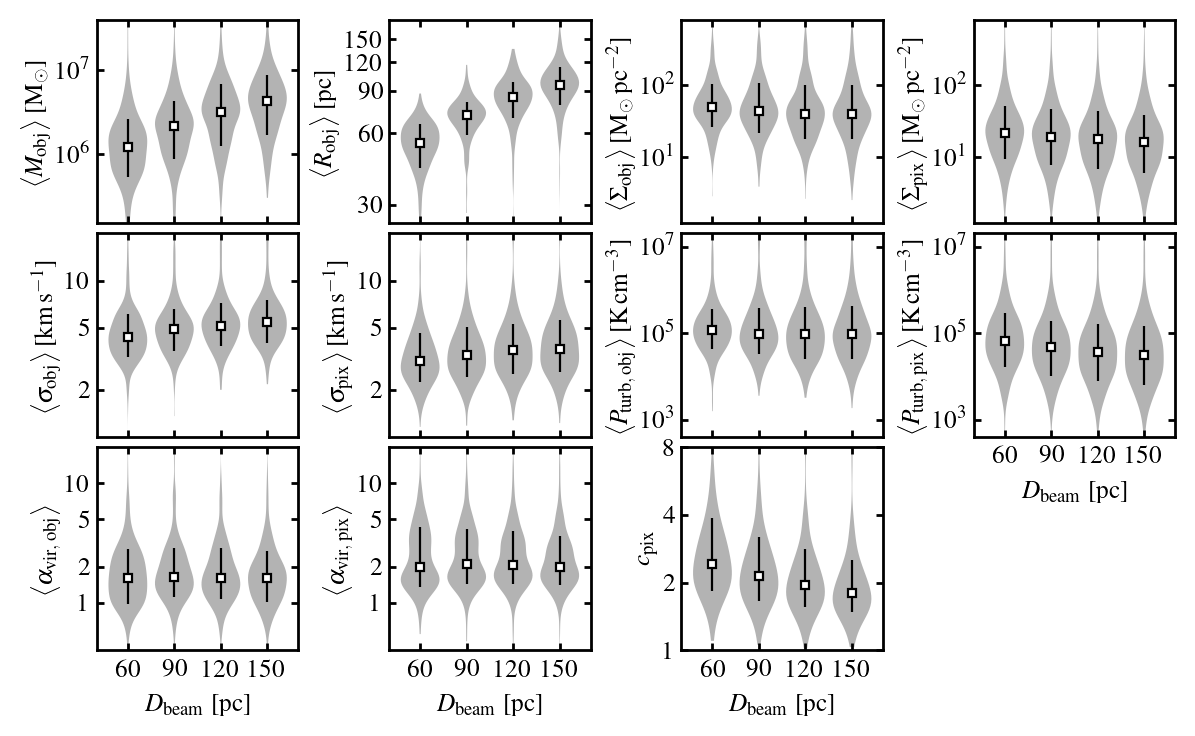}{0.99\textwidth}{}}
\vspace{-2.5\baselineskip}
\caption{
Resolution dependence of the molecular cloud population-averaged properties shown in Figure~\ref{fig:hist_MC}.
In each panel, the four ``violins'' represent histograms of the corresponding cloud properties at 60, 90, 120, and 150~pc resolutions.
These histograms are made from a common set of apertures for which measurements at all four resolutions are available.
The black open square and vertical bar indicate the median value and 16--84 percentile range (assigning equal weight per aperture).
}
\label{fig:resolution}
\end{figure}

The molecular cloud measurements are available for fewer and fewer apertures/galaxies as the resolution goes from 150~pc to 60~pc \citep[see][]{Leroy_etal_2021a}.
To control for changes in the aperture/galaxy sample and any associated selection effects, here we focus on a subset of \apernumallscale\ apertures in \galnumallscale\ galaxies with data at all four spatial scales.
Besides, the data sensitivity also drops as the beam size decreases, which could also leads to systematic biases in our molecular cloud population statistics.
We address this issue by applying completeness corrections to all population-averaged measurements according to the flux completeness and area coverage fraction of the CO moment maps at each resolution (see Section~\ref{sec:analysis:cloud:completeness}).

Figure~\ref{fig:resolution} illustrates that most molecular cloud properties presented in Figure~\ref{fig:hist_MC} show some level of resolution dependence.
In detail, the average molecular cloud mass and radius both increase strongly as the data resolution gets coarser (the former is most likely driven by the latter).
This reinforces the conclusion that the sizes of the CPROPS-identified objects are primarily set by data resolution rather than physical properties of the gas distribution (see Section~\ref{sec:corr:model}).
The slope of the $\brkt{\Rsub{ojb}}$ trend appears sub-linear ($\brkt{\Rsub{obj}} \propto D\sbsc{beam}^{0.6}$), which seems to suggest that the molecular gas structure is not completely scale-free between 60--150~pc.
The average cloud surface density mildly decreases with beam size, as expected from more beam dilution.
The average cloud velocity dispersion increases with beam size, but with a power-law slope of $\sim$0.20--0.23, shallower than the line width--size relation for molecular clouds in the Milky Way disk \citep[e.g.,][]{Solomon_etal_1987}.
The average cloud turbulent pressure mildly declines toward coarser resolution, whereas the average virial parameter largely remains roughly constant.
Both trends are predictable from their functional relations with cloud mass, size, and velocity dispersion.
Finally, as the molecular gas surface density distribution becomes more homogeneous at coarser resolutions, the estimated clumping factor diminishes accordingly.

\section{Machine-readable Tables} 
\label{apdx:mrt}

\renewcommand\thetable{\thesection\arabic{table}}
\setcounter{table}{0}

We publish our high-level measurements in the form of machine-readable data tables via the PHANGS CANFAR storage\footnote{\url{https://www.canfar.net/storage/vault/list/phangs/RELEASES/Sun_etal_2022}\label{canfar}}.
A permanent copy of the version used in this article will also be available via the CANFAR Data Publication Service\footnote{\url{https://doi.org/10.11570/22.0072}}.
This version includes two types of tables.
The first records the hexagonal aperture measurements, and the second presents the radial bin measurements (see Section~\ref{sec:analysis:aperture}).
Here we provide column-by-column descriptions of the hexagonal aperture tables in Table~\ref{tab:database}.
For clarity, we also add links to the relevant sections and equations for each derived quantity.
We do not separately provide column-by-column descriptions for the radial bin tables, but simply note that the set of columns therein are almost identical, except that the radial bin tables lack the \texttt{\detokenize{RA}}, \texttt{\detokenize{DEC}}, and
\texttt{\detokenize{phi_gal}} columns.

We plan to keep improving these data tables and release subsequent versions via the same CANFAR storage\textsuperscript{\ref{canfar}}.
Future versions will cover a larger sample of galaxies and include more measurements derived from other data sets, such as \HII\ regions and stellar populations from the PHANGS--MUSE survey \citep{Emsellem_etal_2021} and star clusters from the PHANGS--HST survey \citep{Lee_etal_2021}.

\input{database}


\bibliography{main.bib}




\allauthors


\end{document}

%% file: stats_env.tex
\begin{deluxetable*}{lcccccc}
\colnumbers
\tabletypesize{\scriptsize}
\tablecaption{Statistics of Galactic Environmental Properties\label{tab:stats_env}}
\tablewidth{0pt}
\tablehead{
\colhead{Quantity} & 
\multicolumn{3}{c}{Weighted by number of apertures} &
\multicolumn{3}{c}{Weighted by molecular gas mass} \\
\cmidrule(lr){2-4} \cmidrule(lr){5-7}
\colhead{} &
\colhead{All} & 
\colhead{Disk} & 
\colhead{Center/bar} & 
\colhead{All} & 
\colhead{Disk} &
\colhead{Center/bar} \\[-2em]
}
\startdata
\\[-2.2em]
$r_\mathrm{gal}$ [$\mathrm{kpc}$] & $5.4^{+3.1}_{-2.7}$ & $5.8^{+3.2}_{-2.4}$ & $3.1^{+2.9}_{-1.5}$ & $3.5^{+3.4}_{-3.5}$ & $5.0^{+3.0}_{-2.1}$ & $1.5^{+2.0}_{-1.5}$ \\
$r_\mathrm{gal}/r_\mathrm{disk}$ & $1.8^{+1.0}_{-0.8}$ & $1.9^{+1.0}_{-0.7}$ & $0.9^{+0.9}_{-0.5}$ & $1.2^{+1.1}_{-1.2}$ & $1.7^{+0.9}_{-0.7}$ & $0.2^{+0.8}_{-0.2}$ \\
$\Omega_\mathrm{circ}$ [$\mathrm{Gyr^{-1}}$] & $34^{+23}_{-9}$ & $32^{+16}_{-8}$ & $53^{+30}_{-21}$ & $45^{+34}_{-16}$ & $37^{+21}_{-11}$ & $66^{+25}_{-25}$ \\
$A_\mathrm{Oort}$ [$\mathrm{km\;s^{-1}\;kpc^{-1}}$] & $15^{+8}_{-4}$ & $14^{+6}_{-4}$ & $18^{+17}_{-6}$ & $18^{+10}_{-6}$ & $17^{+8}_{-5}$ & $21^{+20}_{-6}$ \\
$\Sigma_\star$ [$\mathrm{M_\odot\;pc^{-2}}$] & $77^{+129}_{-42}$ & $65^{+80}_{-34}$ & $190^{+280}_{-110}$ & $200^{+750}_{-130}$ & $110^{+140}_{-60}$ & $550^{+2000}_{-360}$ \\
$\Sigma_\mathrm{mol}$ [$\mathrm{M_\odot\;pc^{-2}}$] & $6.0^{+13.3}_{-4.1}$ & $5.4^{+10.8}_{-3.7}$ & $10^{+24}_{-7}$ & $28^{+118}_{-20}$ & $17^{+29}_{-11}$ & $83^{+449}_{-66}$ \\
$\Sigma_\mathrm{gas}$ [$\mathrm{M_\odot\;pc^{-2}}$] & $13^{+16}_{-7}$ & $13^{+14}_{-7}$ & $14^{+24}_{-9}$ & $35^{+71}_{-21}$ & $28^{+30}_{-15}$ & $71^{+176}_{-52}$ \\
$\Sigma_\mathrm{SFR}$ [$10^{-3}\;\mathrm{M_\odot\;yr^{-1}\;kpc^{-2}}$] & $4.0^{+8.3}_{-2.5}$ & $3.5^{+6.8}_{-2.1}$ & $7.2^{+18.7}_{-5.0}$ & $16^{+92}_{-12}$ & $9.6^{+20.9}_{-6.4}$ & $45^{+469}_{-36}$ \\
$f_\mathrm{gas}$ & $0.14^{+0.13}_{-0.08}$ & $0.16^{+0.14}_{-0.07}$ & $0.07^{+0.07}_{-0.05}$ & $0.16^{+0.13}_{-0.08}$ & $0.18^{+0.13}_{-0.07}$ & $0.11^{+0.08}_{-0.06}$ \\
$f_\mathrm{mol}$ & $0.58^{+0.22}_{-0.31}$ & $0.53^{+0.22}_{-0.30}$ & $0.76^{+0.18}_{-0.27}$ & $0.78^{+0.18}_{-0.21}$ & $0.71^{+0.13}_{-0.20}$ & $0.92^{+0.07}_{-0.15}$ \\
$t_\mathrm{dep,\,mol}$ [$\mathrm{Gyr}$] & $1.5^{+1.1}_{-0.8}$ & $1.5^{+1.0}_{-0.7}$ & $1.5^{+1.3}_{-0.8}$ & $1.7^{+1.0}_{-0.9}$ & $1.8^{+1.0}_{-0.7}$ & $1.6^{+0.9}_{-0.9}$ \\
$t_\mathrm{dep,\,gas}$ [$\mathrm{Gyr}$] & $3.0^{+2.2}_{-1.2}$ & $3.3^{+2.2}_{-1.3}$ & $2.2^{+1.6}_{-1.1}$ & $2.4^{+1.7}_{-1.1}$ & $2.8^{+1.7}_{-1.1}$ & $2.0^{+1.1}_{-1.2}$ \\
\enddata
\tablecomments{
The median value and $\pm1\sigma$ range (i.e., $16{-}84$ percentile range) of the galactic environmental metrics shown in Figure~\ref{fig:hist_env}.
Results in columns (2) to (4) are calculated by weighting each aperture equally, whereas those in columns (5) to (7) are calculated by weighting each aperture by the molecular gas mass it contains.
}
\end{deluxetable*}

%% file: stats_GMC.tex
\begin{deluxetable*}{lcccccccc}
\colnumbers
\tabletypesize{\scriptsize}
\tablecaption{Statistics of Molecular Cloud Population Properties\label{tab:stats_gmc}}
\tablewidth{0pt}
\tablehead{
\colhead{Quantity} & 
\multicolumn{4}{c}{Weighted by number of apertures} &
\multicolumn{4}{c}{Weighted by molecular gas mass} \\
\cmidrule(lr){2-5} \cmidrule(lr){6-9}
\colhead{} &
\colhead{All (uncor.)} & 
\colhead{All} & 
\colhead{Disk} & 
\colhead{Center/bar} & 
\colhead{All (uncor.)} & 
\colhead{All} & 
\colhead{Disk} &
\colhead{Center/bar} \\[-2em]
}
\startdata
\\[-2.2em]
\multicolumn{9}{c}{Object-based Population Statistics (\apernumwobj\ Apertures)} \\
\hline
$\left<M_\mathrm{obj,\,150pc}\right>$ [$10^6\;\mathrm{M_\odot}$] & $4.2^{+6.1}_{-2.3}$ & $3.2^{+6.1}_{-2.0}$ & $2.9^{+4.9}_{-1.8}$ & $4.6^{+12.5}_{-3.0}$ & $11^{+30}_{-7}$ & $10^{+30}_{-7}$ & $6.9^{+10.5}_{-4.5}$ & $27^{+91}_{-21}$ \\
$\left<R_\mathrm{obj,\,150pc}\right>$ [$\mathrm{pc}$] & -- & $92^{+23}_{-21}$ & $91^{+22}_{-22}$ & $96^{+24}_{-19}$ & -- & $95^{+21}_{-16}$ & $90^{+19}_{-15}$ & $100^{+20}_{-20}$ \\
$\left<\Sigma_\mathrm{obj,\,150pc}\right>$ [$\mathrm{M_\odot\,pc^{-2}}$] & $47^{+77}_{-28}$ & $35^{+64}_{-23}$ & $33^{+58}_{-20}$ & $45^{+102}_{-31}$ & $130^{+260}_{-90}$ & $110^{+250}_{-80}$ & $78^{+124}_{-50}$ & $210^{+560}_{-160}$ \\
$\left<\sigma_\mathrm{obj,\,150pc}\right>$ [$\mathrm{km\,s^{-1}}$] & -- & $5.0^{+1.9}_{-1.4}$ & $4.8^{+1.6}_{-1.3}$ & $6.2^{+3.0}_{-1.7}$ & -- & $6.7^{+3.7}_{-1.9}$ & $5.8^{+1.9}_{-1.5}$ & $9.0^{+3.6}_{-3.0}$ \\
$\left<P_\mathrm{turb,\,obj,\,150pc}\right>$ [$10^5\;\mathrm{K\,cm^{-3}}$] & $1.0^{+3.3}_{-0.7}$ & $0.75^{+3.03}_{-0.57}$ & $0.65^{+2.20}_{-0.49}$ & $1.5^{+10.5}_{-1.2}$ & $4.7^{+31.8}_{-3.8}$ & $4.3^{+32.2}_{-3.6}$ & $2.1^{+7.2}_{-1.6}$ & $18^{+68}_{-16}$ \\
$\left<\alpha_\mathrm{vir,\,obj,\,150pc}\right>$ & -- & $1.4^{+1.2}_{-0.6}$ & $1.3^{+1.0}_{-0.6}$ & $1.7^{+1.8}_{-0.8}$ & -- & $1.1^{+0.9}_{-0.5}$ & $1.1^{+0.8}_{-0.4}$ & $1.0^{+1.2}_{-0.6}$ \\
\hline
\multicolumn{9}{c}{Pixel-based Population Statistics (\apernumwpix\ Apertures)} \\
\hline
$\left<\Sigma_\mathrm{pix,\,150pc}\right>$ [$\mathrm{M_\odot\,pc^{-2}}$] & $17^{+29}_{-9}$ & $12^{+27}_{-8}$ & $10^{+21}_{-7}$ & $21^{+64}_{-15}$ & $61^{+316}_{-43}$ & $55^{+294}_{-41}$ & $33^{+60}_{-22}$ & $200^{+840}_{-160}$ \\
$\left<\sigma_\mathrm{pix,\,150pc}\right>$ [$\mathrm{km\,s^{-1}}$] & -- & $3.3^{+2.2}_{-1.0}$ & $3.1^{+1.6}_{-0.9}$ & $4.8^{+6.1}_{-2.0}$ & -- & $5.9^{+14.0}_{-2.5}$ & $4.7^{+2.2}_{-1.7}$ & $16^{+11}_{-10}$ \\
$\left<P_\mathrm{turb,\,pix,\,150pc}\right>$ [$10^5\;\mathrm{K\,cm^{-3}}$] & $0.29^{+1.92}_{-0.22}$ & $0.20^{+1.77}_{-0.16}$ & $0.16^{+0.95}_{-0.13}$ & $0.76^{+16.64}_{-0.68}$ & $3.4^{+230.1}_{-3.1}$ & $3.1^{+216.3}_{-2.8}$ & $1.1^{+5.4}_{-0.9}$ & $65^{+1064}_{-64}$ \\
$\left<\alpha_\mathrm{vir,\,pix,\,150pc}\right>$ & -- & $1.7^{+1.1}_{-0.6}$ & $1.6^{+0.9}_{-0.6}$ & $2.5^{+1.7}_{-0.9}$ & -- & $1.9^{+1.6}_{-0.6}$ & $1.6^{+0.9}_{-0.6}$ & $2.5^{+2.2}_{-1.0}$ \\
$c_\mathrm{pix,\,150pc}$ & $1.4^{+0.4}_{-0.3}$ & $1.9^{+1.1}_{-0.4}$ & $1.9^{+1.0}_{-0.4}$ & $1.9^{+1.2}_{-0.5}$ & $1.7^{+0.6}_{-0.3}$ & $1.9^{+1.0}_{-0.4}$ & $1.8^{+0.8}_{-0.4}$ & $2.1^{+1.2}_{-0.7}$ \\
\enddata
\tablecomments{
The median value and $\pm1\sigma$ range (i.e., $16{-}84$ percentile range) of the population average molecular cloud properties shown in Figure~\ref{fig:hist_MC}.
Results in columns (2) to (5) are calculated by weighting each aperture equally, whereas those in columns (6) to (9) are calculated by weighting each aperture by the molecular gas mass it contains.
Columns (2) and (6) correspond to measurements without applying completeness corrections (see Section~\ref{sec:analysis:cloud:completeness}).
}
\end{deluxetable*}

%% file: lasso_models.tex
\begin{deluxetable*}{lccc}
\tabletypesize{\small}
\tablecaption{Power-law Predictive Models\label{tab:models}}
\tablewidth{0pt}
\tablehead{
\colhead{Model} & 
\colhead{Residual~[dex]} &
\colhead{$R^2$} &
\colhead{$\Delta\mathrm{BIC}$} \\[-1.8em]
}
\startdata
$\frac{\left<M_\mathrm{obj,150pc}\right>}{10^6\rm\;M_\odot}=10^{0.59}\,\left(\frac{\Sigma_\mathrm{mol}}{10\rm\;M_\odot\,pc^{-2}}\right)^{0.75}\,\left(\cos{i}\right)^{-0.50}$ & 0.17 & 0.70 & 678.18 \\
$\frac{\left<R_\mathrm{obj,150pc}\right>}{10\rm\;pc}=10^{0.99}$ & 0.09 & 0.00 & -- \\
$\frac{\left<\Sigma_\mathrm{obj,150pc}\right>}{10^2\rm\;M_\odot\,pc^{-2}}=10^{-0.37}\,\left(\frac{\Sigma_\mathrm{mol}}{10\rm\;M_\odot\,pc^{-2}}\right)^{0.78}\,\left(\frac{\Sigma_\mathrm{SFR}}{10^{-2}\rm\;M_\odot\,yr^{-1}\,kpc^{-2}}\right)^{0.08}$ & 0.19 & 0.74 & 968.54 \\
$\frac{\left<\Sigma_\mathrm{pix,150pc}\right>}{10^2\rm\;M_\odot\,pc^{-2}}=10^{-0.65}\,\left(\frac{\Sigma_\mathrm{mol}}{10\rm\;M_\odot\,pc^{-2}}\right)^{0.77}\,\left(\frac{\Sigma_\mathrm{SFR}}{10^{-2}\rm\;M_\odot\,yr^{-1}\,kpc^{-2}}\right)^{0.13}$ & 0.14 & 0.85 & 1046.80 \\
$\frac{\left<\sigma_\mathrm{obj,150pc}\right>}{10\rm\;km\,s^{-1}}=10^{-0.24}\,\left(\frac{\Sigma_\mathrm{mol}}{10\rm\;M_\odot\,pc^{-2}}\right)^{0.10}\,\left(\frac{\Sigma_\mathrm{SFR}}{10^{-2}\rm\;M_\odot\,yr^{-1}\,kpc^{-2}}\right)^{0.06}$ & 0.10 & 0.32 & 32.82 \\
$\frac{\left<\sigma_\mathrm{pix,150pc}\right>}{10\rm\;km\,s^{-1}}=10^{-0.33}\,\left(\frac{\Sigma_\mathrm{mol}}{10\rm\;M_\odot\,pc^{-2}}\right)^{0.18}\,\left(\frac{\Sigma_\mathrm{SFR}}{10^{-2}\rm\;M_\odot\,yr^{-1}\,kpc^{-2}}\right)^{0.08}$ & 0.11 & 0.52 & 119.87 \\
$\frac{\left<P_\mathrm{turb,obj,150pc}\right>}{10^4\rm\;K\,cm^{-3}}=10^{1.07}\,\left(\frac{\Sigma_\mathrm{mol}}{10\rm\;M_\odot\,pc^{-2}}\right)^{0.93}\,\left(\frac{\Sigma_\mathrm{SFR}}{10^{-2}\rm\;M_\odot\,yr^{-1}\,kpc^{-2}}\right)^{0.28}$ & 0.30 & 0.69 & 764.43 \\
$\frac{\left<P_\mathrm{turb,pix,150pc}\right>}{10^4\rm\;K\,cm^{-3}}=10^{0.87}\,\left(\frac{\Sigma_\mathrm{mol}}{10\rm\;M_\odot\,pc^{-2}}\right)^{1.05}\,\left(\frac{\Sigma_\star}{10^2\rm\;M_\odot\,pc^{-2}}\right)^{0.08}\,\left(\frac{\Sigma_\mathrm{SFR}}{10^{-2}\rm\;M_\odot\,yr^{-1}\,kpc^{-2}}\right)^{0.39}$ & 0.34 & 0.71 & 1104.86 \\
$\left<\alpha_\mathrm{vir,obj,150pc}\right>=10^{0.15}\,\left(\frac{\Sigma_\mathrm{mol}}{10\rm\;M_\odot\,pc^{-2}}\right)^{-0.30}\,\left(\frac{\Omega_\mathrm{circ}}{30\rm\;Gyr^{-1}}\right)^{0.27}\,\left(\frac{A_\mathrm{Oort}}{10\rm\;km\,s^{-1}\,kpc^{-1}}\right)^{0.14}\,\left(\cos{i}\right)^{0.26}$ & 0.20 & 0.35 & 49.23 \\
$\left<\alpha_\mathrm{vir,pix,150pc}\right>=10^{0.25}\,\left(\frac{A_\mathrm{Oort}}{10\rm\;km\,s^{-1}\,kpc^{-1}}\right)^{0.13}$ & 0.19 & 0.09 & 6.72 \\
\enddata
\tablecomments{
The first column lists the power-law predictive models given by the \lasso\ regression and BIC-based model selection (Section~\ref{sec:corr:model}).
The second column shows the residual scatters around these models.
The third column quotes the coefficients of determination, i.e., the fraction of variation in the dependent variable that is explained by the model.
The last column records the BIC difference between the selected model and a null model with only the normalization term.}
\vspace{-1\baselineskip}
\end{deluxetable*}

%% file: stats_timescales.tex
\begin{deluxetable*}{lcccccc}
\colnumbers
\tabletypesize{\scriptsize}
\tablecaption{Statistics of Characteristic Timescales\label{tab:stats_timescales}}
\tablewidth{0pt}
\tablehead{
\colhead{Quantity} & 
\multicolumn{3}{c}{Weighted by number of apertures} &
\multicolumn{3}{c}{Weighted by molecular gas mass} \\
\cmidrule(lr){2-4} \cmidrule(lr){5-7}
\colhead{} &
\colhead{All} & 
\colhead{Disk} & 
\colhead{Center/bar} & 
\colhead{All} & 
\colhead{Disk} &
\colhead{Center/bar} \\[-2em]
}
\startdata
\\[-2.2em]
$\bar{t}_\mathrm{ff,\,obj}$ [Myr] & $9.3^{+4.2}_{-3.4}$ & $9.2^{+4.5}_{-3.3}$ & $9.5^{+3.1}_{-3.9}$ & $6.6^{+3.9}_{-2.0}$ & $6.7^{+3.8}_{-1.9}$ & $6.5^{+3.9}_{-2.2}$ \\
$\bar{t}_\mathrm{ff,\,pix}$ [Myr] & $10.5^{+4.7}_{-3.8}$ & $10.9^{+4.8}_{-3.9}$ & $9.9^{+4.0}_{-4.1}$ & $7.7^{+4.1}_{-2.5}$ & $8.0^{+4.2}_{-2.2}$ & $6.8^{+4.5}_{-2.5}$ \\
$\bar{t}_\mathrm{cr,\,obj}$ [Myr] & $15.6^{+7.1}_{-4.2}$ & $16.1^{+7.0}_{-4.1}$ & $14.0^{+7.3}_{-4.1}$ & $13.1^{+6.6}_{-3.0}$ & $13.8^{+6.8}_{-2.9}$ & $11.5^{+5.4}_{-2.0}$ \\
$\bar{t}_\mathrm{cr,\,pix}$ [Myr] & $15.3^{+6.1}_{-4.6}$ & $15.7^{+6.4}_{-4.1}$ & $13.1^{+4.8}_{-6.2}$ & $12.5^{+4.4}_{-4.1}$ & $13.3^{+4.3}_{-3.1}$ & $9.2^{+6.2}_{-3.6}$ \\
$t_\mathrm{shear}$ [Myr] & $59^{+25}_{-19}$ & $62^{+23}_{-19}$ & $52^{+32}_{-23}$ & $54^{+24}_{-20}$ & $57^{+22}_{-16}$ & $44^{+29}_{-20}$ \\
$t_\mathrm{orb}$ [Myr] & $160^{+70}_{-60}$ & $170^{+70}_{-50}$ & $120^{+70}_{-50}$ & $140^{+60}_{-60}$ & $160^{+60}_{-60}$ & $100^{+60}_{-30}$ \\
$\bar{t}_\mathrm{coll,\,obj}$ [Myr] & $74^{+101}_{-37}$ & $85^{+99}_{-45}$ & $57^{+64}_{-31}$ & $62^{+71}_{-32}$ & $73^{+74}_{-38}$ & $45^{+37}_{-24}$ \\
$\bar{t}_\mathrm{coll,\,pix}$ [Myr] & $90^{+61}_{-36}$ & $92^{+53}_{-35}$ & $77^{+84}_{-29}$ & $80^{+52}_{-31}$ & $84^{+47}_{-32}$ & $68^{+85}_{-27}$ \\
$t_\mathrm{dep,\,mol}$ [Gyr] & $1.8^{+0.9}_{-0.6}$ & $1.7^{+0.8}_{-0.5}$ & $1.9^{+1.0}_{-1.0}$ & $1.9^{+0.9}_{-0.6}$ & $1.8^{+0.8}_{-0.6}$ & $2.0^{+1.0}_{-0.9}$ \\
\enddata
\tablecomments{
The median value and 1$\sigma$ range (i.e., $16{-}84$ percentile range) of the characteristic timescales shown in Figure~\ref{fig:timescales}.
Results in columns (2) to (4) are calculated by weighting each aperture equally, whereas those in columns (5) to (7) are calculated by weighting each aperture by the molecular gas mass it contains.
}
\end{deluxetable*}

%% file: sample.tex
\startlongtable
\begin{deluxetable*}{lDDDDDDDccc}
\colnumbers
\tabletypesize{\footnotesize}
\tablecaption{Galaxy Sample\label{tab:sample}}
\tablewidth{0pt}
\tablehead{
\colhead{Galaxy} & 
\multicolumn2c{$d$} & 
\multicolumn2c{$i$} & 
\multicolumn2c{P.A.} &
\multicolumn2c{$\Riso$} &
\multicolumn2c{$\Rdisk$} &
\multicolumn2c{$M_\star$} &
\multicolumn2c{SFR} &
\colhead{\HI\ Data} & 
\colhead{IR Data} & 
\colhead{UV Data}\\
\colhead{} & 
\multicolumn2c{[$\mathrm{Mpc}$]} & 
\multicolumn2c{[$\deg$]} & 
\multicolumn2c{[$\deg$]} & 
\multicolumn2c{[$\mathrm{kpc}$]} &
\multicolumn2c{[$\mathrm{kpc}$]} &
\multicolumn2c{[$10^{10}\uM$]} &
\multicolumn2c{[$\uM/\mathrm{yr}$]} &
\colhead{} & 
\colhead{} & 
\colhead{}
}
\decimals
\startdata
Circinus & 4.2 & 64.3 & 36.7 & 5.3 & 1.8 & 3.4 & 4.1 & ATCA:LVHIS & WISE & -- \\
IC~1954 & 12.8 & 57.1 & 63.4 & 5.6 & 1.5 & 0.47 & 0.36 & -- & IRAC \& WISE & FUV \& NUV \\
IC~5273 & 14.2 & 52.0 & 234.1 & 6.3 & 1.3 & 0.53 & 0.54 & -- & IRAC \& WISE & FUV \& NUV \\
IC~5332 & 9.0 & 26.9 & 74.4 & 8.0 & 2.8 & 0.47 & 0.41 & -- & IRAC \& WISE & FUV \& NUV \\
NGC~253 & 3.7 & 75.0 & 52.5 & 14.4 & 2.8 & 4.3 & 5.0 & ATCA:LVHIS & IRAC \& WISE & FUV \& NUV \\
NGC~300 & 2.1 & 39.8 & 114.3 & 5.9 & 1.3 & 0.18 & 0.15 & ATCA:LVHIS & IRAC \& WISE & FUV \& NUV \\
NGC~628 & 9.8 & 8.9 & 20.7 & 14.1 & 2.9 & 2.2 & 1.8 & VLA:THINGS & IRAC \& WISE & FUV \& NUV \\
NGC~685 & 19.9 & 23.0 & 100.9 & 8.7 & 3.1 & 1.2 & 0.42 & -- & IRAC \& WISE & -- \\
NGC~1087 & 15.9 & 42.9 & 359.1 & 6.9 & 2.1 & 0.86 & 1.3 & VLA:PHANGS & IRAC \& WISE & FUV \& NUV \\
NGC~1097 & 13.6 & 48.6 & 122.4 & 20.9 & 4.3 & 5.7 & 4.7 & VLA:AH539 & IRAC \& WISE & FUV \& NUV \\
NGC~1300 & 19.0 & 31.8 & 278.0 & 16.4 & 3.7 & 4.1 & 1.2 & VLA:PHANGS & IRAC \& WISE & FUV \& NUV \\
NGC~1317 & 19.1 & 23.2 & 221.5 & 8.5 & 2.4 & 4.2 & 0.48 & -- & WISE & FUV \& NUV \\
NGC~1365 & 19.6 & 55.4 & 201.1 & 34.2 & 13.1 & 9.8 & 17 & -- & IRAC \& WISE & FUV \& NUV \\
NGC~1385 & 17.2 & 44.0 & 181.3 & 8.5 & 2.6 & 0.95 & 2.1 & VLA:PHANGS & IRAC \& WISE & FUV \& NUV \\
NGC~1433 & 18.6 & 28.6 & 199.7 & 16.8 & 6.9 & 7.3 & 1.1 & -- & IRAC \& WISE & FUV \& NUV \\
NGC~1511 & 15.3 & 72.7 & 297.0 & 8.2 & 1.7 & 0.81 & 2.3 & -- & IRAC \& WISE & FUV \& NUV \\
NGC~1512 & 18.8 & 42.5 & 261.9 & 23.1 & 6.2 & 5.2 & 1.3 & VLA:AT285 & IRAC \& WISE & FUV \& NUV \\
NGC~1546 & 17.7 & 70.3 & 147.8 & 9.5 & 2.1 & 2.2 & 0.83 & -- & IRAC \& WISE & FUV \& NUV \\
NGC~1559 & 19.4 & 65.4 & 244.5 & 11.8 & 2.4 & 2.3 & 3.8 & -- & IRAC \& WISE & NUV \\
NGC~1566 & 17.7 & 29.5 & 214.7 & 18.6 & 3.9 & 6.1 & 4.5 & -- & IRAC \& WISE & FUV \& NUV \\
NGC~1637 & 11.7 & 31.1 & 20.6 & 5.4 & 1.8 & 0.88 & 0.64 & VLA:AR351 & IRAC \& WISE & -- \\
NGC~1792 & 16.2 & 65.1 & 318.9 & 13.1 & 2.4 & 4.1 & 3.7 & ATCA:literature & IRAC \& WISE & FUV \& NUV \\
NGC~1809 & 20.0 & 57.6 & 138.2 & 10.9 & 2.4 & 0.59 & 5.7 & -- & IRAC \& WISE & NUV \\
NGC~2090 & 11.8 & 64.5 & 192.5 & 7.7 & 1.7 & 1.1 & 0.41 & -- & WISE & FUV \& NUV \\
NGC~2283 & 13.7 & 43.7 & 355.9 & 5.5 & 1.9 & 0.78 & 0.52 & VLA:PHANGS & WISE & -- \\
NGC~2566 & 23.4 & 48.5 & 312.0 & 14.5 & 4.0 & 5.1 & 8.7 & VLA:PHANGS & WISE & -- \\
NGC~2775 & 23.1 & 41.2 & 156.5 & 14.3 & 4.1 & 12 & 0.87 & VLA:PHANGS & IRAC \& WISE & FUV \& NUV \\
NGC~2835 & 12.2 & 41.3 & 1.0 & 11.4 & 2.2 & 1.0 & 1.2 & VLA:PHANGS & WISE & FUV \& NUV \\
NGC~2903 & 10.0 & 66.8 & 203.7 & 17.4 & 3.5 & 4.3 & 3.1 & VLA:THINGS & IRAC \& WISE & FUV \& NUV \\
NGC~2997 & 14.1 & 33.0 & 108.1 & 21.0 & 4.0 & 5.4 & 4.4 & VLA:PHANGS & WISE & FUV \& NUV \\
NGC~3059 & 20.2 & 29.4 & 345.2 & 11.2 & 3.2 & 2.4 & 2.4 & -- & WISE & -- \\
NGC~3137 & 16.4 & 70.3 & 359.7 & 13.2 & 3.0 & 0.77 & 0.49 & VLA:PHANGS & WISE & FUV \& NUV \\
NGC~3239 & 10.9 & 60.3 & 72.9 & 5.7 & 2.0 & 0.15 & 0.39 & VLA:PHANGS & WISE & FUV \& NUV \\
NGC~3351 & 10.0 & 45.1 & 193.2 & 10.5 & 2.1 & 2.3 & 1.3 & VLA:THINGS & IRAC \& WISE & FUV \& NUV \\
NGC~3489 & 11.9 & 63.7 & 70.0 & 5.9 & 1.4 & 1.9 & 0.023 & -- & IRAC \& WISE & FUV \& NUV \\
NGC~3507 & 23.5 & 21.7 & 55.8 & 10.0 & 2.3 & 2.5 & 0.99 & VLA:PHANGS & IRAC \& WISE & FUV \& NUV \\
NGC~3511 & 13.9 & 75.1 & 256.8 & 12.2 & 2.4 & 1.1 & 0.81 & VLA:PHANGS & IRAC \& WISE & FUV \& NUV \\
NGC~3521 & 13.2 & 68.8 & 343.0 & 16.0 & 4.9 & 11 & 3.7 & VLA:THINGS & IRAC \& WISE & FUV \& NUV \\
NGC~3596 & 11.3 & 25.1 & 78.4 & 6.0 & 2.0 & 0.45 & 0.30 & VLA:EveryTHINGS & IRAC \& WISE & NUV \\
NGC~3599 & 19.9 & 23.0 & 41.9 & 6.9 & 2.0 & 1.1 & 0.047 & -- & IRAC \& WISE & FUV \& NUV \\
NGC~3621 & 7.1 & 65.8 & 343.8 & 9.8 & 2.0 & 1.1 & 0.99 & VLA:THINGS & WISE & FUV \& NUV \\
NGC~3626 & 20.0 & 46.6 & 165.2 & 8.6 & 2.1 & 2.9 & 0.21 & VLA:AJ255 & IRAC \& WISE & NUV \\
NGC~3627 & 11.3 & 57.3 & 173.1 & 16.9 & 3.7 & 6.8 & 3.8 & VLA:THINGS & IRAC \& WISE & FUV \& NUV \\
NGC~4254 & 13.1 & 34.4 & 68.1 & 9.6 & 1.8 & 2.7 & 3.1 & VLA:HERACLES & IRAC \& WISE & FUV \& NUV \\
NGC~4293 & 15.8 & 65.0 & 48.3 & 14.3 & 2.8 & 3.2 & 0.51 & VLA:VIVA & IRAC \& WISE & FUV \& NUV \\
NGC~4298 & 14.9 & 59.2 & 313.9 & 5.5 & 1.6 & 1.0 & 0.46 & VLA:VIVA & IRAC \& WISE & FUV \& NUV \\
NGC~4303 & 17.0 & 23.5 & 312.4 & 17.0 & 3.1 & 3.3 & 5.3 & VLA:AW536 & IRAC \& WISE & FUV \& NUV \\
NGC~4321 & 15.2 & 38.5 & 156.2 & 13.5 & 3.6 & 5.6 & 3.6 & VLA:HERACLES & IRAC \& WISE & FUV \& NUV \\
NGC~4457 & 15.1 & 17.4 & 78.7 & 6.1 & 2.2 & 2.6 & 0.31 & VLA:VIVA & IRAC \& WISE & FUV \& NUV \\
NGC~4459 & 15.9 & 47.0 & 108.8 & 9.6 & 3.3 & 4.8 & 0.22 & -- & WISE & FUV \& NUV \\
NGC~4476 & 17.5 & 60.1 & 27.4 & 4.3 & 1.2 & 0.65 & 0.040 & -- & WISE & FUV \& NUV \\
NGC~4477 & 15.8 & 33.5 & 25.7 & 8.5 & 2.1 & 3.9 & 0.079 & -- & WISE & FUV \& NUV \\
NGC~4496A & 14.9 & 53.8 & 51.1 & 7.3 & 1.9 & 0.34 & 0.61 & VLA:EveryTHINGS & IRAC \& WISE & FUV \& NUV \\
NGC~4535 & 15.8 & 44.7 & 179.7 & 18.7 & 3.8 & 3.4 & 2.2 & VLA:VIVA & IRAC \& WISE & FUV \& NUV \\
NGC~4536 & 16.2 & 66.0 & 305.6 & 16.7 & 2.7 & 2.5 & 3.4 & VLA:HERACLES & IRAC \& WISE & FUV \& NUV \\
NGC~4540 & 15.8 & 28.7 & 12.8 & 5.0 & 1.4 & 0.61 & 0.17 & VLA:PHANGS & IRAC \& WISE & FUV \& NUV \\
NGC~4548 & 16.2 & 38.3 & 138.0 & 13.1 & 3.0 & 4.9 & 0.52 & VLA:VIVA & IRAC \& WISE & FUV \& NUV \\
NGC~4569 & 15.8 & 70.0 & 18.0 & 20.9 & 4.3 & 6.4 & 1.3 & VLA:HERACLES & IRAC \& WISE & FUV \& NUV \\
NGC~4571 & 14.9 & 32.7 & 217.5 & 7.7 & 2.0 & 1.2 & 0.29 & VLA:PHANGS & IRAC \& WISE & FUV \& NUV \\
NGC~4596 & 15.8 & 36.6 & 120.0 & 9.0 & 3.8 & 3.9 & 0.11 & -- & IRAC \& WISE & FUV \& NUV \\
NGC~4689 & 15.0 & 38.7 & 164.1 & 8.3 & 3.0 & 1.6 & 0.40 & VLA:VIVA & IRAC \& WISE & -- \\
NGC~4731 & 13.3 & 64.0 & 255.4 & 12.2 & 3.0 & 0.30 & 0.60 & VLA:PHANGS & IRAC \& WISE & FUV \& NUV \\
NGC~4781 & 11.3 & 59.0 & 290.0 & 6.1 & 1.1 & 0.44 & 0.48 & VLA:PHANGS & IRAC \& WISE & FUV \& NUV \\
NGC~4826 & 4.4 & 59.1 & 293.6 & 6.7 & 1.1 & 1.7 & 0.20 & VLA:THINGS & IRAC \& WISE & FUV \& NUV \\
NGC~4941 & 15.0 & 53.4 & 202.2 & 7.3 & 2.2 & 1.5 & 0.44 & VLA:AM384 & IRAC \& WISE & FUV \& NUV \\
NGC~4951 & 15.0 & 70.2 & 91.2 & 6.9 & 1.9 & 0.62 & 0.35 & VLA:PHANGS & IRAC \& WISE & FUV \& NUV \\
NGC~5042 & 16.8 & 49.4 & 190.6 & 10.2 & 2.4 & 0.80 & 0.60 & VLA:PHANGS & IRAC \& WISE & FUV \& NUV \\
NGC~5068 & 5.2 & 35.7 & 342.4 & 5.7 & 1.3 & 0.25 & 0.28 & VLA:PHANGS & IRAC \& WISE & FUV \& NUV \\
NGC~5128 & 3.7 & 45.3 & 32.2 & 13.7 & 4.1 & 9.4 & 1.2 & -- & WISE & FUV \& NUV \\
NGC~5134 & 19.9 & 22.7 & 311.6 & 7.9 & 2.1 & 2.6 & 0.45 & VLA:PHANGS & IRAC \& WISE & FUV \& NUV \\
NGC~5236 & 4.9 & 24.0 & 225.0 & 9.7 & 2.4 & 3.4 & 4.2 & VLA:THINGS & IRAC \& WISE & FUV \& NUV \\
NGC~5248 & 14.9 & 47.4 & 109.2 & 8.8 & 2.0 & 2.5 & 2.3 & VLA:AS787 & IRAC \& WISE & FUV \& NUV \\
NGC~5530 & 12.3 & 61.9 & 305.4 & 8.6 & 1.7 & 1.2 & 0.33 & -- & WISE & -- \\
NGC~5643 & 12.7 & 29.9 & 318.7 & 9.7 & 1.6 & 2.2 & 2.6 & -- & WISE & -- \\
NGC~6300 & 11.6 & 49.6 & 105.4 & 9.0 & 2.1 & 2.9 & 1.9 & ATCA:literature & WISE & -- \\
NGC~6744 & 9.4 & 52.7 & 14.0 & 21.4 & 4.8 & 5.3 & 2.4 & -- & WISE & FUV \& NUV \\
NGC~7456 & 15.7 & 67.3 & 16.0 & 9.4 & 2.9 & 0.44 & 0.37 & -- & IRAC \& WISE & FUV \& NUV \\
NGC~7496 & 18.7 & 35.9 & 193.7 & 9.1 & 1.5 & 0.99 & 2.3 & -- & IRAC \& WISE & FUV \& NUV \\
NGC~7743 & 20.3 & 37.1 & 86.2 & 7.7 & 1.9 & 2.3 & 0.21 & -- & IRAC \& WISE & FUV \& NUV \\
NGC~7793 & 3.6 & 50.0 & 290.0 & 5.5 & 1.1 & 0.23 & 0.27 & VLA:THINGS & IRAC \& WISE & FUV \& NUV \\
\enddata
\tablecomments{
(2) distance \citep{Anand_etal_2020};
(3) inclination angle \citep{Lang_etal_2020};
(4) position angle \citep{Lang_etal_2020};
(5) isophotal radius at 25~mag/arcsec$^2$ in $B$~band (LEDA);
(6) stellar disk scale length \citep{Leroy_etal_2021a};
(7) global stellar mass \citep{Leroy_etal_2021a};
(9) \HI\ data source
(VLA:PHANGS -- A.~Sardone et al., in preparation;
VLA:EveryTHINGS -- I.~Chiang et al., in preparation;
VLA:THINGS -- \citealt{Walter_etal_2008}; 
VLA:HERACLES -- \citealt{Leroy_etal_2009}; 
VLA:VIVA -- \citealt{Chung_etal_2009}; 
ATCA:LVHIS -- \citealt{Koribalski_etal_2018};
ATCA:literature -- \citealt{Murugeshan_etal_2019});
(10) IR data source (\textit{Spitzer} IRAC -- S$^4$G, \citealt{Sheth_etal_2010},
\textit{WISE} -- $z0$MGS, \citealt{Leroy_etal_2019});
(11) UV data source (\textit{GALEX} FUV and NUV -- $z0$MGS, \citealt{Leroy_etal_2019}).
}
\end{deluxetable*}

%% file: database.tex
\startlongtable
\begin{deluxetable}{lll}
\tabletypesize{\scriptsize}
\tablecaption{Column-by-column Descriptions of the Published Machine-readable Tables\label{tab:database}}
\tablewidth{0pt}
\tablehead{
\colhead{Column name} & 
\colhead{Unit} &
\colhead{Description} \\[-1.5em]
}
\startdata
\texttt{\detokenize{ID}} & $\mathrm{}$ & Aperture ID \\
\texttt{\detokenize{RA}} & $\mathrm{{}^{\circ}}$ & Right Ascension of the aperture center (\S\ref{sec:analysis:env}) \\
\texttt{\detokenize{DEC}} & $\mathrm{{}^{\circ}}$ & Declination of the aperture center (\S\ref{sec:analysis:env}) \\
\texttt{\detokenize{r_gal}} & $\mathrm{kpc}$ & Deprojected galactocentric radius (\S\ref{sec:analysis:env}) \\
\texttt{\detokenize{phi_gal}} & $\mathrm{{}^{\circ}}$ & Deprojected azimuthal angle (0 = receding major axis; \S\ref{sec:analysis:env}) \\
\texttt{\detokenize{frac_CO21_center}} & $\mathrm{}$ & Fraction of \CO21\ flux in the central region (\S\ref{sec:data:environ}, \S\ref{sec:analysis:env}) \\
\texttt{\detokenize{frac_CO21_bars}} & $\mathrm{}$ & Fraction of \CO21\ flux in the bar region (\S\ref{sec:data:environ}, \S\ref{sec:analysis:env}) \\
\texttt{\detokenize{V_circ_CO21_URC}} & $\mathrm{km\,s^{-1}}$ & CO-derived circular velocity (Persic+96 model; \S\ref{sec:data:rotcurve}, \S\ref{sec:analysis:env}) \\
\texttt{\detokenize{e_V_circ_CO21_URC}} & $\mathrm{km\,s^{-1}}$ & Statistical error on CO-derived circular velocity \\
\texttt{\detokenize{beta_CO21_URC}} & $\mathrm{}$ & Logarithmic derivative of CO rotation curve (Persic+96 model; \S\ref{sec:data:rotcurve}, \S\ref{sec:analysis:env}) \\
\texttt{\detokenize{e_beta_CO21_URC}} & $\mathrm{}$ & Statistical error on logarithmic derivative of CO rotation curve \\
\texttt{\detokenize{Zprime}} & $\mathrm{}$ & Gas-phase metallicity relative to solar (Eq.~\ref{eq:MZR}--\ref{eq:Z_grad}) \\
\texttt{\detokenize{alpha_CO21_S20}} & $\mathrm{s\,M_{\odot}\,K^{-1}\,km^{-1}\,pc^{-2}}$ & \CO21-to-H$_2$ conversion factor (fiducial; Sun+20; Eq.~\ref{eq:alphaCO_S20}) \\
\texttt{\detokenize{alpha_CO21_N12}} & $\mathrm{s\,M_{\odot}\,K^{-1}\,km^{-1}\,pc^{-2}}$ & \CO21-to-H$_2$ conversion factor (Narayanan+12; Eq.~\ref{eq:alphaCO_N12}) \\
\texttt{\detokenize{alpha_CO21_B13}} & $\mathrm{s\,M_{\odot}\,K^{-1}\,km^{-1}\,pc^{-2}}$ & \CO21-to-H$_2$ conversion factor (Bolatto+13; Eq.~\ref{eq:alphaCO_B13}) \\
\texttt{\detokenize{<alpha_CO21_G20ICO_Xpc>}} & $\mathrm{s\,M_{\odot}\,K^{-1}\,km^{-1}\,pc^{-2}}$ & Flux-weighted mean \CO21-to-H$_2$ conversion factor (Gong+20 $I_\mathrm{CO}$-based @ $X$\tablenotemark{*} pc; Eq.~\ref{eq:alphaCO_G20}) \\
\texttt{\detokenize{Sigma_mol}} & $\mathrm{M_{\odot}\,pc^{-2}}$ & Region-average molecular gas surface density (PHANGS-ALMA; \S\ref{sec:analysis:env}) \\
\texttt{\detokenize{e_Sigma_mol}} & $\mathrm{M_{\odot}\,pc^{-2}}$ & Statistical error on region-average molecular gas surface density \\
\texttt{\detokenize{Sigma_atom}} & $\mathrm{M_{\odot}\,pc^{-2}}$ & Region-average atomic gas surface density (PHANGS-HI; Eq.~\ref{eq:SigHI}, \S\ref{sec:analysis:env}) \\
\texttt{\detokenize{e_Sigma_atom}} & $\mathrm{M_{\odot}\,pc^{-2}}$ & Statistical error on region-average atomic gas surface density \\
\texttt{\detokenize{MtoL_3p4um}} & $\mathrm{M_{\odot}\,L_{\odot}^{-1}}$ & Stellar mass-to-light ratio at 3.4\,$\mu$m (Leroy+21; \S\ref{sec:data:NIR}) \\
\texttt{\detokenize{Sigma_star}} & $\mathrm{M_{\odot}\,pc^{-2}}$ & Region-average stellar mass surface density (fiducial; \S\ref{sec:data:NIR}, \S\ref{sec:analysis:env}) \\
\texttt{\detokenize{e_Sigma_star}} & $\mathrm{M_{\odot}\,pc^{-2}}$ & Statistical error on region-average stellar mass surface density (fiducial) \\
\texttt{\detokenize{Sigma_star_3p6um}} & $\mathrm{M_{\odot}\,pc^{-2}}$ & Region-average stellar mass surface density (3.6\,$\mu$m + varying M/L; Eq.~\ref{eq:Sigstar_3p6um}, \S\ref{sec:analysis:env}) \\
\texttt{\detokenize{e_Sigma_star_3p6um}} & $\mathrm{M_{\odot}\,pc^{-2}}$ & Statistical error on region-average stellar mass surface density (3.6\,$\mu$m + varying M/L) \\
\texttt{\detokenize{Sigma_star_3p4um}} & $\mathrm{M_{\odot}\,pc^{-2}}$ & Region-average stellar mass surface density (3.4\,$\mu$m + varying M/L; Eq.~\ref{eq:Sigstar_3p4um}, \S\ref{sec:analysis:env}) \\
\texttt{\detokenize{e_Sigma_star_3p4um}} & $\mathrm{M_{\odot}\,pc^{-2}}$ & Statistical error on region-average stellar mass surface density (3.4\,$\mu$m + varying M/L) \\
\texttt{\detokenize{Sigma_SFR}} & $\mathrm{M_{\odot}\,yr^{-1}\,kpc^{-2}}$ & Region-average SFR surface density (fiducial; \S\ref{sec:data:z0MGS}, \S\ref{sec:analysis:env}) \\
\texttt{\detokenize{e_Sigma_SFR}} & $\mathrm{M_{\odot}\,yr^{-1}\,kpc^{-2}}$ & Statistical error on region-average SFR surface density (fiducial) \\
\texttt{\detokenize{Sigma_SFR_FUVW4}} & $\mathrm{M_{\odot}\,yr^{-1}\,kpc^{-2}}$ & Region-average SFR surface density (GALEX FUV + WISE4; Eq.~\ref{eq:SigSFR_FUVW4}, \S\ref{sec:analysis:env}) \\
\texttt{\detokenize{e_Sigma_SFR_FUVW4}} & $\mathrm{M_{\odot}\,yr^{-1}\,kpc^{-2}}$ & Statistical error on region-average SFR surface density (GALEX FUV + WISE4) \\
\texttt{\detokenize{Sigma_SFR_NUVW4}} & $\mathrm{M_{\odot}\,yr^{-1}\,kpc^{-2}}$ & Region-average SFR surface density (GALEX NUV + WISE4; Eq.~\ref{eq:SigSFR_NUVW4}, \S\ref{sec:analysis:env}) \\
\texttt{\detokenize{e_Sigma_SFR_NUVW4}} & $\mathrm{M_{\odot}\,yr^{-1}\,kpc^{-2}}$ & Statistical error on region-average SFR surface density (GALEX NUV + WISE4) \\
\texttt{\detokenize{Sigma_SFR_W4ONLY}} & $\mathrm{M_{\odot}\,yr^{-1}\,kpc^{-2}}$ & Region-average SFR surface density (WISE4 only; Eq.~\ref{eq:SigSFR_W4ONLY}, \S\ref{sec:analysis:env}) \\
\texttt{\detokenize{e_Sigma_SFR_W4ONLY}} & $\mathrm{M_{\odot}\,yr^{-1}\,kpc^{-2}}$ & Statistical error on region-average SFR surface density (WISE4 only) \\
\texttt{\detokenize{Sigma_SFR_HaW4}} & $\mathrm{M_{\odot}\,yr^{-1}\,kpc^{-2}}$ & Region-average SFR surface density (H$\alpha$ + WISE4; Eq.~\ref{eq:SigSFR_HaW4}, \S\ref{sec:analysis:env}) \\
\texttt{\detokenize{e_Sigma_SFR_HaW4}} & $\mathrm{M_{\odot}\,yr^{-1}\,kpc^{-2}}$ & Statistical error on region-average SFR surface density (H$\alpha$ + WISE4) \\
\texttt{\detokenize{fracA_CO21_pix_Xpc}} & $\mathrm{}$ & Area filling fraction of \CO21\ detection @ $X$\tablenotemark{*} pc (\S\ref{sec:analysis:cloud:completeness}) \\
\texttt{\detokenize{fracF_CO21_pix_Xpc}} & $\mathrm{}$ & Flux completeness of \CO21\ detection @ $X$\tablenotemark{*} pc (\S\ref{sec:analysis:cloud:completeness}) \\
\texttt{\detokenize{corr_I_CO21_pix_Xpc}} & $\mathrm{}$ & Completeness correction on flux-weighted mean cloud surface density @ $X$\tablenotemark{*} pc (Eq.~\ref{eq:correction_Sigma}) \\
\texttt{\detokenize{corr_c_CO21_pix_Xpc}} & $\mathrm{}$ & Completeness correction on \CO21\ clumping factor @ $X$\tablenotemark{*} pc (Eq.~\ref{eq:correction_clumping}) \\
\texttt{\detokenize{corr_t_ff^-1_pix_Xpc}} & $\mathrm{}$ & Completeness correction on flux-weighted mean reciprocal of free-fall time @ $X$\tablenotemark{*} pc (Eq.~{\ref{eq:correction_tff}}) \\
\texttt{\detokenize{c_CO21_pix_Xpc}} & $\mathrm{}$ & Clumping factor of \CO21\ emission @ $X$\tablenotemark{*} pc (Eq.~\ref{eq:clumping}) \\
\texttt{\detokenize{e_c_CO21_pix_Xpc}} & $\mathrm{}$ & Statistical error on clumping factor of \CO21\ emission @ $X$\tablenotemark{*} pc \\
\texttt{\detokenize{<Sigma_mol_pix_Xpc>}} & $\mathrm{M_{\odot}\,pc^{-2}}$ & Flux-weighted mean molecular gas surface density @ $X$\tablenotemark{*} pc (Eq.~\ref{eq:Sigma_pix}, \S\ref{sec:analysis:cloud}) \\
\texttt{\detokenize{e_<Sigma_mol_pix_Xpc>}} & $\mathrm{M_{\odot}\,pc^{-2}}$ & Statistical error on flux-weighted mean molecular gas surface density @ $X$\tablenotemark{*} pc \\
\texttt{\detokenize{<vdisp_mol_pix_Xpc>}} & $\mathrm{km\,s^{-1}}$ & Flux-weighted mean molecular gas velocity dispersion @ $X$\tablenotemark{*} pc (Eq.~\ref{eq:vdisp_pix}, \S\ref{sec:analysis:cloud}) \\
\texttt{\detokenize{e_<vdisp_mol_pix_Xpc>}} & $\mathrm{km\,s^{-1}}$ & Statistical error on flux-weighted mean molecular gas velocity dispersion @ $X$\tablenotemark{*} pc \\
\texttt{\detokenize{<P_turb_pix_Xpc>}} & $\mathrm{K\,cm^{-3}}$ & Flux-weighted mean molecular gas turbulent pressure @ $X$\tablenotemark{*} pc (Eq.~\ref{eq:Pturb_pix}, \S\ref{sec:analysis:cloud}) \\
\texttt{\detokenize{e_<P_turb_pix_Xpc>}} & $\mathrm{K\,cm^{-3}}$ & Statistical error on flux-weighted mean molecular gas turbulent pressure @ $X$\tablenotemark{*} pc \\
\texttt{\detokenize{<alpha_vir_pix_Xpc>}} & $\mathrm{}$ & Flux-weighted mean virial parameter @ $X$\tablenotemark{*} pc (Eq.~\ref{eq:alphavir_pix}, \S\ref{sec:analysis:cloud}) \\
\texttt{\detokenize{e_<alpha_vir_pix_Xpc>}} & $\mathrm{}$ & Statistical error on flux-weighted mean virial parameter @ $X$\tablenotemark{*} pc \\
\texttt{\detokenize{<t_cross^-1_pix_Xpc>}} & $\mathrm{Myr^{-1}}$ & Flux-weighted mean reciprocal of crossing time @ $X$\tablenotemark{*} pc (Eq.~\ref{eq:t_cr}, \S\ref{sec:analysis:cloud}) \\
\texttt{\detokenize{e_<t_cross^-1_pix_Xpc>}} & $\mathrm{Myr^{-1}}$ & Statistical error on flux-weighted mean reciprocal of crossing time @ $X$\tablenotemark{*} pc \\
\texttt{\detokenize{<t_ff^-1_pix_Xpc>}} & $\mathrm{Myr^{-1}}$ & Flux-weighted mean reciprocal of free-fall time @ $X$\tablenotemark{*} pc (Eq.~\ref{eq:t_ff}, \S\ref{sec:analysis:cloud}) \\
\texttt{\detokenize{e_<t_ff^-1_pix_Xpc>}} & $\mathrm{Myr^{-1}}$ & Statistical error on flux-weighted mean reciprocal of free-fall time @ $X$\tablenotemark{*} pc \\
\texttt{\detokenize{N_obj_Xpc}} & $\mathrm{}$ & Number of CPROPS objects in each aperture @ $X$\tablenotemark{*} pc \\
\texttt{\detokenize{fracF_CO21_obj_Xpc}} & $\mathrm{}$ & Flux completeness of CPROPS objects @ $X$\tablenotemark{*} pc (\S\ref{sec:analysis:cloud:completeness}) \\
\texttt{\detokenize{<M_mol_obj_Xpc>}} & $\mathrm{M_{\odot}}$ & Flux-weighted mean object molecular gas mass @ $X$\tablenotemark{*} pc (Eq.~\ref{eq:M_obj}, \S\ref{sec:analysis:cloud}) \\
\texttt{\detokenize{e_<M_mol_obj_Xpc>}} & $\mathrm{M_{\odot}}$ & Statistical error on flux-weighted mean object molecular gas mass @ $X$\tablenotemark{*} pc \\
\texttt{\detokenize{<Sigma_mol_obj_Xpc>}} & $\mathrm{M_{\odot}\,pc^{-2}}$ & Flux-weighted mean object molecular gas surface density @ $X$\tablenotemark{*} pc (Eq.~\ref{eq:Sigma_obj}, \S\ref{sec:analysis:cloud}) \\
\texttt{\detokenize{e_<Sigma_mol_obj_Xpc>}} & $\mathrm{M_{\odot}\,pc^{-2}}$ & Statistical error on flux-weighted mean object molecular gas surface density @ $X$\tablenotemark{*} pc \\
\texttt{\detokenize{<vdisp_mol_obj_Xpc>}} & $\mathrm{km\,s^{-1}}$ & Flux-weighted mean object velocity dispersion @ $X$\tablenotemark{*} pc (Eq.~\ref{eq:vdisp_obj}, \S\ref{sec:analysis:cloud}) \\
\texttt{\detokenize{e_<vdisp_mol_obj_Xpc>}} & $\mathrm{km\,s^{-1}}$ & Statistical error on flux-weighted mean object velocity dispersion @ $X$\tablenotemark{*} pc \\
\texttt{\detokenize{<R_3d_obj_Xpc>}} & $\mathrm{pc}$ & Flux-weighted mean object 3d radius @ $X$\tablenotemark{*} pc (Eq.~\ref{eq:R_obj}, \S\ref{sec:analysis:cloud}) \\
\texttt{\detokenize{e_<R_3d_obj_Xpc>}} & $\mathrm{pc}$ & Statistical error on flux-weighted mean object 3d radius @ $X$\tablenotemark{*} pc \\
\texttt{\detokenize{<P_turb_obj_Xpc>}} & $\mathrm{K\,cm^{-3}}$ & Flux-weighted mean object molecular gas turbulent pressure @ $X$\tablenotemark{*} pc (Eq.~\ref{eq:Pturb_obj}, \S\ref{sec:analysis:cloud}) \\
\texttt{\detokenize{e_<P_turb_obj_Xpc>}} & $\mathrm{K\,cm^{-3}}$ & Statistical error on flux-weighted mean object molecular gas turbulent pressure @ $X$\tablenotemark{*} pc \\
\texttt{\detokenize{<alpha_vir_obj_Xpc>}} & $\mathrm{}$ & Flux-weighted mean object virial parameter @ $X$\tablenotemark{*} pc (Eq.~\ref{eq:alphavir_obj}, \S\ref{sec:analysis:cloud}) \\
\texttt{\detokenize{e_<alpha_vir_obj_Xpc>}} & $\mathrm{}$ & Statistical error on flux-weighted mean object virial parameter @ $X$\tablenotemark{*} pc \\
\texttt{\detokenize{<t_cross^-1_obj_Xpc>}} & $\mathrm{Myr^{-1}}$ & Flux-weighted mean reciprocal of object crossing time @ $X$\tablenotemark{*} pc (Eq.~\ref{eq:t_cr}, \S\ref{sec:analysis:cloud}) \\
\texttt{\detokenize{e_<t_cross^-1_obj_Xpc>}} & $\mathrm{Myr^{-1}}$ & Statistical error on flux-weighted mean reciprocal of object crossing time @ $X$\tablenotemark{*} pc \\
\texttt{\detokenize{<t_ff^-1_obj_Xpc>}} & $\mathrm{Myr^{-1}}$ & Flux-weighted mean reciprocal of object free-fall time @ $X$\tablenotemark{*} pc (Eq.~\ref{eq:t_ff}, \S\ref{sec:analysis:cloud}) \\
\texttt{\detokenize{e_<t_ff^-1_obj_Xpc>}} & $\mathrm{Myr^{-1}}$ & Statistical error on flux-weighted mean reciprocal of object free-fall time @ $X$\tablenotemark{*} pc \\
\enddata
\tablenotetext{*}{$X$ = 60, 90, 120, and 150.}
\end{deluxetable}